\documentclass[a4paper,onecolumn,11pt,accepted=2023-03-16]{quantumarticle}
\pdfoutput=1

\usepackage[utf8]{inputenc}
\usepackage{graphicx}
\usepackage{color}
\usepackage{physics}
\usepackage{braket}
\usepackage{comment}
\usepackage{mathrsfs}
\usepackage{enumerate}
\usepackage{amsmath,amssymb}
\usepackage[version=4]{mhchem}
\usepackage[breaklinks=true]{hyperref}
\newcommand{\mr}[1]{\mathrm{#1}}
\newcommand{\ms}[1]{\mathscr{#1}}
\newcommand{\mcl}[1]{\mathcal{#1}}
\newcommand{\bbC}{\mathbb{C}}
\newcommand{\bbZ}{\mathbb{Z}}
\newcommand{\bbN}{\mathbb{N}}
\newcommand{\eff}{\mathrm{eff}}
\newcommand{\imax}{\mathrm{max}}

\newcommand{\poly}[1]{\mathrm{poly} \left( #1 \right)}
\usepackage{amsthm}
\theoremstyle{definition}
\newtheorem{theorem}{Theorem}[]

\newtheorem{proposition}[theorem]{Proposition}
\newtheorem{lemma}[theorem]{Lemma}
\newtheorem*{theorem*}{Theorem}
\newtheorem*{proposition2*}{Proposition}

\begin{document}
\title{Optimal Hamiltonian simulation for time-periodic systems}

\author{Kaoru Mizuta}
\email{kaoru.mizuta@riken.jp}
\affiliation{RIKEN Center for Quantum Computing (RQC), Hirosawa 2-1, Wako, Saitama 351-0198, Japan}

\author{Keisuke Fujii}
\affiliation{Graduate School of Engineering Science, Osaka University, 1-3 Machikaneyama, Toyonaka, Osaka 560-8531, Japan.}
\affiliation{Center for Quantum Information and Quantum Biology, Osaka University, Japan.}
\affiliation{RIKEN Center for Quantum Computing (RQC), Hirosawa 2-1, Wako, Saitama 351-0198, Japan}
\affiliation{Fujitsu Quantum Computing Joint Research Division at QIQB,
Osaka University, 1-2 Machikaneyama, Toyonaka 560-0043, Japan}

\maketitle

\begin{abstract}
The implementation of time-evolution operators $U(t)$, called Hamiltonian simulation, is one of the most promising usage of quantum computers.
For time-independent Hamiltonians, qubitization has recently established efficient realization of time-evolution $U(t)=e^{-iHt}$, with achieving the optimal computational resource both in time $t$ and an allowable error $\varepsilon$.
In contrast, those for time-dependent systems require larger cost due to the difficulty of handling time-dependency.
In this paper, we establish optimal/nearly-optimal Hamiltonian simulation for generic time-dependent systems with time-periodicity, known as Floquet systems.
By using a so-called Floquet-Hilbert space equipped with auxiliary states labeling Fourier indices, we develop a way to certainly obtain the target time-evolved state without relying on either time-ordered product or Dyson-series expansion.
Consequently, the query complexity, which measures the cost for implementing the time-evolution, has optimal and nearly-optimal dependency respectively in time $t$ and inverse error $\varepsilon$, and becomes sufficiently close to that of qubitization.
Thus, our protocol tells us that, among generic time-dependent systems, time-periodic systems provides a class accessible as efficiently as time-independent systems despite the existence of time-dependency. 
As we also provide applications to simulation of nonequilibrium phenomena and adiabatic state preparation, our results will shed light on nonequilibrium phenomena in condensed matter physics and quantum chemistry, and quantum tasks yielding time-dependency in quantum computation.

\end{abstract}

\section{Introduction}\label{Sec:Introduction}
Simulating quantum many-body systems is one of the most promising usages of quantum computers that can fully exploit their computational powers, as R. Feynman's proposal in the beginning of quantum computers \cite{feynman1982}.
The implementation of a time-evolution operator under Hamiltonians, called Hamiltonian simulation, has been the most important problem for efficiently and accurately accomplishing quantum simulation.
In fact, the application of Hamiltonian simulation nowadays ranges from condensed matter physics to quantum chemistry, such as quantum dynamics \cite{Lloyd1996-ko,Smith2019-zz,Arute2020-qr} and quantum phase estimation \cite{Yu_Kitaev1995-oi,Du2010-hl,Lanyon2010-wk,OMalley2016-dh}.
For long time, Trotterization has been a standard way of Hamiltonian simulation for both time-independent and time-dependent systems, which can provide simple realization available in today's quantum computers \cite{Lloyd1996-ko,Abrams1997-jb,Sornborger1999-oq,Campbell2019-prl,Childs2021-trotter}.
Instead, it requires a huge number of elementary gates up to $\poly{1/\varepsilon}$ to achieve an acceptable error $\varepsilon$.

For time-independent systems, various efficient Hamiltonian simulation algorithms have appeared in the past decade, which yield fewer resources than Trotterization to implement the time-evolution $U(t)=e^{-iHt}$ \cite{Childs2012-qh,Berry2015-gt,Berry2015-taylor,Berry2017-oblivious,Low2017-QSP,Low2019-qubitization}.
Among them, the qubitization technique \cite{Low2019-qubitization} achieves the best cost in that its query complexity (a measure of cost) has optimal scaling both in time $t$ and inverse error $1/\varepsilon$ in an additive way. \cite{Low2019-qubitization}.
In contrast, for time-dependent systems, while there exist several efficient Hamiltonian simulation for constructing time-evolution $U(t)$ \cite{Low2018-dyson,Kieferova2019-dyson,Berry2020-time-dep,Haah2021-time-dep,Chen2021-time-dep,Watkins2022-time-dep}, they largely rely on discretizing the time.
Although the truncated Dyson-series algorithm \cite{Low2018-dyson,Kieferova2019-dyson}, which is versatile among them, achieves the query complexity optimal or nearly-optimal both in $t$ and $1/\varepsilon$, its scaling is multiplicative with requiring much larger cost than qubitization. 
This originates from the difficulty of efficiently dealing with continuous-time modulation.
It is nontrivial whether we can simulate time-dependent Hamiltonians with much fewer cost close to that of the qubitization.

Among time-dependent Hamiltonians, one of the most important targets of quantum simulation is a time-periodic Hamiltonian satisfying $H(t+T)=H(t)$ with $T$ being a period.
In fact, quantum systems under time-periodic Hamiltonians are called Floquet systems, and have been platforms of various nonequilibrium phenomena in condensed matter physics and quantum chemistry; for instance, they can host nonequilibrium phases of matter absent in time-independent systems, such as topological phases \cite{Kitagawa2010-ct,Rudner2013-yk,Harper2019-pk} and time crystals \cite{Khemani2016-dq,Else2016-mg,Khemani2019-pf,Mei2020-te,Randall2021-ke,Mi2022-lw}.
Time-periodic Hamiltonians also cover optical responses in solids and molecules \cite{Potvliege1989-xk,Faisal1997-mv,Oka2009photovoltaic,Oka2019review}, exemplified by high-harmonic generation and photo-chemical reactions.
In addition, adiabatic quantum dynamics such as Thouless pumping \cite{Thouless1983-xj,Lohse2015-sl,Nakajima2016-tj} and adiabatic state preparation for quantum computation \cite{Aspuru-Guzik2005-nr,Du2010-hl,Albash2018RevModPhys} can be regarded as a part of time-periodic Hamiltonian dynamics under sufficiently large period $T$.
Despite fundamental significance and potential application of time-periodic Hamiltonians in broad fields, there is no Hamiltonian simulation protocol 
which can efficiently handle their time-dependency, while their simulation itself is possible by the truncated Dyson-series algorithm.

In this paper, we develop an optimal and/or nearly-optimal quantum algorithm for time-periodic Hamiltonian simulation.
The key idea of our formalism is mapping time-dependent systems to effective time-independent systems in the infinite-dimensional Floquet-Hilbert space, obtained by the Fourier series expansion in time.
Although the infinite-dimensionality prohibits its simulation on quantum computers, we resolve this by formulating the Lieb-Robinson bound \cite{Lieb1972-uo}, amplitude amplification, and the qubitization technique.
Consequently, we can simulate the target time-evolved state with arbitrary small error and failure probability with a finite-dimensional Hilbert space by preparing ancillary states labeling Fourier indices.

The resulting query complexity for time-periodic Hamiltonian simulation has optimal and nearly-optimal scaling respectively in time $t$ and inverse error $1/\varepsilon$, and significantly, these contributions appear in an additive way. 
This implies that the time-periodic Hamiltonian simulation can be executed with the cost sufficiently close to qubitization, which provides the theoretically best scaling, despite the existence of time-dependency.
We also note that the auxiliary states for Fourier indices accurately reproduce the exact dynamics with fewer degrees of freedom than those for discretized time, which is employed for generic time-dependent systems.
This leads to smaller query complexity and simpler oracles of our algorithm compared to the truncated Dyson-series algorithm.
In addition, we provide simulation of the Fermi-Hubbard model under light and adiabatic state preparation as potential applications. 
Thus, our protocol will shed light on the promising usage of quantum computers for nonequilibrium quantum many-body phenomena in condensed matter physics and quantum chemistry, and the optimal control in quantum computation.

The rest of this paper is organized as follows.
In Section \ref{Sec:Preliminaries}, we briefly review Floquet theory and the qubitization for this paper to be self-contained.
We provide our main results in Sections \ref{Sec:Summary_paper}-\ref{Sec:Discussion}, with firstly summarizing them in Section \ref{Sec:Summary_paper}.
In Section \ref{Sec:truncated_Floquet_Hilbert}, we derive the Lieb-Robinson bound in time-periodic systems, and the way to accurately reproduce the exact dynamics with the Floquet-Hilbert space.
Sections \ref{Sec:Amplification} and \ref{Sec:Block_encoding_effective} are devoted to deriving the subroutines of the algorithm. They respectively provide the amplification protocol that can realize the time-evolved state with sufficiently high success probability (Section \ref{Sec:Amplification}) and efficient implementation of the time-evolution operator in the Floquet-Hilbert space (Section \ref{Sec:Block_encoding_effective}).
Section \ref{Sec:Algorithm_cost} completes the optimal time-periodic Hamiltonian simulation, and compares its resource with other algorithms for time-independent / -dependent Hamiltonian simulation.
We provide some promising applications in Section \ref{Sec:examples}, and concludes our paper in Section \ref{Sec:Discussion}.
\section{Preliminaries}\label{Sec:Preliminaries}

\subsection{Review on Floquet theory}\label{Subsec:preli_Floquet}
\begin{figure}[t]
    \begin{center}
    \includegraphics[height=4.7cm, width=8.5cm]{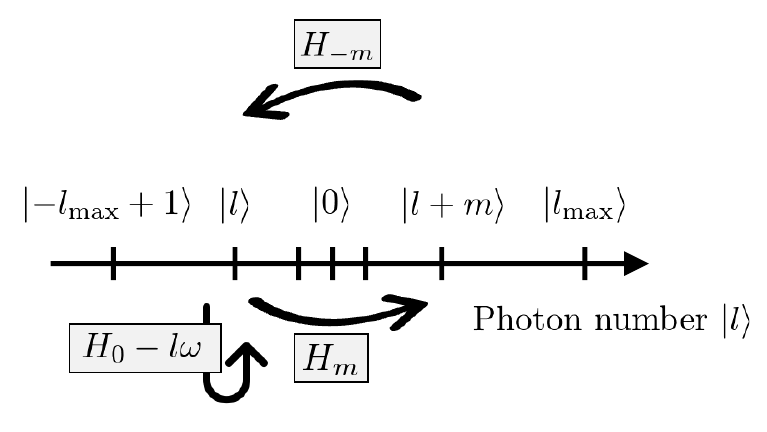}
    \caption{Intuitive understanding of the effective Hamiltonian $\ms{H}_\eff$ [Eq. (\ref{Eq2A:def_effective_Hamiltonian})] in the Floquet-Hilbert space. It can be seen as a single-particle quantum system with the potential $H_0-l\omega$ and the hopping $H_{-m}$ on a infinite one-dimensional chain. We should introduce the cutoffs $\ket{-l_\imax + 1}$ and  $\ket{l_\imax}$ to embed it on quantum computers, which leads to $\ms{H}_\eff^{l_\imax}$ [Eq. (\ref{Eq3A:def_effective_H_l_max})]. } 
    \label{Fig2A:effective_Hamiltonian}
    \end{center}
\end{figure}

We briefly review Floquet theory to analyze Schr\"{o}dinger equation under a time-periodic Hamiltonian,
\begin{equation}
    i \dv{t} \ket{\psi(t)} = H(t) \ket{\psi(t)}, \quad H(t+T) = H(t),
\end{equation}
and its solution
\begin{equation}
    \ket{\psi(t)} = U(t) \ket{\psi(0)}, \quad U(t) = \mcl{T} \exp \left( -i \int_0^t \dd{t^\prime} H(t^\prime) \right).
\end{equation}
Here, the time-evolved state of the system of interest is represented by $\ket{\psi(t)}$, and we assume that it is defined on a finite-dimensional Hilbert space $\mcl{H}$.
The set of states $\{ \ket{\psi_j} \}_{j=1}^{\mr{dim}(\mcl{H})}$ denotes a certain choice of the basis of $\mcl{H}$.
Floquet theorem says that the solution $\ket{\psi(t)}$ is always written in the form of
\begin{equation}
    \ket{\psi(t)} = \sum_{\alpha=1}^{\mr{dim}(\mcl{H})} c_\alpha e^{-i \varepsilon_\alpha t} \ket{\phi_\alpha (t)}, \quad \ket{\phi_\alpha (t+T)}=\ket{\phi_\alpha (t)},
\end{equation}
similar to Bloch theorem for spatially-periodic systems. Here, $\varepsilon_\alpha \in [-\pi/T, \pi/T )$ and $\ket{\phi_\alpha(t)} \in \mcl{H}$ are respectively called quasienergy and Floquet state.
Time-periodicity of $H(t)$ and $\ket{\phi_\alpha(t)}$ allows the Fourier series expansion as
\begin{equation}
    H(t) = \sum_{m \in \bbZ} H_m e^{-im\omega t}, \quad \ket{\phi_\alpha(t)} = \sum_{m \in \bbZ} \ket{\phi_\alpha^m} e^{-im\omega t},
\end{equation}
with the frequency $\omega = 2\pi /T$. The hermiticity of $H(t)$ implies $H_{-m}=H_m^\dagger$. We assume $\norm{H(t)} < \infty$ ($\norm{\cdot}$ ; operator norm), which results in $\sum_{m \in \bbZ} \norm{H_m}^2 < \infty$. Introducing an auxiliary degree of freedom $\{ \ket{l} \}_{l \in \bbZ}$ (sometimes called photon number) to relate $\ket{\phi_\alpha^l} \leftrightarrow \ket{l}\ket{\phi_\alpha^l}$, every pair of $\varepsilon_\alpha, \ket{\phi_\alpha(t)}$ can be obtained by the time-independent eigenvalue equation,
\begin{equation}
    \ms{H}_\eff \left( \sum_{l \in \bbZ} \ket{l} \ket{\phi_\alpha^l} \right) = \varepsilon_\alpha \left( \sum_{l \in \bbZ} \ket{l} \ket{\phi_\alpha^l} \right),
\end{equation}
with the effective Hamiltonian defined by
\begin{equation}\label{Eq2A:def_effective_Hamiltonian}
    \ms{H}_\eff = \sum_{l \in \bbZ} \ket{l}\bra{l} \otimes (H_0-l\omega) + \sum_{l,m \in \bbZ} \ket{l}\bra{l+m} \otimes H_{-m}.
\end{equation}
Intuitively, this time-independent Hamiltonian $\ms{H}_\eff$ describes a static one-dimensional system, where each $l$-th site has potential energy $H_0-l\omega$, as Fig. \ref{Fig2A:effective_Hamiltonian}. 
Off-diagonal terms $H_{-m}$ represent hopping by $-m$ sites, which can be recognized as either emission or absorption of $|m|$ photons. 
We remark that the difficulty of time-dependence does not vanish by this mapping; it is translated into  infinite dimensionality of the space spanned by $\{ \ket{l} \ket{\psi_j} \, | \, l \in \bbZ, \, j=1,2,\hdots, \mr{dim}(\mcl{H}) \}$, which is called Floquet-Hilbert space or Sambe space \cite{Sambe1973}.
While the effective Hamiltonian $\ms{H}_\eff$ is often used for identifying $\varepsilon_\alpha, \ket{\phi_\alpha(t)}$, it can be employed also for directly determining the dynamics as follows \cite{Levante1995-tx};
\begin{equation}\label{Eq2A:psi_t_infinite}
    \ket{\psi(t)} = \sum_{l \in \bbZ} e^{-il\omega t} \braket{l |e^{-i \ms{H}_\eff t}| 0}\ket{\psi(0)}.
\end{equation}
We can extract the dynamics under a time-dependent Hamiltonian $H(t)$ from that under the time-independent one $\ms{H}_\mr{eff}$ by preparing auxiliary systems labeled by $\ket{l}$.
However, we note that applying optimal algorithms for time-independent systems to simulating this extended dynamics is not straightforward, since the Floquet-Hilbert space is infinite-dimensional.

\subsection{Review on qubitization technique}\label{Subsec:preli_qubitization}

Here, we briefly review so-called qubitization technique to efficiently implement $e^{-iHt}$ for time-independent Hamiltonian $H$ on the Hilbert space $\mcl{H}$ \cite{Low2019-qubitization}.
It begins with assuming the existence of block-encoding, that is, we suppose an $n_a$-qubit auxiliary state $\ket{G}_a$ (called an oracle state) and a unitary gate $O$ on the space $\bbC^{2^{n_a}} \otimes \mcl{H}$ (called an oracle gate) satisfying
\begin{equation}\label{Eq2B:block_encode}
    \braket{G|O|G}_a = \frac{H}{\alpha}, \quad \alpha \geq \norm{H}.
\end{equation}
The inequality $\alpha \geq \norm{H}$ comes from the norm of the unitary gate as $\norm{O}=1$, and the parameter $\alpha$ represents a typical energy scale of the whole system.
Here, the oracles $O$ and $\ket{G}_a$ are supposed to be efficiently implemented; the unitary gates $O$ and $G$, which realizes the oracle state from a reference state $\ket{0}_a$ as $G\ket{0}_a = \ket{G}_a$, require at most $C \in \poly{N}$ elementary gates.

The explicit construction of the oracles has been explored for certain classes of static Hamiltonians.
For instance, when the Hamiltonian $H$ is given by a linear combination of unitary (LCU) as
\begin{equation}\label{Eq3B:def_LCU}
    H = \sum_{j=1}^{j_\imax} \alpha_j U_j, \quad \alpha_j \geq 0, \quad \text{$U_j$; unitary},
\end{equation}
a possible choice of the oracles is
\begin{equation}\label{Eq3B:block_encode_LCU}
    O = \sum_{j=1}^{j_\imax} \ket{j}\bra{j}_a \otimes U_j, \quad \ket{G}_a = \sum_{j=1}^{j_\imax} \sqrt{\frac{\alpha_j}{\alpha}} \ket{j}_a,
\end{equation}
with $\alpha = \sum_{j=1}^{j_\imax} \alpha_j$. 
The number of ancillary qubits $n_a$ scales as $\log j_\imax$.
Since $j_\imax$ is $\poly{N}$ for typical $N$-site systems, the number of elementary gates for the oracles amounts to $C \in \poly{N}$.
LCUs cover various types of Hamiltonians exemplified by quantum spin systems composed of local Pauli operators and fermionic systems in condensed matter physics and quantum chemistry \cite{Babbush2018-hz}.
The block-encoding of other Hamiltonians, such as sparse-access matrices and purified density matrices, has been also revealed \cite{Low2019-qubitization,Gilyen2019-qsvt,MartynPRXQ-grand-unif}.

Once we find out the oracles $O$ and $\ket{G}_a$, we can organize a unitary gate $W_q$, implemented by $\order{1}$ additional ancillary qubits and $\order{q}$-times usage of $O$, $G$, and $\order{n_a}$ elementary gates, which satisfies
\begin{equation}
    W_q \ket{0}^{\otimes \left\{ n_a + \order{1} \right\}} \ket{\psi} = \ket{0}^{\otimes \left\{ n_a + \order{1} \right\} }  e^{-i H t} \ket{\psi} + \order{\varepsilon_q},
\end{equation}
for every state $\ket{\psi} \in \mcl{H}$.
Here, the error $\varepsilon_q$ comes from approximating $e^{-iHt}$ by a certain degree-$q$ polynomial of $H$, and decays as $\varepsilon_q \in \order{(\alpha t / q)^q}$.
The scaling of $q$ required for making the error $\varepsilon_q$ smaller than an acceptable error $\varepsilon$ is given by the following formula;
\begin{equation}\label{Eq2B:query_qubitization}
    q \in \Theta \left( \alpha t + \frac{\log (1/\varepsilon)}{\log (e+(\alpha t)^{-1} \log (1/\varepsilon))}\right).
\end{equation}
In order to determine the explicit form of $W_q$, we resort to quantum signal processing \cite{Low2017-QSP}.
It dictates that only $\poly{q}$-time classical computation is required for this purpose.

To summarize the qubitization technique, we require the following computational resources to implement the time-evolution operator $e^{-iHt}$ with an acceptable error $\varepsilon$ (See also Table \ref{Table:comparison_algorithms});
\begin{itemize}
    \item Number of ancillary qubits; $n_a+\order{1}$
    \item Query complexity; $q$ [Eq. (\ref{Eq2B:query_qubitization})]
    \item Number of elementary gates;
    \begin{equation}\label{Eq2B:qubitization_cost}
        \order{\left\{ \alpha t + \frac{\log (1/\varepsilon)}{\log (e + (\alpha t)^{-1} \log (1/\varepsilon) )}\right\} (C+n_a)}.
    \end{equation}
\end{itemize}
Here, the query complexity is defined by the complexity counted by the number of oracles, corresponding to $q$ in this case, and gives the scaling of the number of elementary gates below.
Significantly, the query complexity takes an additive form as
\begin{equation}\label{Eq2:qubitization_additive_scale}
    \alpha t + o(\log (1/\varepsilon)) \in \poly{N} t + o(\log (1/\varepsilon)).
\end{equation}
The $o(\log (1/\varepsilon))$ term scales as $\log (1/\varepsilon) / \log \log (1/\varepsilon)$ under $1/\varepsilon \to \infty$ with fixed time $t$.
Its scaling is known to be optimal both in $t$ and $1/\varepsilon$ for simulating generic time-independent Hamiltonians \cite{Gilyen2019-qsvt}.

In our algorithm for time-periodic Hamiltonians, we exploit qubitization as a subroutine to implement time evolution under an effective Hamiltonian in Floquet theory. 
We can achieve the query complexity for time-periodic systems in the additive form like Eq. (\ref{Eq2:qubitization_additive_scale}), while the conventional truncated Dyson-series algorithm for generic time-dependent systems needs the query complexity in a multiplicative form $\alpha t \times o(\log (\alpha t / \varepsilon))$.

\section{Summary of this paper}\label{Sec:Summary_paper}
In this section, we overview our main results on efficient quantum simulation of time-periodic Hamiltonians.
We will provide their detailed derivation in the following Sections \ref{Sec:truncated_Floquet_Hilbert}-\ref{Sec:Algorithm_cost}.

\subsection{Overview of Algorithm}\label{Subsec:Overview_algorithm}
We first show an overview of the algorithm for time-evolution under a time-periodic Hamiltonian $H(t+T)=H(t)$.
Throughout the main text, we assume that the Fourier components of the Hamiltonian $H(t)$ vanishes at a certain cutoff $m_\imax \in \order{1}$;
\begin{equation}
    H_m = 0, \quad \text{if} \quad |m| > m_\imax.
\end{equation}
(This discussion can be extended to cases where $H_m$ exponentially decays in $|m|$. See Appendix \ref{Asec:Extension_expo}.).

The central attempt in the algorithm is to efficiently simulate Eq. (\ref{Eq2A:psi_t_infinite}),
\begin{equation}\label{Eq3A:psi_t_infinite}
    \ket{\psi(t)}  = \sum_{l = - \infty}^\infty e^{-il\omega t} \bra{l} e^{-i \ms{H}_\eff t} \ket{0} \ket{\psi(0)},
\end{equation}
on quantum circuits.
The time-dependency is erased by mapping the dynamics to the one on the Floquet-Hilbert space, which results in the following two benefits. 
First, we do not discretize the time with infinitesimal intervals.
We instead rely on Fourier indices $l \in \bbZ$, which are originally discrete, and actually they lead to much better accuracy with the same auxiliary degrees of freedom.
Second, we can use various Hamiltonian simulation algorithms for time-independent systems.
Since the qubitization technique has achieved the best query complexity in the time $t$ and the inverse error $1/\varepsilon$, exploiting it for $\ms{H}_\eff$ accelerate the simulation of time-periodic systems.

On the other hand, in order to simulate the time-evolved state $\ket{\psi(t)}$ via Eq. (\ref{Eq3A:psi_t_infinite}) on quantum circuits, we also have several problems to be tackled.
First, the Floquet-Hilbert space is infinite dimensional. We have to introduce truncation at a certain Fourier index $l_\imax$, and consider a finite set of indices defined by
\begin{equation}
    D^{l_\imax} = \{ -l_\imax +1, -l_\imax+2, \hdots, l_\imax \} \subset \bbZ.
\end{equation}
Then, we focus on the approximate dynamics by
\begin{eqnarray}
    \ms{H}_\eff^{l_\imax} &=& \sum_{l \in D^{l_\imax}} \ket{l}\bra{l} \otimes (H_0-l\omega) + \sum_{l \in D^{l_\imax}} \sum_{m \neq 0; l+m \in D^{l_\imax}} \ket{l}\bra{l+m} \otimes H_{-m}, \nonumber \\
    && \label{Eq3A:def_effective_H_l_max}\\
    \ket{\psi^{l_\imax}(t)}  &=& \sum_{l \in D^{l_\imax}} e^{-il\omega t} \bra{l} e^{-i \ms{H}_\eff^{l_\imax} t} \ket{0}\ket{\psi(0)}. \label{Eq3A:def_psi_l_max}
\end{eqnarray}
The finite-dimensional space, spanned by $2l_\imax \times \mr{dim}(\mcl{H})$ states $\{ \ket{l} \ket{\psi_j} \, | \, l \in D^{l_\imax}, \, j=1,\hdots,\mr{dim}(\mcl{H})\}$, is called the truncated Floquet-Hilbert space. 
While $l_\imax \to \infty$ reproduces $\ket{\psi^{l_\imax}(t)} \to \ket{\psi(t)}$, it is nontrivial how we should choose $l_\imax$ to achieve the accuracy $\norm{\ket{\psi(t)} - \ket{\psi^{l_\imax}(t)}} \leq \varepsilon$.
The second problem is the small success probability of post-selecting the ancillary state.
Even if we succeed in approximation with finite $l_\imax$, Eq. (\ref{Eq3A:def_psi_l_max}) requires a projection to a unnormalized ancillary state $\sum_{l \in D^{l_\imax}} \ket{l}$ for the state in the truncated Floquet-Hilbert space,
\begin{equation}\label{Eq3A:def_large_psi_l_max}
    \ket{\Psi^{l_\imax}(t)} = e^{-i \omega t \sum_l \ket{l} \bra{l}} e^{-i \ms{H}_\eff^{l_\imax} t} \ket{0} \ket{\psi(0)}.
\end{equation}
In the actual computation, we need post-selection to an ancillary state given by
\begin{equation}\label{Eq3A:def_a_l_max}
   \ket{a^{l_\imax}} = \frac{1}{\sqrt{2l_\imax}}\sum_{l \in D^{l_\imax}} \ket{l}.
\end{equation}
Success probability of the projection is given by $\norm{\braket{a^{l_\imax}|\Psi^{l_\imax}(t)}}^2 \simeq \braket{\psi(t)|\psi(t)}/(2l_\imax) = 1/(2l_\imax)$.
As $l_\imax$ increases to ensure the accuracy, the expected time to successfully obtain $\ket{\psi(t)}$ becomes longer in proportion to it.
The final problem is about implementation of $e^{-i\ms{H}_\eff^{l_\imax}t}$.
While the effective Hamiltonian $\ms{H}_\eff$ is time-independent, its structure is complicated due to the additional degrees of freedom labeled by $\{\ket{l}\}$.
It is nontrivial whether the optimal Hamiltonian simulation algorithm, i.e. the qubitization technique, provides the optimality for time-periodic systems.

Our algorithm relying on the truncated Floquet-Hilbert space efficiently simulates the time-evolved state $\ket{\psi(t)}$ with resolving the above problems.
The significant developments are composed of the following steps;
\begin{enumerate}[(a)]
    \item Decision of the Fourier index for truncation, $l_\imax$, to achieve an allowable error $\varepsilon$.
    \item Amplification by symmetry and oblivious amplitude amplification to enhance the success probability up to $1-\order{\varepsilon}$.
    \item Efficient implementation of $\exp ( -i \ms{H}_\eff^{l_\imax} t)$ by the qubitization technique.
\end{enumerate}
In Step (a), we explicitly derive the upper bound of $\norm{\ket{\psi(t)}-\ket{\psi^{l_\imax}(t)}}$ in a similar way to the Lieb-Robinson bound. 
We show that the choice of $l_\imax$, based on
\begin{eqnarray}
    l_\imax &\in& \Theta \left( \gamma t + \frac{\log (1/\varepsilon)}{\log ( e + (\gamma t)^{-1} \log (1/\varepsilon) )} \right), \\
    \gamma &=& \sup_t (\norm{H(t)-H_0}) \label{Eq3A:def_gamma},
\end{eqnarray}
is sufficient to make the error smaller than $\varepsilon$.
Here, $\gamma$ gives the scale of the time-dependent terms in $H(t)$.
The second step (b) plays a role in amplifying the success probability of the post selection from $1/(2l_\imax)$ to $1-\order{\varepsilon}$. 
The first protocol, which we call amplification by symmetry, exploits the symmetry of $\ms{H}_\eff$ which is always present and inherent in time-periodic systems.
It brings the success probability from $1/(2l_\imax)$ to $1/4-\order{\varepsilon}$ only with small cost of $\order{\log l_\imax}$ elementary gates.
Following this, we apply the oblivious amplitude amplification \cite{Berry2017-oblivious}, reminiscent of Grover's search algorithm \cite{Grover1997-prl}. The success probability is further amplified from $1/4-\order{\varepsilon}$ to $1-\order{\varepsilon}$.
Exploiting these two kinds of amplification, we can obtain the target state $\ket{\psi(t)}$ only with a little additional resource that does not change the scaling.
The final step (c) constructs conversion of the effective Hamiltonian so that $\exp ( -i \ms{H}_\eff^{l_\imax} t)$  can be efficiently implemented.
The original effective Hamiltonian $\ms{H}_\eff^{l_\imax}$ is not suitable for the qubitization technique, since the additional degrees of freedom $\ket{l}$ makes its block-encoding inefficient.
To resolve this problem, we derive an alternative effective Hamiltonian such that the desired evolution $\exp ( -i \ms{H}_\eff^{l_\imax} t)$ is accurately reproduced.
The block-encoding for the alternative effective Hamiltonian requires $\order{1}$-times queries to the oracles for each Fourier component $\{ H_m \}$, which is important for achieving optimal or nearly-optimal dependence in $t$ and $1/\varepsilon$. 

\subsection{Main results}\label{Subsec:Main_results}

Here, we summarize the computational resources for computing the dynamics $\ket{\psi(t)}$ under a time-periodic Hamiltonian $H(t)$. 
We construct two different quantum algorithms depending on the time scale of the dynamics. 
The first case is an adiabatic-like case, where we are interested in the dynamics over $\order{1}$ periods as $\omega t \in \order{1}$.
We call it ``adiabatic" since adiabatic dynamics under sufficiently large period $T$ is a typical target, while $T$ is not required to be large.
The other case is generic long-time regime, in which we focus on dynamics over multiple periods as $\omega t \in \Omega (1)$.
The algorithms for both cases follow Steps (a)-(c) of Section \ref{Subsec:Overview_algorithm}; the number of ancillary qubits is determined by the truncation order $l_\imax$ of Step (a), and the number of elementary gates comes mainly from the cost of implementing $\exp ( -i \ms{H}_\eff^{l_\imax} t)$  in Step (c).

We assume that the Fourier component $ H_m $ becomes zero for $|m|>m_\imax$ with an $\order{1}$ constant $m_\imax$. The block-encoding for each $H_m$ is supposed to be given by an oracle unitary gate $O_m$ and an $n_a$-qubit oracle state $\ket{G_m}$ as
\begin{equation}\label{Eq3B:def_block_encode_H_m}
    \braket{G_m |O_m| G_m} = \frac{H_m}{\alpha_m}, \quad \alpha_m > 0.
\end{equation}
Each oracle state $\ket{G_m}$ is generated by trivial states as $\ket{G_m}=G_m\ket{0}$.
We define the energy scale of the Hamiltonian $H(t)$ by
\begin{equation}\label{Eq3B:def_alpha}
    \alpha = \sum_{|m| \leq m_\imax} \alpha_m,
\end{equation}
and suppose that $\{ \alpha_m \}_m$ can be embedded into an $\order{1}$-qubit quantum system by
\begin{equation}\label{Eq3B:def_G_coef}
    G_\mr{coef} \ket{0} = \sum_{|m| \leq m_\imax} \sqrt{\frac{\alpha_m}{\alpha}} \ket{m} \in \bbC^{2m_\imax+1}.
\end{equation}
The query complexity is defined by the number of queries to the oracles $O_m$, $G_m$, and $G_\mr{coef}$, where each of them is supposed to require at-most $C$ elementary gates.
We summarize the computational resources for simulating $\ket{\psi(t)}$ in the adiabatic-like regime and the generic long-time regime respectively by the following theorems.

\begin{theorem}\label{Thm3:resource_adiabatic}
\textbf{(Resource for adiabatic-like regime)}

Suppose that we are interested in the time-evolved state $\ket{\psi(t)}$ from an arbitrary initial state $\ket{\psi(0)}$ over $\order{1}$ periods.
The computational resources to obtain it with the acceptable error and the failure probability smaller than $\order{\varepsilon}$ are summarized as follows;
\begin{itemize}
    \item Number of ancillary qubits;
    \begin{equation}
        n_a + \order{\log (\gamma t) + \log \log (1/\varepsilon)}.
    \end{equation}
    
    \item Scaling of query complexity;
    \begin{equation}\label{Eq3B:query_complexity_adiabatic}
        \alpha t  + \frac{\log (1/\varepsilon)}{\log ( e + \{ \alpha t + o(\log (1/\varepsilon))\}^{-1} \log (1/\varepsilon) )}.
    \end{equation}
    The $o(\log (1/\varepsilon))$ term in the above denominator scales as
    \begin{equation}\label{Eq3B:o_log_term_adiabatic}
        \frac{\log (1/\varepsilon)}{\log (e+ (\gamma t)^{-1} \log (1/\varepsilon))}.
    \end{equation}
    \item Additional gates per query;
    \begin{equation}
        \order{n_a + \log (\gamma t) + \log \log (1/\varepsilon)}.
    \end{equation}
\end{itemize}
The query complexity has optimal scaling in the time $t$ and nearly-optimal scaling in the inverse error $1/\varepsilon$.
\end{theorem}

\begin{theorem}\label{Thm3:resource_long_time}
\textbf{(Resource for long-time regime)}

Suppose that we are interested in the time-evolved state $\ket{\psi(t)}$ from arbitrary initial states $\ket{\psi(0)}$ over multiple periods $\omega t \in \Omega (1)$.
The computational resources to obtain it with the acceptable error and the failure probability smaller than $\order{\varepsilon}$ are summarized as follows;
\begin{itemize}
    \item Number of ancillary qubits;
    \begin{equation}
        n_a + \order{\log (\gamma /\omega) + \log \log (\omega t/\varepsilon)}.
    \end{equation}
    
    \item Scaling of query complexity;
    \begin{equation}
         \alpha t + \frac{ \omega t \log (\omega t/\varepsilon)}{\log (e+\{ \alpha /\omega + o(\log (\omega t /\varepsilon))\}^{-1} \log (\omega t/\varepsilon))}.
    \end{equation}
    The $o(\log (\omega t /\varepsilon))$ term in the above denominator scales as
    \begin{equation}\label{Eq3B:o_log_term_long}
        \frac{\log (\omega t /\varepsilon)}{\log (e + (\gamma / \omega)^{-1} \log (\omega t /\varepsilon))}.
    \end{equation}
    
    \item Additional gates per query;
    \begin{equation}
        \order{n_a + \log (\gamma /\omega) + \log \log (\omega t /\varepsilon)}
    \end{equation}
\end{itemize}
The scaling of the query complexity is optimal in time $t$ for practical problems up to $\poly{N}$-time, while it is formally nearly-optimal.
It is nearly-optimal in the inverse error $1/\varepsilon$.
\end{theorem}

The parameters $\alpha$ and $\gamma$, defined by Eqs. (\ref{Eq3B:def_alpha}) and (\ref{Eq3A:def_gamma}), respectively represent energy scales of the overall terms and the time-dependent terms in $H(t)$. 
They typically scale as $\alpha, \gamma \in \mr{poly}(N)$ with the system size $N$.
In contrast, the frequency $\omega=2 \pi / T$ is typically an  $\order{N^0}$ value much smaller than $\alpha$ and $\gamma$ (The high-frequency cases where $\omega$ are comparable to or larger than $\alpha, \gamma$ are trivial, as discussed in Appendix \ref{Asec:high_frequency}).
While the above theorems are about time-periodic Hamiltonians with vanishing Fourier components at $m_\imax \in \order{1}$ as $H_m= 0$ ($|m| > m_\imax$), we obtain similar results for Hamiltonians with exponentially-decaying Fourier components $\norm{H_m} \lesssim e^{-\order{|m|}}$.
See Appendix \ref{Asec:Extension_expo} for its derivation.

\section{Building an appropriate truncated Floquet-Hilbert space}\label{Sec:truncated_Floquet_Hilbert}

This section is devoted to deriving the proper truncation order for the Floquet-Hilbert space, following (a) in Section \ref{Subsec:Overview_algorithm}.
In order to achieve desirable accuracy for the exact time-evolved state with the truncated Floquet-Hilbert space, we should obtain an exact upper bound on its error.
Here, we prove the Lieb-Robinson bound in the Floquet-Hilbert space in Section \ref{Subsec:Lieb_Robinson_Floquet_Hilbert}, and determine a proper truncation order based on it in Section \ref{Subsec:Truncation_order}. 

\subsection{Lieb-Robinson bound in Floquet-Hilbert space}\label{Subsec:Lieb_Robinson_Floquet_Hilbert}

The proper choice of the truncation order $l_\imax$ is determined so that the approximate state $\ket{\psi^{l_\imax}(t)}$ [See Eq. (\ref{Eq3A:def_psi_l_max})] can reproduce the time-evolved state $\ket{\psi(t)}$ with satisfying the relation $\norm{\ket{\psi(t)}-\ket{\psi^{l_\imax}(t)}} \leq \varepsilon$.
From the explicit formula Eq. (\ref{Eq3A:def_psi_l_max}), we can see that it is important to observe how the transition amplitude $\braket{l|e^{-i \ms{H}_\eff^{l_\imax}t}|0}$ behaves for sufficiently large $l$.
We first show the upper bound on this transition amplitude in Theorem \ref{Thm4:Bound_transition_rate}. 
Since it is reminiscent of the Lieb-Robinson bound in single-particle quantum systems \cite{Gong2019bound}, we call it the Lieb-Robinson bound in the Floquet-Hilbert space.

\begin{theorem}\label{Thm4:Bound_transition_rate}
\textbf{(Bound on transition amplitude)}

We assume $H_m = 0$ for $|m| > m_\imax$. Then, for indices $l, l^\prime \in \bbZ$ satisfying $|l-l^\prime| \geq 2 m_\imax \gamma t$, the transition amplitude is bounded by
\begin{eqnarray}
    \norm{\bra{l} e^{-i \ms{H}_\eff^{l_\imax} t} \ket{l^\prime}} &\leq& 2 \frac{(\gamma t)^{\lceil |l-l^\prime| /m_\imax \rceil}}{(\lceil |l-l^\prime| /m_\imax \rceil)!} \label{Eq4A:thm_Lieb_Robinson_1} \\
    &\leq& \left( \frac{\gamma t}{\lceil |l-l^\prime| /m_\imax \rceil}\right)^{\lceil |l-l^\prime| /m_\imax \rceil}, \label{Eq4A:them_Lieb_Robinson_2}
\end{eqnarray}
where the parameter $\gamma$ is defined by Eq. (\ref{Eq3A:def_gamma}).
\end{theorem}

\textbf{Proof.---}
We will omit the superscripts $l_\imax$ for some operators introduced here since they are not important. We first employ the interaction picture. With the unitary operation defined by
\begin{equation}\label{Eq4A:def_reference_hamiltonian}
    \ms{U}_0(t) = e^{-i \ms{H}_0 t}, \quad \ms{H}_0 = \sum_{l \in D^{l_\imax}} \ket{l}\bra{l} \otimes (H_0-l \omega ),
\end{equation}
the time evolution operator $e^{-i\ms{H}_\eff^{l_\imax}t}$ is represented as
\begin{eqnarray}
   e^{-i\ms{H}_\eff^{l_\imax} t} &=& \ms{U}_0(t) \ms{U}_I (t), \\
   \ms{U}_I(t) &=& \mcl{T} \exp \left( -i \int_0^t \dd t^\prime \ms{H}_I(t^\prime )\right).
\end{eqnarray}
Here, the Hamiltonian in the interaction picture is defined by
\begin{eqnarray}
    \ms{H}_I(t) &=& \ms{U}_0(t)^\dagger (\ms{H}_\eff^{l_\imax}-\ms{H}_0) \ms{U}_0(t) \nonumber \\
    &=& \sum_{l} \sum_{m>0} (e^{im\omega t}\ket{l}\bra{l+m} \otimes H_{-m}^I(t) + \text{h.c.} ), \label{Eq4B:effective_H_int_pic}\\
    H_{-m}^I(t) &=& e^{i H_0 t} H_{-m} e^{-iH_0 t},
\end{eqnarray}
where the summation in the second line is taken over $l,m$ such that $l, l+m \in D^{l_\imax}$.
Using the Dyson series expansion, we obtain
\begin{eqnarray}
    \norm{\bra{l} e^{-i \ms{H}_\eff^{l_\imax} t} \ket{l^\prime}} &=& \norm{\braket{l|\ms{U}_I(t)|l^\prime}} \nonumber \\
    &\leq& \sum_{n=0}^\infty \int_0^t \dd t_n \hdots \int_0^{t_2} \dd t_1 \norm{\braket{l|\ms{H}_I(t_n) \hdots \ms{H}_I(t_1)|l^\prime}} \nonumber \\
    &=& \sum_{n=0}^\infty \int_0^t \dd t_n \hdots \int_0^{t_2} \dd t_1 \norm{\sum_{\{l_i\}}\prod_{i=1}^n \braket{l_i|\ms{H}_I(t_i)|l_{i-1}}}.  \label{Eq4A:Dyson_transition_path}
\end{eqnarray}
In the last equality, we employ the identity $\sum_{l_i \in D^{l_\imax}} \ket{l_i}\bra{l_i}=I$ for $n-1$ times.
The summation $\sum_{\{l_i \}}$ is taken over $l_i \in D^{l_\imax}$ for $i=1,2,\hdots,n-1$ under the fixed $l_0=l^\prime$ and $l_n=l$.
Each product $\prod_{i=1}^n \braket{l_i|\ms{H}_I(t_i)|l_{i-1}}$ represents a complex transition amplitude from $\ket{l^\prime}$ to $\ket{l}$ along the path $\ket{l^\prime} \to \ket{l_1} \to \hdots \to \ket{l_{n-1}} \to \ket{l}$ under $\ms{H}_I(t_i)$.
Since the Hamiltonian $\ms{H}_I(t_i)$ shifts the Fourier index $\ket{l_i}$ by at-most $m_\imax$ due to Eq. (\ref{Eq3A:def_effective_H_l_max}), the low order terms labeled by $n < |l-l^\prime|/m_\imax$ disappears.
This results in
\begin{eqnarray}
    \norm{\bra{l} e^{-i \ms{H}_\eff^{l_\imax} t} \ket{l^\prime}} &\leq& \sum_{n=\lceil |l-l^\prime|/m_\imax \rceil}^\infty \int_0^t \dd t_n \hdots \int_0^{t_2} \dd t_1 \norm{\braket{l|\prod_{i=1}^n \ms{H}_I(t_i) |0}} \nonumber \\
    &\leq& \sum_{n=\lceil |l-l^\prime|/m_\imax \rceil}^\infty \frac{t^n}{n!} \left( \sup_t (\norm{\ms{H}_I(t)})\right)^n 
\end{eqnarray}

Here, $\ms{H}_I(t)$ is a Toeplitz matrix that satisfies $\braket{l|\ms{H}_I(t)|l^\prime} = e^{-i(l-l^\prime)\omega t} H_{l-l^\prime}^I(t) - H_0 \delta_{l-l^\prime,0}$.
This leads to the upper bound of its operator as follows \cite{Gray2006-da},
\begin{eqnarray}
    \norm{\ms{H}_I(t)} &\leq& \sup_{t^\prime} \left( \norm{\sum_{m \in \bbZ} (e^{im\omega t} H_{-m}^I(t) - H_0 \delta_{m,0}) e^{im\omega t^\prime}}\right) \nonumber \\
    &=& \sup_{t^\prime} \left( \norm{H(t-t^\prime) - H_0} \right) = \gamma. \label{Eq4A:H_eff_bound_gamma}
\end{eqnarray}
We also use the following inequality;
\begin{equation}\label{Eq4A:sum_Taylor}
    \sum_{n=n_0}^\infty \frac{x^n}{n!} \leq 2 \frac{x^{n_0}}{n_0 !}, \quad \text{if} \quad  n_0 \geq 2x \geq 0.
\end{equation}
This relation can be easily confirmed by
\begin{eqnarray}
    \sum_{n=n_0}^\infty \frac{x^n}{n!} &=& \frac{x^{n_0}}{n_0 !} \sum_{n=n_0}^\infty \frac{x^{n-n_0}}{n (n-1) \hdots (n_0+1)} \leq  \frac{x^{n_0}}{n_0 !} \sum_{n=0}^\infty \left(\frac12\right)^n.
\end{eqnarray}
For the indices $l, l^\prime$ satisfying $|l-l^\prime| \geq 2m_\imax \gamma t$, we can substitute $x=\gamma t$ and $n_0=\lceil |l-l^\prime| /m_\imax \rceil$. This results in the bound,
\begin{equation}
    \norm{\bra{l} e^{-i \ms{H}_\eff^{l_\imax} t} \ket{l^\prime}} \leq 2 \frac{(\gamma t)^{\lceil |l-l^\prime| /m_\imax \rceil}}{(\lceil |l-l^\prime| /m_\imax \rceil)!}.
\end{equation}
The inequality from the Stirling formula,
\begin{equation}\label{Eq4A:Stirling_inequality}
    n_0 ! \geq 2 \left( \frac{n_0}{e}\right)^{n_0},
\end{equation}
ensures the inequality Eq. (\ref{Eq4A:them_Lieb_Robinson_2}). $\quad \square$

The Lieb-Robinson bound in the transition amplitude provides a guide for choosing a proper truncation order.
By setting $l^\prime=0$, Eqs. (\ref{Eq4A:thm_Lieb_Robinson_1}) and (\ref{Eq4A:them_Lieb_Robinson_2}) say that the contributions to $\ket{\psi^{l_\imax}(t)}$ from indices satisfying $|l| \geq 2 m_\imax \gamma t$ rapidly decay as $l^{-l}$.
Thus, $\order{\gamma t}$ becomes a possible truncation order.
However, we note that the change in $l_\imax$ affects $\ket{\psi^{l_\imax}(t)}$ also via the change in the support of the effective Hamiltonian $\ms{H}_\eff^{l_\imax}$.
By taking it into account with using the Lieb-Robinson bound, we obtain the exact upper bound on the error between $\ket{\psi(t)}$ and $\ket{\psi^{l_\imax}(t)}$ as follows.

\begin{theorem}\label{Thm4:Floquet_Hilbert_truncation}
\textbf{(Floquet-Hilbert space truncation)}

We consider the approximate time-evolved state obtained from the truncated Floquet-Hilbert space, $\ket{\psi^{l_\imax}(t)}$, given by Eq. (\ref{Eq3A:def_psi_l_max}). 
Then, its deviation from the exact one $\ket{\psi}$ is bounded by
\begin{equation}\label{Eq4B:deviation_truncation}
    \norm{\ket{\psi(t)} - \ket{\psi^{l_\imax}(t)}} \leq 20 m_\imax \frac{(\gamma t)^{\lceil l_\imax/m_\imax \rceil}}{(\lceil l_\imax/m_\imax \rceil)!},
\end{equation}
if the truncation order $l_\imax$ satisfies $l_\imax \geq 2 m_\imax \gamma t$.
\end{theorem}
\textbf{Proof.---}
We evaluate the convergence of $\ket{\psi^{l_\imax}(t)}$.
For different orders $l_\imax, l_\imax^\prime$ satisfying $l_\imax^\prime > l_\imax$, we compute the difference,
\begin{equation}\label{Eq4A:thm_truncation_proof_1}
    \norm{\ket{\psi^{l_\imax^\prime}(t)}-\ket{\psi^{l_\imax}(t)}} \leq \varepsilon_1 +\varepsilon_2,
\end{equation}
\begin{eqnarray}
    \varepsilon_1 &=& \sum_{ l \in (D^{l_\imax^\prime} \backslash D^{l_\imax})} \norm{\bra{l} e^{i\ms{H}_\eff^{l_\imax^\prime}t} \ket{0}}, \label{Eq4A:thm_truncation_proof_2}\\
    \varepsilon_2 &=& \sum_{l \in D^{l_\imax}} \norm{\bra{l} e^{i\ms{H}_\eff^{l_\imax^\prime}t}-e^{i\ms{H}_\eff^{l_\imax}t} \ket{0}}. \label{Eq4A:thm_truncation_proof_3}
\end{eqnarray}
The first error comes from truncating the order of the post-selected state $\ket{a^{l_\imax}}$.
Using Theorem \ref{Thm4:Bound_transition_rate} directly implies
\begin{eqnarray}
    \varepsilon_1 \leq \sum_{ l \in (D^{l_\imax^\prime} \backslash D^{l_\imax})} \frac{(\gamma t)^{\lceil |l|/m_\imax \rceil}}{(\lceil |l|/m_\imax \rceil)!} \leq 2 \sum_{l = l_\imax}^\infty  \frac{(\gamma t)^{\lceil l/m_\imax \rceil}}{(\lceil l/m_\imax \rceil)!} .
\end{eqnarray}
The summation $\sum_{l=l_\imax}^\infty$ can be divided based on $l$ $(\text{mod. } m_\imax)$, and each summation corresponds to the left hand side of Eq. (\ref{Eq4A:sum_Taylor}) with $n_0 \geq \lceil l_\imax / m_\imax \rceil$.
In other words, for $l_\imax \geq 2 m_\imax \gamma t$, we obtain the following inequality,
\begin{equation}\label{Eq4A:sum_Taylor_modulo}
    \sum_{l = l_\imax}^\infty  \frac{(\gamma t)^{\lceil l/m_\imax \rceil}}{(\lceil l/m_\imax \rceil)!} \leq 2 m_\imax \frac{(\gamma t)^{\lceil l_\imax/m_\imax \rceil}}{(\lceil l_\imax/m_\imax \rceil)!}.
\end{equation}
As a result, the first error $\varepsilon_1$ is bounded by $4 m_\imax (\gamma t)^{\lceil l_\imax/m_\imax \rceil}/(\lceil l_\imax/m_\imax \rceil)!$.

The second error $\varepsilon_2$ comes from truncating the order of the effective Hamiltonian.
In a similar way to the proof of Theorem \ref{Thm4:Bound_transition_rate}, each term is bounded by
\begin{equation}
    \norm{\bra{l} e^{i\ms{H}_\eff^{l_\imax^\prime}t}-e^{i\ms{H}_\eff^{l_\imax}t} \ket{0}} \leq \sum_{n=0}^\infty \int_0^t \dd t_n \hdots \int_0^{t_2} \dd t_1 \norm{\braket{l|\prod_{i=1}^n \ms{H}_I^{l_\imax^\prime}(t_i) - \prod_{i=1}^n \ms{H}_I^{l_\imax}(t_i)|0}}, 
\end{equation}
where we explicitly write the size $l_\imax,l_\imax^\prime$ for $H_I(t)$ [See Eq. (\ref{Eq4B:effective_H_int_pic})].
In a similar manner to Eq. (\ref{Eq4A:Dyson_transition_path}), $\braket{l|\prod_{i=1}^n \ms{H}_I^{l_\imax^\prime}(t_i)|0}$ and $\braket{l|\prod_{i=1}^n \ms{H}_I^{l_\imax}(t_i)|0}$ respectively represent transition amplitudes summed over all possible paths $\ket{0} \to \ket{l_1} \to \hdots \to \ket{l_{n-1}} \to \ket{l}$ under $\ms{H}_\eff^{l_\imax^\prime}(t)$ and $\ms{H}_\eff^{l_\imax}(t)$.
Their differences survive only when the path goes across the domain $D^{l_\imax^\prime} \backslash D^{l_\imax}$, where these two Hamiltonians have different actions.
In other words, low order terms with $n < \{ (l_\imax-0) + (l_\imax - |l|)/m_\imax$ should vanish.
Considering that the norm of $\ms{H}_\eff^{l_\imax}(t)$ and $\ms{H}_\eff^{l_\imax^\prime}(t)$ is both bounded by $\gamma$, we obtain
\begin{eqnarray}
    \varepsilon_2 &\leq& \sum_{l \in D^{l_\imax}} \sum_{n=\lceil (2l_\imax-|l|)/m_\imax \rceil}^\infty \frac{t^n}{n!} 2 \gamma^n \nonumber \\
    &\leq& 4 \sum_{l \in D^{l_\imax}} \frac{(\gamma t)^{\lceil (2l_\imax-|l|)/m_\imax \rceil}}{(\lceil (2l_\imax-|l|)/m_\imax \rceil)!} \nonumber \\
    &\leq& 16 m_\imax \frac{(\gamma t)^{\lceil l_\imax/m_\imax \rceil}}{(\lceil l_\imax/m_\imax \rceil)!}.
\end{eqnarray}
In the last inequality, we again use the relation Eq. (\ref{Eq4A:sum_Taylor_modulo}). 
Taking the limit $l_\imax^\prime \to \infty$ for $\varepsilon_1 + \varepsilon_2$ reproduces the result of Theorem \ref{Thm4:Floquet_Hilbert_truncation}. $\quad \square$

\subsection{Truncation order of Floquet-Hilbert space}\label{Subsec:Truncation_order}

We hereby determine the truncation order $l_\imax$ so that $\ket{\psi^{l_\imax}(t)}$ can reproduce the exact time-evolved state with a desirable error up to $\order{\varepsilon}$.
By using the inequality from the Stirling formula, Eq. (\ref{Eq4A:Stirling_inequality}) to Theorem \ref{Thm4:Floquet_Hilbert_truncation}, we obtain the error bounded by
\begin{equation}
    \norm{\ket{\psi(t)}-\ket{\psi^{l_\imax}(t)}} \leq 10 m_\imax \left( \frac{em_\imax \gamma t}{l_\imax}\right)^{l_\imax/m_\imax},
\end{equation}
for $l_\imax \geq 2 m_\imax \gamma t$.
The truncation order $l_\imax$ is chosen so that the right hand side can be smaller than $\varepsilon$.
To this aim, we should evaluate a function $f(x)=(\kappa/x)^x$ [$\kappa > 0$], which is known to be dealt with the Lambert W function $W(x)$ satisfying $W(x)e^{W(x)}=x$ \cite{Corless1996-lambert}. 
Here, we rely on the resulting proposition \cite{Gilyen2019-qsvt}.
\begin{proposition}\label{Prop:Lambert_function} $\qquad$

The function $f(x)=(\kappa/x)^x$ [$\kappa>0$] is monotonically decreasing in $x \geq \kappa /e$, and satisfies the following inequality for $0 < \eta < 1$;
\begin{equation}
    f(x) \leq \eta, \quad^\forall x \geq e \kappa + \frac{4 \log (1/\eta)}{\log (e + \kappa^{-1} \log (1/\eta))}.
\end{equation}
\end{proposition}

See Lemma 59 in the full version of Ref. \cite{Gilyen2019-qsvt} for the proof. 
Based on the above proposition, we choose $l_\imax$ by
\begin{eqnarray}
    l_\imax  = m_\imax + \left\lceil e^2 m_\imax \gamma t + \frac{4 m_\imax \log (10m_\imax / \varepsilon )}{\log (e + (e\gamma t)^{-1} \log (10m_\imax / \varepsilon))}\right\rceil, \label{Eq4B:l_max_choice}
\end{eqnarray}
so that the error can be bounded from above as
\begin{equation}\label{Eq4B:error_condition}
    10 m_\imax \left( \frac{em_\imax \gamma t}{l_\imax-m_\imax}\right)^{(l_\imax-m_\imax)/m_\imax} \leq \varepsilon.
\end{equation}
We note that this choice does not violate the condition $l_\imax \geq 2 m_\imax \gamma t$, which is required for Theorem \ref{Thm4:Floquet_Hilbert_truncation}.
The monotonicity of $f(x)$ ensures $\norm{\ket{\psi(t)}-\ket{\psi^{l_\imax}(t)}} \leq \varepsilon$ under the above choice (The additional term $m_\imax$ is attached for later calculation, especially for Appendix \ref{Asubsec2:amplification_symmetry}) and \ref{Asubsec2:refine_effective_Hamiltonian}).

Let us discuss how $l_\imax$ increases in the time $t$ and the inverse error $1/\varepsilon$.
The form of the error in $l_\imax$, given by Eq. (\ref{Eq4B:deviation_truncation}), is the same as that for qubitization in the query complexity $q$, given by $\varepsilon_q \in \order{(\alpha t /q)^q}$.
Therefore, its scaling can be determined in a similar way, which results in 
\begin{equation}\label{Eq4B:l_max_scaling}
    l_\imax \in \Theta \left( \gamma t + \frac{\log (1/\varepsilon)}{\log ( e+ (\gamma t)^{-1} \log (1/\varepsilon) )} \right).
\end{equation}
To reproduce the truncated Floquet-Hilbert space, we should prepare an ancillary system labeled by $\{ \ket{l} \}_{l\in D^{l_\imax}}$.
The number of qubits for such ancillary system is at-most
\begin{equation}
    \order{\log l_\imax} \subset \order{\log (\gamma t) + \log \log (1/\varepsilon)}.
\end{equation}

\section{Amplitude amplification of the dynamics}\label{Sec:Amplification}

In this section, we show two kinds of amplification to achieve sufficiently high success probability of extracting the time-evolved state $\ket{\psi(t)}$.

As we stated in Section \ref{Subsec:Overview_algorithm}, the approximate state $\ket{\psi^{l_\imax}(t)}$ cannot be directly realized on quantum circuits. 
After simulating the dynamics in the truncated Floquet-Hilbert space to get $\ket{\Psi^{l_\imax}(t)}$ defined by Eq. (\ref{Eq3A:def_large_psi_l_max}), we make a projection to $\ket{a^{l_\imax}}$.
Although the resulting renormalized state is sufficiently close to $\ket{\psi^{l_\imax}(t)}$ and also the time-evolved state $\ket{\psi(t)}$ as
\begin{equation}
    \frac{\braket{a^{l_\imax}|\Psi^{l_\imax}(t)}}{ \norm{\braket{a^{l_\imax}|\Psi^{l_\imax}(t)}}} = \frac{\ket{\psi^{l_\imax}(t)}}{\norm{\ket{\psi^{l_\imax}(t)}}} = \ket{\psi(t)} + \order{\varepsilon}
\end{equation}
for $l_\imax$ given by Eq. (\ref{Eq4B:l_max_choice}), we should be careful of the low success probability; it becomes $\norm{\braket{a^{l_\imax}|\Psi^{l_\imax}(t)}}^2 \in \order{(l_\imax)^{-1}}$.

We need approximately $\order{\gamma t}$-times trials of the post selection, and every trial is expected to require at-least $\order{t}$ complexity for implementing $\ket{\Psi^{l_\imax}(t)} = e^{-i \ms{H}_\eff^{l_\imax} t} \ket{0}\otimes \ket{\psi(0)}$.
Therefore, the naive implementation based on Eq. (\ref{Eq3A:def_psi_l_max}) is not efficient for the time-evolved state $\ket{\psi(t)}$ in that the expected computational time reaches $\order{t^2}$.

We resolve this problem by the amplification of the success probability up to $1-\order{\varepsilon}$ in the next section.
The first one, which exploits the symmetry of the effective Hamiltonian $\ms{H}_\eff$, amplifies from $\order{l_\imax^{-1}}$ to $\order{1}$. 
The latter one following this, which is reminiscent of the Grover's search algorithm, allows the success probability $1-\order{\varepsilon}$.
As discussed later, using only either one fails to efficiently compute $\ket{\psi(t)}$.

\subsection{Amplification by symmetry}\label{Subsec:amplification_symmetry}

We introduce the amplification exploiting the symmetry of the effective Hamiltonian $\ms{H}_\eff$.
First of all, we specify the symmetry here; it is about the translation of the photon number from $\ket{l}$ to $\ket{l + m}$.
When we define the translation operator on the Floquet-Hilbert space by $\ms{T}_m = \sum_{l \in \bbZ} \ket{l}\bra{l+m} \otimes I$, $\ms{H}_\eff$ satisfies the translation symmetry,
\begin{eqnarray}
    \ms{T}_m^\dagger \ms{H}_\eff \ms{T}_m &=& \ms{H}_\eff + m\omega, \label{Eq5A:symmetry_effective_H} \\
    \braket{l|e^{-i \ms{H}_\eff t}|l^\prime} &=& e^{il^\prime \omega t} \braket{l-l^\prime|e^{-i \ms{H}_\eff t}|0}. \label{Eq5A:symmetry_transition_amplitude}
\end{eqnarray}
Since the symmetry is present as long as the time-periodicity $H(t+T)=H(t)$ holds, the amplification discussed here is always available in our algorithm.

Here, we provide two tasks for the amplification by symmetry. 
First, we slightly extend the truncated Floquet-Hilbert space to $\bbC^{8l_\imax} \otimes \mcl{H}$, where the Fourier index $\ket{l}$ is chosen from $l \in D^{4l_\imax} = \{-4l_\imax+1, -4 l_\imax+2, \hdots, 4l_\imax\}$.
The number of additional qubits required for this extension from $\bbC^{l_\imax} \otimes \mcl{H}$ is three, and hence this does not affect the scaling of the computational resources.
The second task is to preprocess the initial state by
\begin{equation}
    \ms{U}_\mr{ini}^{l_\imax} \ket{0}\ket{\psi(0)} = \ket{a^{l_\imax}} \ket{\psi(0)},
\end{equation}
where the unitary operator $\ms{U}_\mr{ini}^{l_\imax}$ is expressed by
\begin{equation}
    \ms{U}_\mr{ini}^{l_\imax} = \left( \ket{a^{l_\imax}} \bra{0} + \hdots \right) \otimes I.
\end{equation}
The unitary operator $\ms{U}_\mr{ini}^{l_\imax}$ nontrivially acts only on the ancillary space, and it can be implemented with $\order{\log l_\imax}$ gates
(e.g. for $l_\imax$ such that $\log_2 l_\imax \in \bbN$, we have $\ms{U}_\mr{ini}^{l_\imax} = \mr{Had}^{\otimes \log_2 (2l_\imax)} \otimes I$ with the Hadamard gate $\mr{Had}$).
This process plays a role of making the initial state approximately translation-invariant with the width of Fourier indices, $2l_\imax$.
The remaining procedures after the above two tasks is the same as those for the original protocol in Section \ref{Subsec:Overview_algorithm}.
Reflecting that the ancillary Hilbert space is extended to $\bbC^{8l_\imax}$, we evolve the above uniform state $\ket{a^{l_\imax}} \ket{\psi(0)}$ by $\ms{H}_\eff^{4l_\imax}$ and $\sum_{l \in D^{4l_\imax}} l\omega \ket{l} \bra{l} \otimes I$, and make a projection to $\ket{a^{4l_\imax}}$.
The resulting state $\ket{\bar{\psi}^{l_\imax}(t)}$ is calculated as follows.
\begin{equation}\label{Eq5A:dynamics_uniform_initial_state}
    \ket{\bar{\psi}^{l_\imax}(t)} \equiv \bra{a^{4l_\imax}} e^{-it \sum_l l \omega \ket{l}\bra{l} }e^{-i\ms{H}_\eff^{4l_\imax}t} \ket{a^{l_\imax}} \ket{\psi(0)}.
\end{equation}

The point of this process is that the above state $\ket{\bar{\psi}^{l_\imax}(t)}$ can reproduce $\ket{\psi(t)}$ with $\order{1}$ amplitude owing to the translation symmetry of the effective Hamiltonian, Eqs. (\ref{Eq5A:symmetry_effective_H}) and (\ref{Eq5A:symmetry_transition_amplitude}).
We give its brief explanation in this section, while the rigorous derivation is provided in Appendix \ref{Asubsec2:amplification_symmetry}.
Although we do not have the exact translation symmetry represented by Eqs. (\ref{Eq5A:symmetry_transition_amplitude}) due to the finite-size effect of $l_\imax$, it is expected to approximately hold as 
\begin{equation}\label{Eq5A:approximate_symmetry}
     \bra{l} e^{-i\ms{H}_\eff^{4l_\imax}t} \ket{l^\prime} \simeq e^{il^\prime \omega t}\bra{l-l^\prime} e^{-i\ms{H}_\eff^{4l_\imax}t} \ket{0}.
\end{equation}
if the truncation order $l_\imax$ is sufficiently large.
As a matter of fact, we can derive the exact upper bound on the difference between the left- and right-hand sides based on the Lieb-Robinson bound (See Lemma \ref{LemmaA2:translation_symmetry} in Appendix \ref{Asubsec2:amplification_symmetry}).
We proceed the discussion with assuming the approximate relation Eq. (\ref{Eq5A:approximate_symmetry}) here.
The resulting state of the process $\ket{\bar{\psi}^{l_\imax}(t)}$ can be roughly computed as follows;
\begin{eqnarray}
    \ket{\bar{\psi}^{l_\imax}(t)} &=& \frac{1}{4l_\imax}\sum_{l^\prime \in D^{l_\imax}}  \sum_{l \in D^{4l_\imax}} e^{-il\omega t} \bra{l} e^{-i\ms{H}_\eff^{4l_\imax}t} \ket{l^\prime} \ket{\psi(0)} \nonumber \\
    &\simeq& \frac{1}{4l_\imax}\sum_{l^\prime \in D^{l_\imax}}   \sum_{l \in D^{4l_\imax}} e^{-i(l-l^\prime)\omega t} \bra{l-l^\prime} e^{-i\ms{H}_\eff^{4l_\imax}t} \ket{0} \ket{\psi(0)} \nonumber \\
    &=& \frac{1}{4l_\imax}\sum_{l^\prime \in D^{l_\imax}} (\ket{\psi(t)} + \order{\varepsilon}) \nonumber \\
    &=& \frac12 \ket{\psi(t)} + \order{\varepsilon}. \label{Eq5A:amp1_calculation}
\end{eqnarray}
The second equality comes from Theorem \ref{Thm4:Floquet_Hilbert_truncation}, considering that the summation of $l-l^\prime$ over $l \in D^{4l_\imax}$ is sufficient to suppress the error up to $\order{\varepsilon}$.
As a result, we obtain
\begin{equation}
    \ket{\bar{\psi}^{l_\imax}(t)} \simeq  \frac{1}{2}(\ket{\psi(t)}+\order{\varepsilon}).
\end{equation}

\begin{figure}
\begin{center}
    \includegraphics[height=6cm, width=15cm]{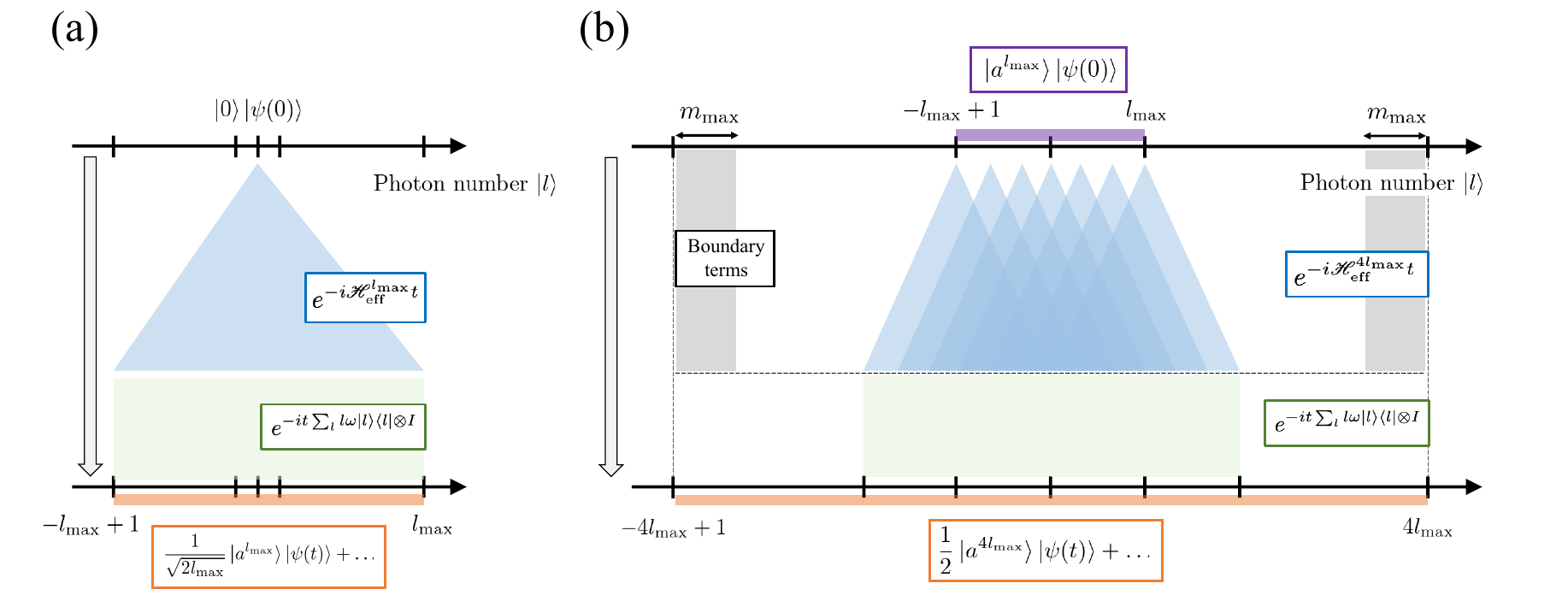}
    \caption{Schematic picture of how the dynamics in the truncated Floquet-Hilbert space provides the time-evolved state $\ket{\psi(t)}$. (a) The dynamics when we begin with the initial state $\ket{0}\ket{\psi(0)}$ [See Eq. (\ref{Eq3A:def_large_psi_l_max})]. Following the Lieb-Robinson bound, Theorem \ref{Thm4:Bound_transition_rate}, the state spreads out within the region $|l| \lesssim l_\imax$. This spread results in the low amplitude $\order{1/\sqrt{l_\imax}}$ of the target output $\ket{a^{l_\imax}}\ket{\psi(t)}$. (b) The dynamics under the amplification by symmetry, where we employ the uniform initial state $\ket{a^{l_\imax}}\ket{\ket{\psi(0)}}$ as Eq. (\ref{Eq5A:dynamics_uniform_initial_state}). Interference takes place within the region $|l| \lesssim l_\imax$, and thereby enhances the amplitude of the uniform target state $\ket{a^{4l_\imax}}\ket{\psi(t)}$ up to $\order{1}$. The gray region represents the support of boundary terms in the refined effective Hamiltonian $\ms{H}_{\eff,\mr{pbc}}^{4l_\imax}$ [See Eq. (\ref{Eq6B:def_effective_H_pbc})]. It is sufficiently far from the light-blue region relevant to the dynamics, which is determined by the Lieb-Robinson bound. This results in the validity of the refined effective Hamiltonian as Theorem \ref{Thm6:refined_H_eff} says. } 
    \label{fig5A:dynamics}
 \end{center}
 \end{figure}

Let us focus on the success probability of the projection onto $\ket{a^{4l_\imax}}$. 
It is provided by $\braket{\bar{\psi}^{l_\imax}(t)|\bar{\psi}^{l_\imax}(t)} = 1/4 + \order{\varepsilon}$, which is much larger than the original one $1/(2l_\imax)$.
Therefore, the amplification protocol by symmetry enables us to avoid $\order{t}$-times repetition of the time evolution $e^{-i \ms{H}_\eff t}$ in contrast to the original protocol.
Intuitively, this drastic improvement can be understood as a result of the interference under the translation symmetry, as Fig. \ref{fig5A:dynamics}.
As we can see from Eq. (\ref{Eq3A:def_large_psi_l_max}), the target state $\ket{\psi(t)}$ is extracted from $\ket{\Psi^{l_\imax}(t)}$ via the accompanying ancillary state $\ket{a^{l_\imax}}$, which is uniform in the Fourier index $\ket{l}$.
When we begin with the non-uniform initial state $\ket{0}\ket{\psi(0)}$, the resulting state after the time evolution is also non-uniform.
Since it involves $(1/\sqrt{2l_\imax}) \sum_{l \in D^{l_\imax}} e^{ikl} \ket{l}$ for $k \in (2\pi / 2l_\imax) \bbZ$ (different eigenstates of the translation operator) with approximate equal weight, the amplitude of the desirable component $\ket{a^{l_\imax}}$ is comparably small as $1/\sqrt{2l_\imax}$.
In contrast, when we employ the uniform initial state $\ket{a^{l_\imax}}\ket{\psi(0)}$, only the uniform components that includes $\ket{a^{l_\imax}}$ are amplified while the other components cancels one another.
It can be viewed also as the interference of different initial states $\ket{l^\prime}\ket{\psi(0)}$ in $\ket{a^{l_\imax}}\ket{\psi(0)}$ as shown in Fig. \ref{fig5A:dynamics}, and this is why the amplification protocol achieves $\order{1}$ amplitude of $\ket{a^{l_\imax}}\ket{\psi(0)}$.

We summarize the amplification by symmetry with adding the exact results obtained in Appendix \ref{Asubsec2:amplification_symmetry}.
We prepare the truncated Floquet-Hilbert space $\bbC^{8 l_\imax} \otimes \mcl{H}$, and make the initial state uniform as $\ket{a^{l_\imax}}\ket{\psi(0)}$.
As a result, this process is described by a unitary gate
\begin{equation}\label{Eq5A:U_amp1_def}
     \ms{U}_\mr{amp1}^{l_\imax} (t) = (\ms{U}_\mr{ini}^{4l_\imax})^\dagger e^{-i t \sum_l l \omega \ket{l}\bra{l}} e^{-i \ms{H}_\eff^{4l_\imax} t} \ms{U}_\mr{ini}^{l_\imax},
\end{equation}
where the last unitary gate $(\ms{U}_\mr{ini}^{4l_\imax})^\dagger$ is added to replace the projection to $\ket{a^{4l_\imax}}$ by the one to $\ket{0}$.
This approximately realizes the time-evolved state $\ket{\psi(t)}$ with $1/4+\order{\varepsilon}$ success probability as
\begin{equation}\label{Eq5A:approximate_amp1_result}
    \bra{0} \ms{U}_\mr{amp1}^{l_\imax} (t) \ket{0} \ket{\psi(0)} \simeq \frac{1}{2} (\ket{\psi(t)}+\order{\varepsilon}).
\end{equation}
The probability $1/4$ comes from the ratio of the width of the initial state $\ket{a^{l_\imax}}$ to that of the projected ancillary state $\ket{a^{4l_\imax}}$.
The exact version, derived based on the Lieb-Robinson bound in Appendix \ref{Asubsec2:amplification_symmetry}, is stated as follows, and ensures the validity of the discussion here.

\begin{theorem}\label{Thm5:amplification_symmetry}
\textbf{(Amplification by symmetry)}

We designate the truncation order $l_\imax \in \Theta (\gamma t + \log (1/\varepsilon) / \log \log (1/\varepsilon))$ by Eq. (\ref{Eq4B:l_max_choice}). Then, 
\begin{equation}
    \norm{\bra{0} \ms{U}_\mr{amp1}^{l_\imax} (t) \ket{0} \ket{\psi(0)} - \frac{1}{2}\ket{\psi(t)}} \leq \frac{\varepsilon}{3}.
\end{equation}
is satisfied for an arbitrary initial state $\ket{\psi(0)} \in \mcl{H}$.
\end{theorem}

We finally note that the amplification solely by this process cannot achieve the success probability $1-\order{\varepsilon}$ with keeping the efficiency.
Analogous to the above formulation, when we set the widths of the supports of the initial ancillary state, the effective Hamiltonian, and the projected ancillary state to $p l_\imax$, $q l_\imax$, $q l_\imax$ ($p, q \in \bbN$, $q \geq p +2$) respectively, the success probability becomes $p/q$.
However, the probability larger than $1 - \varepsilon$ demands the relation,
\begin{equation}
    1- \varepsilon < \frac{p}{q} \leq \frac{p}{p+2}.
\end{equation}
This implies $p \geq 2(1-\varepsilon)/\varepsilon$, and hence the number of ancillary qubits reaches at-least $\log (p l_\imax) \in \order{\log (\gamma t/\varepsilon)}$.
In addition, the query complexity includes a term proportional to the dimension of the ancillary system, $q l_\imax$ as discussed later in Section \ref{Sec:Block_encoding_effective}.
This implies that the query complexity increases linearly in $1/\varepsilon$, destroying the original logarithmic scaling.
Thus, relying solely on this amplification protocol is not suitable for efficient simulation of the dynamics.
We resolve this problem by the oblivious amplitude amplification below.

\subsection{Oblivious amplitude amplification}\label{Subsec:Oblivious_amplification}
The above amplification based on the translation symmetry of $\ms{H}_\eff$ enhances the amplitude of $\ket{\psi(t)}$ from $1/\sqrt{2l_\imax}$ to $1/2-\order{\varepsilon}$.
Here, we introduce another amplification, called the oblivious amplitude amplification \cite{Berry2017-oblivious}, to achieve the amplitude (or the success probability) $1-\order{\varepsilon}$.

The starting point of this amplification is the result of the previous section \ref{Subsec:amplification_symmetry}.
Equation (\ref{Eq5A:approximate_amp1_result}), or equivalently Theorem \ref{Thm5:amplification_symmetry}, indicates that the consequence of the amplification by symmetry is written as
\begin{equation}\label{Eq5B:action_amp1}
    \ms{U}_\mr{amp1}^{l_\imax} (t) \ket{0}\ket{\psi(0)} = \frac{1}{2} \ket{0}  (\ket{\psi(t)}+\order{\varepsilon}) + \ket{\Psi^\perp},
\end{equation}
with an additional term $\ket{\Psi^\perp} \in \bbC^{8l_\imax} \otimes \mcl{H}$ satisfying 
\begin{equation}
    (\ket{0}\bra{0} \otimes I)\ket{\Psi^\perp} = 0.
\end{equation}
The state $\ket{\Psi^\perp}$ generally depends on $\ket{\psi(0)}$ and $t$.
The oblivious amplitude amplification takes a similar strategy to that of the Grover's search algorithm, in which we compose the following two unitary operators;
\begin{eqnarray}
    \ms{R} &=& (2\ket{0}\bra{0}-I) \otimes I, \label{Eq5B:def_reflection} \\
    \ms{U}_\mr{amp2}^{l_\imax} (t) &=& - \ms{U}_\mr{amp1}^{l_\imax} (t) \ms{R} [\ms{U}_\mr{amp1}^{l_\imax} (t)]^\dagger \ms{R} \ms{U}_\mr{amp1}^{l_\imax} (t). \label{Eq5B:def_amp2}
\end{eqnarray}
The first one $\ms{R}$ reverses the sign of $\ket{l}$ for $l \neq 0$, and it is implemented with $\order{\log l_\imax}$ gates.
The second one $\ms{U}_\mr{amp2}^{l_\imax} (t)$ plays a role in enhancing the amplitude of $\ket{\psi(t)}$ up to $1-\order{\varepsilon}$. 
Its action on any initial state $\ket{0}\ket{\psi(0)} $ is computed as follows;
\begin{eqnarray}
    \ms{U}_\mr{amp2}^{l_\imax} \ket{0}\ket{\psi(0)} &=& \ms{U}_\mr{amp1}^{l_\imax}\ms{R} (\ms{U}_\mr{amp1}^{l_\imax})^\dagger \left( - \frac{1}{2} \ket{0}\ket{\psi(t)} + \ket{\Psi^\perp} \right) + \order{\varepsilon} \nonumber \\
    &=& \ms{U}_\mr{amp1}^{l_\imax}\ms{R} \left\{ \ket{0}\ket{\psi(0)} - (\ms{U}_\mr{amp1}^{l_\imax})^\dagger \ket{0}\ket{\psi(t)}  \right\} + \order{\varepsilon}. \label{Eq5B:amp2_middle_1}
\end{eqnarray}
In the second equality, we use the relation obtained by applying $(\ms{U}_\mr{amp1}^{l_\imax})^\dagger$ to Eq. (\ref{Eq5B:action_amp1}).
Next, we evaluate 
\begin{eqnarray}
\ms{R} (\ms{U}_\mr{amp1}^{l_\imax})^\dagger \ket{0,\psi(t)} = 2 \ket{0} \left( \bra{0} (\ms{U}_\mr{amp1}^{l_\imax})^\dagger \ket{0} \right) \ket{\psi(t)} - (\ms{U}_\mr{amp1}^{l_\imax})^\dagger \ket{0,\psi(t)}. 
\end{eqnarray}
Theorem \ref{Thm5:amplification_symmetry} indicates that the time evolution operator $U(t)$ is approximated as 
\begin{equation}
    \norm{\braket{0|\ms{U}_\mr{amp1}^{l_\imax}(t)|0} - \frac{1}{2} U(t)} \leq \frac{\varepsilon}{3},
\end{equation}
and hence the relation
\begin{equation}
    \norm{\braket{0|[\ms{U}_\mr{amp1}^{l_\imax}(t)]^\dagger|0} - \frac{1}{2} U(t)^\dagger} \leq \frac{\varepsilon}{3}
\end{equation}
is also satisfied. 
This provides the relation,
\begin{eqnarray}
    \ms{R} (\ms{U}_\mr{amp1}^{l_\imax})^\dagger \ket{0}\ket{\psi(t)} = \ket{0}\ket{\psi(0)} - (\ms{U}_\mr{amp1}^{l_\imax})^\dagger \ket{0}\ket{\psi(t)} + \order{\varepsilon},
\end{eqnarray}
and substituting this into Eq. (\ref{Eq5B:amp2_middle_1}) results in 
\begin{equation}
    \ms{U}_\mr{amp2}^{l_\imax}(t) \ket{0} \ket{\psi(0)} = \ket{0} \ket{\psi(t)} + \order{\varepsilon},
\end{equation}
for an arbitrary initial state $\ket{\psi(0)}$.

This result indicates that the operation $\ms{U}_\mr{amp2}^{l_\imax}(t)$ generates the time-evolved state $\ket{\psi(t)}$ with the amplitude $1-\order{\varepsilon}$.
Or equivalently, as an exact bound for the error, we can derive the inequality,
\begin{equation}\label{Eq5B:deviation_amp2}
    \norm{\ms{U}_\mr{amp2}^{l_\imax}(t) \ket{0} \ket{\psi(0)} - \ket{0}  \ket{\psi(t)}} \leq \varepsilon.
\end{equation}
The coefficient comes from the fact that an error bounded by $\varepsilon /3$ appears due to Theorem \ref{Thm5:amplification_symmetry} every time we call Eq. (\ref{Eq5B:action_amp1}).
The amplification protocol $\ms{U}_\mr{amp2}^{l_\imax}(t)$ employs $3$ times queries to $\ms{U}_\mr{amp1}^{l_\imax}(t)$. 
Reflecting that the operations $\ms{U}_\mr{ini}^{l_\imax}$ and $\ms{R}$ require relatively a little resource (at-most $\order{\log l_\imax}$ elementary gates and complexity), the resource for $\ms{U}_\mr{amp2}^{l_\imax}(t)$ has the same scaling as the one for implementing the time evolution operators $e^{-it \sum_l l\omega \ket{l} \bra{l}}$ and $e^{-i \ms{H}_\eff^{4l_\imax}t}$.

As well as the amplification by symmetry, relying only on the oblivious amplitude amplification fails to efficiently enhances the success probability to obtain $\ket{0} \otimes \ket{\psi(t)}$ from $\order{l_\imax^{-1}}$ to $1-\order{\varepsilon}$.
When we do not use the first amplification, the protocol of the oblivious amplitude amplification is given by
\begin{equation}
    \tilde{\ms{U}}_{\mr{amp2}, p}^{l_\imax} = - \left\{ \ms{U}_\mr{orig}^{l_\imax} \ms{R} (\ms{U}_\mr{orig}^{l_\imax})^\dagger \ms{R} \right\}^p \ms{U}_\mr{orig}^{l_\imax},
\end{equation}
where the operation $\ms{U}_\mr{orig}^{l_\imax}$ represents the time evolution without the amplification by symmetry, defined by
\begin{equation}
    \ms{U}_\mr{orig}^{l_\imax}(t) = (\ms{U}_\mr{ini}^{l_\imax})^\dagger e^{-it \sum_l l\omega \ket{l} \bra{l}} e^{-i \ms{H}_\eff^{l_\imax}t}.
\end{equation}
Applying $\tilde{\ms{U}}_{\mr{amp2}, p}^{l_\imax}$ to the initial state $\ket{0} \otimes \ket{\psi(t)}$ returns $\ket{0} \otimes \ket{\psi(t)}$ whose amplitude increases from $1/\sqrt{2l_\imax}$ approximately in proportion to $p$.
The integer $p$ should be $\order{\sqrt{l_\imax}}$ to achieve the amplitude $1-\order{\varepsilon}$, reminiscent of the Grover's search algorithm.
In other words, $\order{\sqrt{l_\imax}}$ times call of $e^{-it \sum_l l\omega \ket{l} \bra{l}}$ and $e^{-i \ms{H}_\eff^{l_\imax}t}$ is required when we use $\tilde{\ms{U}}_{\mr{amp2}, p}^{l_\imax}$.
This is why we suggest the combination of the two amplification protocols, with which $\order{1}$ times call of $e^{-it \sum_l l\omega \ket{l} \bra{l}}$ and $e^{-i \ms{H}_\eff^{l_\imax}t}$ can amplify the amplitude of $\ket{0} \otimes \ket{\psi(t)}$ from $1/\sqrt{2l_\imax}$ to $1-\order{\varepsilon}$.

\section{Block-encoding of Effective Floquet Hamiltonian}\label{Sec:Block_encoding_effective}
In the previous section, we show that the combination of the two kinds of amplification protocols, implemented by $\ms{U}_\mr{amp2}^{l_\imax}(t)$ [See Eq. (\ref{Eq5B:deviation_amp2})], provides the target time-evolved state $\ket{\psi(t)}$ with an arbitrarily small error $\order{\varepsilon}$.
The computational resource for $\ms{U}_\mr{amp2}^{l_\imax}(t)$ is mostly determined by that for $e^{-it \sum_l l\omega \ket{l} \bra{l}}$ and $e^{-i \ms{H}_\eff^{4l_\imax}t}$.
Our strategy is to implement these two time evolution operators by the qubitization technique \cite{Low2019-qubitization}.
As discussed in Section \ref{Subsec:Oblivious_amplification}, the computational resources are determined by how to introduce the block-encoding of the two static Hamiltonians, $\ms{H}_\mr{LP}^{l_\imax}=\sum_{l \in D^{l_\imax}} l \omega \ket{l}\bra{l} \otimes I$ (linear potential Hamiltonian) and $\ms{H}_\eff^{l_\imax}$ (effective Hamiltonian).
The aim of this section is to obtain an efficient block-encoding of them and to evaluate the costs required to implement $e^{-it \sum_l l\omega \ket{l} \bra{l}}$ and $e^{-i \ms{H}_\eff^{4l_\imax}t}$.

\subsection{Block-encoding of linear potential Hamiltonian}\label{Subsec:Block_encoding_linear_potential}
We compose block-encoding of the linear potential Hamiltonian,
\begin{equation}
    \ms{H}_\mr{LP}^{4l_\imax} = \sum_{l \in D^{4l_\imax}} l\omega \ket{l} \bra{l} \otimes I.
\end{equation}
By simple calculation, it can be written in the form of an LCU,
\begin{eqnarray}
    \ms{H}_\mr{LP}^{4l_\imax} &=& \sum_{l \in D^{4l_\imax}} \frac{\omega}{2} V_l^{4l_\imax} \otimes I, \label{Eq6A:LCU_linear_potential} \\
    V_l^{4l_\imax} &=& \sum_{l^\prime = l}^{4l_\imax} \ket{l^\prime} \bra{l^\prime} - \sum_{l^\prime = -4 l_\imax + 1}^{l-1} \ket{l^\prime} \bra{l^\prime}. \label{Eq6A:Unitaries_linear_potential}
\end{eqnarray}
Using the block-encoding formalism of LCUs by Eq. (\ref{Eq3B:block_encode_LCU}), we immediately obtain the oracle unitary gate and the oracle state by
\begin{eqnarray}
    && \ms{O}_\mr{LP}^{4l_\imax} = \sum_{l \in D^{4l_\imax}} (\ket{l}\bra{l})_b \otimes V_l^{4l_\imax} \otimes I, \label{Eq6A:Oracle_gate_linear_potential} \\
    && \braket{a^{4l_\imax}|\ms{O}_\mr{LP}^{4l_\imax}|a^{4l_\imax}}_b = \frac{\ms{H}_\mr{LP}^{4l_\imax}}{4l_\imax \omega }. \label{Eq6A:Oracle_state_linear_potential} 
\end{eqnarray}
The subscript $b$ for the states $\ket{l}_b$ and $\ket{a^{4l_\imax}}_b$ represents a new ancillary system introduced for the block-encoding, requiring the number of qubits $n_b \in \order{\log l_\imax}$.
Combining Eqs. (\ref{Eq6A:Unitaries_linear_potential}) and (\ref{Eq6A:Oracle_gate_linear_potential}), the oracle unitary gate is rewritten by
\begin{equation}
    \ms{O}_\mr{LP}^{4l_\imax} = \left (\sum_{l,l^\prime ; l^\prime-l \geq 0} \ket{l,l^\prime}\bra{l,l^\prime} - \sum_{l,l^\prime ; l^\prime-l < 0} \ket{l,l^\prime}\bra{l,l^\prime} \right) \otimes I . 
\end{equation}
To implement this oracle, we use a comparator defined by
\begin{equation}\label{Eq6A:comparator}
    \mr{Comp} \ket{l,l^\prime} \ket{0}_{b^\prime} = \begin{cases}
    \ket{l,l^\prime} \ket{0}_{b^\prime} & \text{if $l^\prime \geq l$}, \\
    \ket{l,l^\prime} \ket{1}_{b^\prime} & \text{if $l^\prime < l$},
    \end{cases}
\end{equation}
with a single-qubit ancillary system $b^\prime$ \cite{Cuccaro2004-jp}.
We can immediately confirm the relation,
\begin{equation}
    \mr{Comp}^\dagger Z_{b^\prime} \mr{Comp} \ket{l,l^\prime} \ket{0}_{b^\prime} \ket{\psi} = \ms{O}_\mr{LP}^{4l_\imax}  \ket{l,l^\prime} \ket{0}_{b^\prime} \ket{\psi},
\end{equation}
for arbitrary inputs $l,l^\prime \in D^{4l_\imax}$ and $\ket{\psi} \in \mcl{H}$, where $Z_{b^\prime}$ denotes a Pauli $Z$ operator on the system $b^\prime$.
Since the comparator on $n$ qubits can be composed of $\order{n}$ elementary gates, the oracle unitary gate $\ms{O}_\mr{LP}^{4l_\imax}$ requires at-most $\order{\log l_\imax}$ gates.
The oracle state $\ket{a^{4l_\imax}}_b$, which has equal weights in $\ket{l}$, can be prepared by $\order{\log l_\imax}$ elementary gates.

We implement the time-evolution $e^{-i \ms{H}_\mr{LP}^{4l_\imax} t}$ with accuracy $1-\order{\varepsilon}$ by qubitization.
The number of additional gates, other than the block-encoding, is proportional to the number of ancillary qubits for the oracle $\ms{O}_\mr{LP}^{4l_\imax}$, that is, we need $\order{\log l_\imax}$ additional gates per query.
Based on Section \ref{Subsec:preli_qubitization}, it can be executed with the following resources;
\begin{itemize}
    \item Number of ancillary qubits; $\order{\log l_\imax}$
    \item Scaling of overall complexity;
    \begin{equation}\label{Eq6A:gate_count_linear_potential}
        \left\{ l_\imax \omega t + \frac{\log (1/\varepsilon)}{\log ( e + (l_\imax \omega t)^{-1} \log (1/\varepsilon) )} \right\} \log l_\imax.
    \end{equation}
\end{itemize}

\subsection{Block-encoding of effective Hamiltonian}\label{Subsec:block_encode_H_eff}
We hereby provide a way to efficiently implement $e^{-i \ms{H}_\eff^{4l_\imax}t}$ by the qubitization technique.
We find out that naive block-encoding of the effective Hamiltonian $\ms{H}_\eff^{4l_\imax}$, given by Eq. (\ref{Eq3A:def_effective_H_l_max}), faces at the severe increase of the complexity in $l_\imax$.
This originates from the fact that the number of nontrivial terms in $\ms{H}_\eff^{4l_\imax}$ increases linearly in $l_\imax$.
To avoid this problem, we take an alternative approach which yields only $\order{\log l_\imax}$ complexity as follows;
\begin{enumerate}
    \item Find a refined effective Hamiltonian $\ms{H}$ on the truncated Floquet-Hilbert space $\bbC^{8l_\imax} \otimes \mcl{H}$, such that $e^{-i \ms{H} t}$ accurately reproduces $e^{-i \ms{H}_\eff^{4l_\imax}t}$ with an arbitrarily small error $\order{\varepsilon}$.
    \item Construct block-encoding of $\ms{H}$ with composing of efficient implementation of its oracle unitary gate and oracle state.
\end{enumerate}

First, we specify the assumption for the time-periodic Hamiltonian $H(t)$.
Here, we suppose two kinds of the oracles that can be accessed.
The first one is about the block-encoding of each Fourier component $H_m$ ($|m| \leq m_\imax$), given by
\begin{equation}\label{Eq6B:block_encoding_H_m}
    \bra{G_m}_a O_m \ket{G_m}_a  = \frac{H_m}{\alpha_m}, \quad \alpha_m > 0,
\end{equation}
with an $n_a$-qubit ancillary system. 
Each oracle unitary gate on $\bbC^{2^{n_a}} \otimes \mcl{H}$ is represented by $O_m$, and each oracle state $\ket{G_m}_a \in \bbC^{2^{n_a}}$ is generated by a quantum circuit $G_m$ as $\ket{G_m}_a = G_m \ket{0}^{\otimes n_a}$.
The second oracle is a query to coefficients of Fourier components, represented by $\alpha_m$.
We define the oracle unitary circuit $G_\mr{coef}$, acting on $\bbC^{2m_\imax+1}$ [i.e. $\order{1}$-qubit system], by
\begin{equation}\label{Eq6B:oracl_coefficient}
    G_\mr{coef} \ket{0} = \sum_{|m| \leq m_\imax} \sqrt{\frac{\alpha_m}{\alpha}} \ket{m}, \quad \alpha = \sum_{|m| \leq m_\imax} \alpha_m.
\end{equation}
The parameter $\alpha$, defined by $\alpha = \sum_{|m| \leq m_\imax} \alpha_m$, characterizes the energy scale of the Hamiltonian $H(t)$.
In fact, considering $\norm{H_m} \leq \alpha_m$ for each $m$, it is bounded by $\norm{H(t)} \leq \sum_{|m| \leq m_\imax} \norm{H_m} \leq \alpha$.
The parameter $\alpha$ also provides the bound on the energy scale of time-dependent terms $\gamma$, as
\begin{equation}\label{Eq6B:gamma_alpha}
    \gamma \leq \sum_{m \neq 0; |m| \leq m_\imax} \norm{H_m} \leq \alpha.
\end{equation}
Implementation of the oracle $G_\mr{coef}$ is equivalent to embedding probability distributions in quantum states \cite{Grover2002-ry}, which also appears as the oracle state for the qubitization technique in the case of a LCU [See Eq. (\ref{Eq3B:block_encode_LCU})].
Stimulated by its various applications covering linear equation solver \cite{Harrow2009hhl} and quantum singular value transformation \cite{Gilyen2019-qsvt,MartynPRXQ-grand-unif}, there have been various efficient implementations \cite{Grover2000-qg,Grover2002-ry,Babbush2018-hz,Sanders2019-je,Rattew2022-uh}.
Here, we suppose that the number of elementary gates required for the oracles $O_m$, $G_m$, and $G_\mr{coef}$ is at-most $C$.
As discussed later in Section \ref{Sec:examples}, the form of time-periodic Hamiltonians $H(t)$ designated by Eq. (\ref{Eq6B:block_encoding_H_m}) is reasonable in that they involve various important classes such as LCUs.

\subsubsection{Construction of a refined effective Hamiltonian}
Let us compose a refined effective Hamiltonian, which accurately reproduces the dynamics of $e^{-i\ms{H}_\eff^{4l_\imax}t}$.
To come to the point, such a Hamiltonian equipped with suitability for efficient block-encoding is given by
\begin{equation}\label{Eq6B:def_effective_H_pbc}
    \ms{H}_{\eff,\mr{pbc}}^{4l_\imax} = \ms{H}_\eff^{4l_\imax} + \sum_{(l,m) \in \partial F^{4l_\imax}} (\ket{l}\bra{l \oplus m} \otimes H_{-m} + \text{h.c.}),
\end{equation}
with $\partial F^{4 l_\imax} = \{ (l,m) \, | \, 4l_\imax-m_\imax+1 \leq l \leq 4l_\imax, \, 4l_\imax - l +1 \leq m \leq m_\imax \}$.
The integer $l \oplus m \in D^{4l_\imax}$ is defined modulo $8l_\imax$.
The additional terms in $\ms{H}_{\eff,\mr{pbc}}^{4l_\imax}$ connects the boundary of $D^{4l_\imax}$ so that the hopping terms, $\ket{l}\bra{l+m}$, induced by $H_{-m}$ can be translation-invariant under the shift of $\ket{l}$.
The subscript ``pbc" comes from the periodic boundary conditions (PBC) for the hopping terms.

In our algorithm, we employ the time evolution $e^{-i \ms{H}_{\eff,\mr{pbc}}^{4l_\imax} t}$ instead of the original one $e^{-i \ms{H}_{\eff}^{4l_\imax} t}$, which is derived by Floquet theory.
In other words, we organize the amplification protocols with the refined effective Hamiltonian $\ms{H}_{\eff,\mr{pbc}}^{4l_\imax}$ as 
\begin{eqnarray}
    \ms{U}_\mr{amp1,pbc}^{l_\imax} (t) &=& (\ms{U}_\mr{ini}^{4l_\imax})^\dagger e^{-i \ms{H}_\mr{LP}^{4l_\imax}t} e^{-i \ms{H}_\mr{eff,pbc}^{4l_\imax}t} \ms{U}_\mr{ini}^{l_\imax}, \label{Eq6B:def_amp1_pbc} \\
    \ms{U}_\mr{amp2,pbc}^{l_\imax} (t) &=& - \ms{U}_\mr{amp1,pbc}^{l_\imax}\ms{R} (\ms{U}_\mr{amp1,pbc}^{l_\imax})^\dagger \ms{R} \ms{U}_\mr{amp1,pbc}^{l_\imax}, \label{Eq6B:def_amp2_pbc}
\end{eqnarray}
and apply them to the initial state $\ket{0}\ket{\psi(0)}$.
The refined effective Hamiltonian is valid in a sense that it can also provide the exact time-evolved state as
\begin{equation}\label{Eq6B:approximate_amp2_pbc_result}
    \ms{U}_\mr{amp2,pbc}^{l_\imax} (t) \ket{0}\ket{\psi(0)} = \ket{0}\ket{\psi(t)} + \order{\varepsilon}.
\end{equation}
The rigorous upper bound on the error is provided by the following theorem.

\begin{theorem}\label{Thm6:refined_H_eff}
\textbf{(Refined effective Hamiltonian)}

We choose $l_\imax \in \order{\gamma t + \log (1/\varepsilon)/\log \log (1/\varepsilon)}$ by Eq. (\ref{Eq4B:l_max_choice}).
Then, the refined effective Hamiltonian $\ms{H}_{\eff,\mr{pbc}}^{4l_\imax}$ well reproduces the dynamics by the original one $\ms{H}_\eff^{4l_\imax}$ as
\begin{eqnarray}
    \norm{\bra{0} \ms{U}_\mr{amp1,pbc}^{l_\imax} (t) \ket{0} \ket{\psi(0)} - \frac12 \ket{\psi(t)}} &\leq& \frac{\varepsilon}{3}, \label{Eq6B:thm_amp1_pbc}\\
    \norm{ \ms{U}_\mr{amp2,pbc}^{l_\imax} (t) \ket{0} \ket{\psi(0)} - \ket{0}\ket{\psi(t)}} &\leq& \varepsilon, \label{Eq6B:thm_amp2_pbc}
\end{eqnarray}
for an arbitrary initial state $\ket{\psi(0)} \in \mcl{H}$.
\end{theorem}

We deliver its detailed proof in Appendix \ref{Asubsec2:refine_effective_Hamiltonian}, and instead we briefly explain why the refined effective Hamiltonian is valid.
The proof relies mainly on the Lieb-Robinson bound, stated by Theorem \ref{Thm4:Bound_transition_rate}.
Focusing on the amplification protocols $\ms{U}_\mr{amp1}^{l_\imax} (t)$ and $\ms{U}_\mr{amp1,pbc}^{l_\imax} (t)$, the resulting difference caused by them originates from their actions on the uniform initial state $\ket{a^{l_\imax}}\ket{\psi(0)}$, as Eq. (\ref{Eq5A:U_amp1_def}).
As we can see from Fig. \ref{fig5A:dynamics} (b), the Lieb-Robinson bound dictates that the dynamics during $\ms{U}_\mr{amp1}^{l_\imax} (t)$ is almost closed within $- 2l_\imax \lesssim l \lesssim 2l_\imax$ in the Fourier indices.
In contrast, the support of the additional terms of the refined effective Hamiltonian in Eq. (\ref{Eq6B:def_effective_H_pbc}) is located at $\{ \ket{l} \}_l$ for $l \simeq \pm 4l_\imax$, which is sufficiently far from the Fourier indices relevant for the dynamics.
As a result, the refined protocol $\ms{U}_\mr{amp1,pbc}^{l_\imax} (t)$ transforms the state $\ket{a^{l_\imax}}\ket{\psi(0)}$ in almost the same way as the original one $\ms{U}_\mr{amp1}^{l_\imax} (t)$, and hence it also outputs the target state $\ket{\psi(t)}$ as Eq. (\ref{Eq6B:thm_amp1_pbc}).
Discussion similar to Section \ref{Subsec:Oblivious_amplification} ensures that the oblivious amplitude amplification under the refined effective Hamiltonian, represented by $\ms{U}_\mr{amp2,pbc}^{l_\imax} (t)$, provides $\ket{\psi(t)}$ with a sufficiently small error $\order{\varepsilon}$ as Eq. (\ref{Eq6B:thm_amp2_pbc}).

\subsubsection{Block-encoding of a refined effective Hamiltonian}
The benefit of employing the refined effective Hamiltonian $\ms{H}_{\eff,\mr{pbc}}^{4l_\imax}$ instead of the original one $\ms{H}_{\eff}^{4l_\imax}$ is reduction of resources for implementing block-encoding.
We construct an oracle unitary gate $\ms{O}_\eff^{4l_\imax}$ and an oracle state $\ket{G_\eff^{4l_\imax}}$ for it. 
With the help of the additional terms in Eq. (\ref{Eq6B:def_effective_H_pbc}), the Hamiltonian is written in the following form;
\begin{eqnarray}
    \ms{H}_{\eff,\mr{pbc}}^{4l_\imax} &=& \sum_{|m| \leq m_\imax} \mr{Add}_m^{4l_\imax} \otimes H_m - \ms{H}_\mr{LP}^{4l_\imax}, \\
    \mr{Add}_m^{4l_\imax} &=& \sum_{l \in D^{4l_\imax}} \ket{l \oplus m}\bra{l}. \label{Eq6B:def_adder}
\end{eqnarray}
The unitary gate $\mr{Add}_m^{4l_\imax}$ is a full quantum adder that translates an index $l$ by $m$ modulo $8l_\imax$.
It can be implemented by $\order{\log l_\imax}$ complexity and elementary gates.

Then, to construct the block-encoding, we prepare four kinds of auxiliary systems labeled by $a,b,c,d$. 
The system $a$ is an $n_a$-qubit system prepared for the block-encoding of $H_m$ [See Eq. (\ref{Eq6B:block_encoding_H_m})], and the system $b$ is the one for $\ms{H}_\mr{LP}^{4l_\imax}$ having $8l_\imax$ dimension [See Section \ref{Subsec:Block_encoding_linear_potential}].
The other systems $c$ and $d$ have the dimension $2m_\imax +1$ and $2$ respectively, yielding the number of qubits $n_{cd} \in \order{\log m_\imax}$ (it is a constant here).
We define a unitary gate $\ms{O}_\eff^{4l_\imax}$ by
\begin{eqnarray}
    \ms{O}_\eff^{4l_\imax} = \ket{0}\bra{0}_d \otimes \sum_{|m| \leq m_\imax} \ket{m}\bra{m}_c \otimes \mr{Add}_m^{4l_\imax}\otimes O_m - \ket{1}\bra{1}_d  \otimes \ms{O}_\mr{LP}^{4l_\imax},  \label{Eq6B:Oracle_gate_effective_H}
\end{eqnarray}
where we omit the identity operators.
We also provide a state $\ket{G_\eff^{4l_\imax}}$ by
\begin{eqnarray}
\ket{G_\eff^{4l_\imax}} = \frac{ \sum_{m} \sqrt{\alpha_m} \ket{0}_d \ket{m}_c \ket{a^{4l_\imax}}_b \ket{G_m}_a + \sqrt{4l_\imax \omega} \ket{1}_d \ket{0}_c \ket{a^{4l_\imax}}_b \ket{0}_a}{\sqrt{\alpha + 4l_\imax \omega}}. \label{Eq6B:Oracle_state_effective_H}
\end{eqnarray}
In the above formula, we implicitly include ancillary qubits to efficiently implement the oracles [e.g. the comparator for $\ms{O}_\mr{LP}^{4l_\imax}$ in Eq. (\ref{Eq6A:comparator})].
The number of such ancillary qubits is smaller than $\order{\log l_\imax}$, and hence we neglect it below.
We can confirm that they provide the block-encoding of $\ms{H}_\eff^{4l_\imax}$ as follows;
\begin{equation}\label{Eq6B:block_encode_H_eff_pbc}
    \braket{G_\eff^{4l_\imax}|\ms{O}_\eff^{4l_\imax}|G_\eff^{4l_\imax}} = \frac{\ms{H}_{\eff,\mr{pbc}}^{4l_\imax}}{\alpha + 4 l_\imax \omega},
\end{equation}
where we use the block-encoding for $H_m$ and $\ms{H}_\mr{LP}^{4l_\imax}$ represented by Eqs. (\ref{Eq6B:block_encoding_H_m}) and (\ref{Eq6A:Oracle_state_linear_potential}).

We evaluate the resources for implementing the oracles.
The oracle unitary gate $\ms{O}_\eff^{4l_\imax}$ given by Eq. (\ref{Eq6B:Oracle_gate_effective_H}) requires elementary gates at-most $\order{m_\imax (\log l_\imax + C) + \log l_\imax}$, since the unitary operators $\mr{Add}_m^{4l_\imax}$, $O_m$, and $\ms{O}_\mr{LP}^{4l_\imax}$ respectively yield $\order{\log l_\imax}$, $C$, $\order{\log l_\imax}$ gates.
On the other hand, the oracle state $\ket{G_\eff^{4l_\imax}}$ can be prepared by
\begin{equation}
    \ket{G_\eff^{4l_\imax}} = G_\eff^{4l_\imax} \ket{w}_d (G_\mr{coef}\ket{0}_c) \ket{a^{4l_\imax}}_b \ket{0}_a^{\otimes n_a}, \\
\end{equation}
with using the quantum circuit $G_\eff^{4l_\imax}$ and the state $\ket{w}_d$ defined by
\begin{eqnarray}
    G_\eff^{4l_\imax}  &=& I_d \otimes \sum_{|m| \leq m_\imax} \ket{m}\bra{m}_c \otimes I_b \otimes (G_m)_a, \\
    \ket{w}_d &=& \sqrt{\frac{\alpha}{\alpha + 4l_\imax \omega }} \ket{0}_d + \sqrt{\frac{4l_\imax \omega}{\alpha+4l_\imax \omega}} \ket{1}_d. 
\end{eqnarray}
The quantum circuit $G_\eff^{4l_\imax}$ can be composed of $\order{m_\imax C}$ elementary gates. 
The cost of the state preparation $\ket{w}_d$ is negligible compared to others since it requires only a single-qubit rotation.
The cost for preparing $\ket{a^{4l_\imax}}_b$ is at-most $\order{\log l_\imax}$ complexity and elementary gates.
To summarize, the cost for the oracles $\ms{O}_\eff^{4l_\imax}$ and $\ket{G_\eff^{4l_\imax}}$, which embody the refined effective Hamiltonian $\ms{H}_{\eff,\mr{pbc}}^{4l_\imax}$ via Eq. (\ref{Eq6B:block_encode_H_eff_pbc}), is at-most $\order{m_\imax (C+\log l_\imax)}$ elementary gates.
The additional gates per query, other than the block-encoding, amounts to $\order{n_a + \log (l_\imax)}$.

As a result, preparing a unitary circuit corresponding to $e^{-i \ms{H}_\eff^{4l_\imax} t}$ with an allowable error up to $\order{\varepsilon}$ is summarized as follows;
\begin{itemize}
    \item Number of ancillary qubits; 
    \begin{equation}\label{Eq6B:number_ancillary}
        n_a (l_\imax) = n_a + \order{\log l_\imax + \log m_\imax}.
    \end{equation}
    \item Scaling of query complexity $q(l_\imax)$;
    \begin{equation}\label{Eq6B:query_complexity_effective_H}
        (\alpha + l_\imax \omega) t + \frac{\log (1/\varepsilon)}{\log (e + \{ (\alpha + l_\imax \omega) t \}^{-1} \log (1/\varepsilon))}.
    \end{equation}
    
    \item Number of overall gates;
    \begin{equation}
        \order{q(l_\imax) m_\imax \{ C + n_a + \log l_\imax \} }
        \label{Eq6B:gate_count_effective_H}
    \end{equation}
\end{itemize}
Since $m_\imax$ is supposed to be a $\order{1}$ constant, we will omit it from the cost in the rest of the paper.

\section{Algorithm and Its computational cost}\label{Sec:Algorithm_cost}
This section provides the main result of this paper; we compose the efficient quantum algorithm for simulating the time-evolved state $\ket{\psi(t)}=U(t)\ket{\psi(0)}$ under the Hamiltonian $H(t+T)=H(t)$.
We note that we take two different approaches depending on the time scale of interest.
The first case is the adiabatic-like case, in which we are interested in $\order{1}$-period dynamics with $t/T \in \order{1}$.
We call it ``adiabatic-like" since long-time dynamics during $0 \leq t \leq T$ under the sufficiently large period $T$, exemplified by Thouless pumping and adiabatic state preparation, is a typical target.
The second case is the generic long-time case, where we are interested in multiple-period dynamics at $t/T \in \Omega (1)$. 
In that case, the the period $T$ is not so large and we often consider long-time dynamics at $t \gg T$, exemplified by laser-irradiated materials.

\begin{table}[]
    \begin{center}
    \scalebox{0.78}{
    \begin{tabular}{|c|c|c|}
         &   Ancillary qubits &  Query complexity \\ \hline \hline
        \begin{tabular}{c}  Trotterization \\  (for $H$ and $H(t)$) \end{tabular} & 0 & $\alpha t \cdot  (\alpha t/\varepsilon)^{1/p}$, \, ($p$-th order) \\ \hline
       \begin{tabular}{c}  Time-independent $H$ \\   (Qubitization \cite{Low2019-qubitization}) \end{tabular} & $n_a+\order{1}$ (Independent of $t$, $\varepsilon$) &$ \alpha t + \frac{\log (1/\varepsilon)}{\log (e + (\alpha t)^{-1} \log (1/\varepsilon) )}  $ \\ \hline
       \begin{tabular}{c}  Time-periodic $H(t)$ \\  (Adiabatic, Theorem \ref{Thm3:resource_adiabatic}) \end{tabular} & $n_a + \order{\log (\gamma t) + \log \log (1/\varepsilon)}$ & $\alpha t + \frac{\log (1/\varepsilon)}{\log (e + \{ \alpha t + o(\log (1/\varepsilon))\}^{-1} \log (1/\varepsilon))}$ \\ \hline
       \begin{tabular}{c}  Time-periodic $H(t)$ \\  (Generic, Theorem \ref{Thm3:resource_long_time}) \end{tabular} & $\quad n_a + \order{\log (\gamma/\omega) + \log \log (\omega t/\varepsilon)} \quad $ & $\quad \alpha t + \frac{ \omega t \log (\omega t/\varepsilon)}{\log ( e + \{ \alpha/\omega + o(\log (\omega t/\varepsilon)) \}^{-1} \log (\omega t/\varepsilon))} \quad $
       \\ \hline
       \begin{tabular}{c}  Time-dependent $H(t)$ \\  (Dyson series \cite{Low2018-dyson,Kieferova2019-dyson}) \end{tabular} & $n_a + \order{\log \{ (\gamma \omega t / \alpha + \alpha t) / \varepsilon \} }$ & $\alpha t \frac{\log(\alpha t/\varepsilon)}{\log \log (\alpha t /\varepsilon)}$ 
      \\ \hline
    \end{tabular}
    }
    \end{center}
    \caption{Scaling of the computational resources for Hamiltonian simulation. The definitions for several parameters are a little different between the above studies. In that case, for simplicity, we replace them by those which have the similar scales (e.g. $\norm{H(t)} \leftrightarrow \alpha$, $\norm{\dv{t}H(t)} \leftrightarrow \omega \gamma$). We also note that the query complexity is measured by the oracles, but their definitions depends on the algorithms. See Eq. (\ref{Eq6C:oracle_trotter}) [for Trotterization], Eq. (\ref{Eq2B:block_encode}) [for qubitization], Eq. (\ref{Eq3B:def_block_encode_H_m}) and Eq. (\ref{Eq3B:def_G_coef}) [for our algorithm], and Eq. (\ref{Eq6C:oracle_dyson}) [for the truncated Dyson-series algorithm]. }
    \label{Table:comparison_algorithms}
\end{table}

\subsection{Adiabatic-like cases}\label{Subsec:algorithm_adiabatic}
We consider the adiabatic-like cases, where long-time dynamics over $\order{1}$-periods is of interest.
For simplicity, we first consider the dynamics within one period at $t \in [0,T]$. 
The algorithm in this case is composed of the following steps;
\begin{enumerate}
    \item Determine the truncation order of the Fourier index $l_\imax$ by Eq. (\ref{Eq4B:l_max_choice}), for the given time $t$.
    
    \item Compose two unitary gates $\ms{U}_\mr{LP}^{4l_\imax}(t)$ and $\ms{U}_\eff^{4l_\imax}(t)$, corresponding to the time evolution operators $e^{-i \ms{H}_\mr{LP}t}$ and $e^{-i \ms{H}_{\eff,\mr{pbc}}^{4l_\imax} t}$ by the qubitization technique; We can find such operators satisfying
    \begin{eqnarray}
        \ms{U}_\mr{LP}^{4l_\imax} (t) \ket{0}^{\otimes n_a (4l_\imax)} \ket{l} \ket{\psi} &=& \ket{0}^{\otimes n_a (4l_\imax)} e^{-il\omega t}\ket{l} \ket{\psi} + \order{\varepsilon}, \\
        \ms{U}_\mr{eff}^{4l_\imax} (t) \ket{0}^{ n_a (4l_\imax)} \ket{l} \ket{\psi}
         &=& \ket{0}^{\otimes  n_a (4l_\imax) }e^{-i \ms{H}_{\eff,\mr{pbc}}^{4l_\imax}t} \ket{l}\ket{\psi} + \order{\varepsilon}, 
    \end{eqnarray}
    for arbitrary inputs $l \in D^{4l_\imax}$ and $\ket{\psi} \in \mcl{H}$.
    
    \item Execute the amplification protocols with substituting $\ms{U}_\mr{LP}^{4l_\imax}(t)$ and $\ms{U}_\eff^{4l_\imax}(t)$ respectively for $e^{-i \ms{H}_\mr{LP}t}$ and $e^{-i \ms{H}_{\eff,\mr{pbc}}^{4l_\imax} t}$ [See Section \ref{Sec:Amplification}];
    \begin{eqnarray}
        \bar{\ms{U}}_\mr{amp1}^{l_\imax} &=& (\ms{U}_\mr{ini}^{4l_\imax})^\dagger \ms{U}_\mr{LP}^{4l_\imax} \ms{U}_\eff^{4l_\imax} \ms{U}_\mr{ini}^{l_\imax}, \label{Eq7A:amp1_adiabatic}\\
        \bar{\ms{U}}_\mr{amp2}^{l_\imax} &=& - \bar{\ms{U}}_\mr{amp1}^{l_\imax} \ms{R} (\bar{\ms{U}}_\mr{amp1}^{l_\imax})^\dagger \ms{R} \ms{U}_\mr{amp1}^{l_\imax}. \label{Eq7A:amp2_adiabatic}
    \end{eqnarray}
    
    \item Apply the unitary operation $\bar{\ms{U}}_\mr{amp2}^{l_\imax}(t)$ to the initial state $\ket{0}^{\otimes n_a(4l_\imax)} \ket{0} \ket{\psi(0)}$;
    \begin{eqnarray}
        \bar{\ms{U}}_\mr{amp2}^{l_\imax} (t) \ket{0}^{\otimes n_a (4l_\imax)} \ket{0}\ket{\psi(0)} = \ket{0}^{\otimes n_a (4l_\imax)} \ket{0}\ket{\psi(t)} + \order{\varepsilon}.
    \end{eqnarray}
\end{enumerate}

Step 1. determines the dimension of the truncated Floquet-Hilbert space as $\bbC^{8l_\imax} \otimes \mcl{H}$ so that it can accurately reproduce $\ket{\psi(t)}$ with the precision $1-\order{\varepsilon}$, as discussed in Section \ref{Sec:truncated_Floquet_Hilbert}.
In Step 2., we employ the qubitization technique to realize the time evolution in the truncated Floquet-Hilbert space, as discussed in Section \ref{Sec:Block_encoding_effective}.
We note that $\ms{U}_\mr{LP}^{4l_\imax} (t)$ and $\ms{U}_\mr{eff}^{4l_\imax} (t)$ can exploit a common $\order{\log (l_\imax)}$-qubit ancillary system.
It is sufficient to prepare an $\{ n_a(4l_\imax) \}$-qubit ancillary state $\ket{0}^{\otimes n_a(4l_\imax)}$ [See Eq. (\ref{Eq6B:number_ancillary})].
Steps 3. and 4. execute the amplification of the time-evolved state $\ket{\psi(t)}$ as discussed in Section \ref{Sec:Amplification}.
By projecting the ancillary state to $\ket{0}^{\otimes n_a (4l_\imax)}\ket{0}$, we succeed in preparing the target state $\ket{\psi(t)}$ with precision $1-\order{\varepsilon}$, where the success probability is $1-\order{\varepsilon}$.
These steps do not change the scaling of the required resource from Eqs. (\ref{Eq7A:amp1_adiabatic}) and (\ref{Eq7A:amp2_adiabatic}), reflecting that the unitary operations $\ms{U}_\mr{ini}$ and $\ms{R}$ yield at-most $\order{\log l_\imax}$ gates.

We finally determine the resource for simulating time-periodic Hamiltonians in adiabatic-like regimes.
The ancillary system should involve the degree of freedom for Fourier indices $\{ \ket{l} \}_{l \in D^{4l_\imax}}$ and $n_a(4l_\imax)$ qubits for qubitization.
The total number of ancillary qubits is
\begin{equation}
    n_a(4l_\imax) + \lceil \log_2 (8l_\imax) \rceil \in n_a + \order{\log (\gamma t) + \log \log (1/\varepsilon)}.
\end{equation}
The numbers of the oracles $O_m$ and $G_\mr{coef}$ are dominated by the qubitization under the refined effective Hamiltonian.
Considering the assumption $\omega t \in \order{1}$, the query complexity given by Eq. (\ref{Eq6B:query_complexity_effective_H}), scales as
\begin{eqnarray}
    && (\alpha + \gamma) t + \frac{\log (1/\varepsilon)}{\log (e + (\gamma t)^{-1} \log (1/\varepsilon))} + \frac{\log (1/\varepsilon)}{\log (e + \{(\alpha + \gamma) t + o(\log (1/\varepsilon))\}^{-1} \log (1/\varepsilon))}, \nonumber \\
    && \quad \in \order{\alpha t + \frac{\log (1/\varepsilon)}{\log (e + \{\alpha t + o(\log (1/\varepsilon)) \}^{-1} \log (1/\varepsilon))} },
\end{eqnarray}
where the $o(\log (1/\varepsilon))$ term scales as Eq. (\ref{Eq3B:o_log_term_adiabatic}) originating from the scaling of $l_\imax$.
The scaling dependent on $\gamma$ can be absorbed into that of $\alpha$, due to the relation $\gamma \leq \alpha$.
We summarize the results in Theorem \ref{Thm3:resource_adiabatic} and Table \ref{Table:comparison_algorithms}.
The number of overall elementary gates there is determined by the query complexity and the set of additional quantum gates per query, each of which at most yields $\order{\log l_\imax}$ gates.
By the $\order{1}$-times repetition of this procedure for $t \in [0,T]$, the same result for $\order{1}$-period dynamics is obtained.

\subsection{Generic long-time cases}\label{Subsec:algorithm_long_time}
We next consider the other generic cases where we are interested in long-time dynamics over multiple periods as $\omega t \in \Omega (1)$. 
In this regime, we take a different strategy from that for the adiabatic cases; We split the time $t$ by $t = (n+\delta) T$  with $n \in \bbN$ and $\delta \in [0,1 )$.
Following this separation, we implement the time-evolution operator $U(t)$ by $n$-times operation of $U(T)$ and single operation of $U(\delta T)$.
The algorithm is composed of the following steps;
\begin{enumerate}
    \item Split the time by $t=(n+\delta) T$ with $n \in \bbN$ and $\delta \in [0,1)$. Determine the truncation order $l_\imax^T$ by substituting $T$ and $\varepsilon/n$ into $t$ and $\varepsilon$ of Eq. (\ref{Eq4B:l_max_choice});
    \begin{eqnarray}
        l_\imax^T &\in& \Theta \left( \gamma T + \frac{\log (n/\varepsilon)}{\log (e + (\gamma T)^{-1} \log (n/\varepsilon))}\right) \nonumber \\
        &=& \Theta \left( \gamma / \omega  + \frac{\log (\omega t/\varepsilon)}{\log (e + (\gamma/\omega)^{-1} \log (\omega t/\varepsilon) )}\right). 
    \end{eqnarray}
    
    \item Compose a unitary gate $\bar{\ms{U}}^{l_\imax^T}_\mr{amp2}(T)$, which satisfies 
    \begin{eqnarray}
        \bar{\ms{U}}_\mr{amp2}^{l_\imax^T} (T) \ket{0}^{\otimes n_a (4l_\imax^T)}\ket{0}\ket{\psi(0)} = \ket{0}^{\otimes n_a (4l_\imax^T)} \ket{0}\ket{\psi(T)} + \order{\varepsilon/n},
    \end{eqnarray}
    by the qubitization and the amplitude amplification protocols (Follow Steps 2.-4. of Section \ref{Subsec:algorithm_adiabatic} wih substituting $T$, $\varepsilon/n$, and $l_\imax^T$ into $t$, $\varepsilon$, and $l_\imax$ respectively). 
    
    \item Apply the unitary gate  $\bar{\ms{U}}^{l_\imax^T}_\mr{amp2}(T)$ to the initial state $n$ times, which results in
    \begin{eqnarray}
        [\bar{\ms{U}}_\mr{amp2}^{l_\imax^T} (T)]^n \ket{0}^{\otimes n_a (4l_\imax^T)} \ket{0}\ket{\psi(0)} = \ket{0}^{\otimes n_a (4l_\imax^T)}\ket{0}\ket{\psi(nT)} + \order{\varepsilon}.
    \end{eqnarray}
    
    \item Prepare the unitary gate $\bar{\ms{U}}^{l_\imax^T}_\mr{amp2}(\delta T)$ by substituting $\delta T$, $\varepsilon$, and $l_\imax^T$ into $t$, $\varepsilon$, and $l_\imax$ respectively in Steps 2.-4. of Section \ref{Subsec:algorithm_adiabatic}. Applying it once to the above state results in
    \begin{eqnarray}
        \bar{\ms{U}}^{l_\imax^T}_\mr{amp2}(\delta T) [\bar{\ms{U}}_\mr{amp2}^{l_\imax^T} (T)]^n \ket{0}^{\otimes n_a (4l_\imax^T)} \ket{0}\ket{\psi(0)} = \ket{0}^{\otimes n_a (4l_\imax^T)} \ket{0}\ket{\psi(t)} + \order{\varepsilon}.
    \end{eqnarray}
    for arbitrary initial states $\ket{\psi(0)}$.
\end{enumerate}

We remark several points in each step.
Steps 1. and 2. are executed to apply $U(T)$, giving the time-evolution over one period $T$.
Here, we set the acceptable error to $\order{\varepsilon /n}$ so that we can obtain the time-evolved state $\ket{\psi(nT)}$ with an error up to $\order{\varepsilon}$ after the $n$-times repetition in Step 3.
The cost for the unitary gate $\bar{\ms{U}}^{l_\imax^T}_\mr{amp2}(T)$ is dominated only by that of $\ms{U}_\eff^{4l_\imax^T}$, which reproduces the time evolution $\exp ( -i \ms{H}_{\eff,\mr{pbc}}^{4l_\imax^T} T)$ by the qubitization.
This comes from the fact that the time evolution $e^{-i \ms{H}_\mr{LP}^{4l_\imax^T} T}$ is trivial due to
\begin{equation}
    e^{-i \ms{H}_\mr{LP}^{4l_\imax^T} T} = \sum_{l \in D^{4l_\imax^T}} e^{i l \omega T} \ket{l}\bra{l} \otimes I = 1,
\end{equation}
and we do not need $\ms{U}_\mr{LP}^{4l_\imax^T}$ in Eq. (\ref{Eq7A:amp1_adiabatic}).
As a result, the query complexity for $n$-times implementation of $\bar{\ms{U}}^{l_\imax^T}_\mr{amp2}(T)$ during Step 3. is  proportional to 
\begin{eqnarray}
    && n \left\{ (\alpha + l_\imax^T \omega) T + \frac{\log (n/\varepsilon)}{\log ( e + \{ (\alpha + l_\imax^T \omega)T \}^{-1} \log (n/\varepsilon) )} \right\} \nonumber \\
    && \quad \in \order{\alpha t + \frac{\omega t \log (\omega t /\varepsilon)}{\log (e+ \{ \alpha/\omega + o(\log (\omega t/\varepsilon))\}^{-1} \log (\omega t/\varepsilon))}}. 
    \label{Eq7B:cost_amp2_long_time}
\end{eqnarray}
We use the fact $n \in \order{\omega t}$ in the above relation.
Step 4. realizes the remaining micromotion $U(\delta T)$ for the duration $\delta T$ with an error up to $\order{\varepsilon}$.
We remark that the choice of the truncation order $l_\imax^T$ so far is sufficient to achieve the precision $1-\order{\varepsilon}$ for $U(\delta T)$ due to $l_\imax^T > l_\imax^{\delta T}$.
This implies that we can reuse the ancillary state $\ket{0}^{\otimes n_a (4l_\imax^T)}$ for the qubitization approach to the time-evolution $U(\delta T)$ in Step 4.
The cost for implementing $\bar{\ms{U}}^{l_\imax^T}_\mr{amp2}(\delta T)$ once is obtained by setting $t=\delta T$ in adiabatic-like cases.
It is smaller than the cost for implementing the time-evolution $U(T)$, and does not affect the scaling of the computational resource.

Finally, we provide the computational resource for time-periodic Hamiltonian dynamics in generic long-time regimes.
The number of ancillary qubits is given by $n_a(4l_\imax)+\lceil \log_2 (8l_\imax^T) \rceil$, which is bounded by
\begin{equation}\label{Eq7B:ancillary_long_time}
    n_a + \order{\log (\gamma /\omega) + \log \log (\omega t /\varepsilon)}.
\end{equation}
The query complexity is dominated by Eq. (\ref{Eq7B:cost_amp2_long_time}).
We summarize them in Theorem \ref{Thm3:resource_long_time} and Table \ref{Table:comparison_algorithms}.

\subsection{Comparison with other algorithms}
Let us compare our algorithm on time-periodic Hamiltonians with other quantum algorithms for Hamiltonian simulation, based on Table \ref{Table:comparison_algorithms}.
We pick up the qubitization technique \cite{Low2019-qubitization} for time-independent Hamiltonian $H$ and the truncated Dyson-series algorithm \cite{Low2018-dyson,Kieferova2019-dyson} for generic time-dependent Hamiltonian $H(t)$, whose resources have the best scaling in $t$ and $1/\varepsilon$ as far as we know.
We also consider the standard way, Trotterization, which covers generic time-independent, time-periodic, and generic time-dependent Hamiltonians.

Let us first compare our algorithm with Trotterization.
When the Hamiltonian can be divided into $H(t)=\sum_{r=1}^\Gamma H_r(t)$, where every term in $H_r(t)$ commutes with one another at every time, the first-order Trotterization roughly approximates the time-evolution $U(t)$ by
\begin{equation}
    U(t) = \prod_{l=0}^{M-1} \left[ \prod_{r=1}^\Gamma \left( \mcl{T} e^{-i \int_{t_l}^{t_{l+1}} d t^\prime H_r(t^\prime)} \right) \right] + \order{\frac{(\alpha t)^2}{M}},
\end{equation}
with $M$ partitions of the time $t$ as $t_l=l t/M$ \cite{Lloyd1996-ko}.
Reflecting that the Trotterization error polynomially decreases in the partition number $M$, we need queries to a layered quantum circuit
\begin{equation}\label{Eq6C:oracle_trotter}
    \prod_{r=1}^\Gamma \left( \mcl{T} e^{-i \int_{t_l}^{t_{l+1}} d t^\prime H_r(t^\prime)} \right)
\end{equation}
for $\order{(\alpha t)^2/\varepsilon}$ times to achieve the allowable error $\varepsilon$.
While the coefficient $\alpha$ and the powers in $t$ and $1/\varepsilon$ can be improved by higher-order Trotterization \cite{Childs2021-trotter}, its resource increases polynomially in $1/\varepsilon$.
Our algorithm has better scaling of elementary gates than Trotterization in that the resource increases logarithmically in $1/\varepsilon$.

The comparison with the qubitization and the truncated Dyson-series algorithm is instructive for evaluating the efficiency of our algorithm due to the inclusion relation,
\begin{eqnarray}
    [\text{Set of time-independent $H$}] &\subset& [\text{Set of time-periodic $H(t)=H(t+T)$}] \nonumber \\
    &\subset& [\text{Set of generic time-dependent $H(t)$}].
\end{eqnarray}
The qubitization technique achieves the least number of ancillary qubits in a sense that it is independent of $t$ and $1/\varepsilon$.
It also has the best query complexity, in which the optimal scaling both in $t$ and $1/\varepsilon$ appears in an additive way as Eq. (\ref{Eq2:qubitization_additive_scale}).
The number of ancillary qubits and the query complexity for the qubitization provide the best bound for time-periodic Hamiltonians, while it is nontrivial whether or not it is actually achievable.
Comparison with the truncated Dyson-series algorithm tells us how efficiently we deal with the time-dependent Schr\"{o}dinger equation.
The cost reduction compared to them can be interpreted as improvement of efficiency due to the time-periodicity.

Before going to the comparison respectively for the adiabatic-like regime and the generic long-time regime, we remark several points in common.
First, we replace some parameters by those which have similar scales to simply compare these algorithms.
For instance, the parameters $\alpha$ and $\gamma$ respectively give the energy scales of the whole Hamiltonian $H(t)$ and the time-dependent terms $H(t)-H_0$ according to Eqs. (\ref{Eq3B:def_alpha}) and (\ref{Eq3A:def_gamma}).
We substitute $\alpha$ and $\omega \gamma$ respectively for $\sup_t (\norm{H(t)})$ and $\sup_t (\norm{\dv{t} H(t)})$, which are characteristic values in the truncated Dyson-series algorithm.

The second point is the complexity of the oracles themselves. 
In the qubitization technique, the query complexity is measured by the oracles to a static Hamiltonian $H$ as Eq. (\ref{Eq2B:block_encode}).
In contrast, the truncated Dyson-series algorithm for generic cases employs the oracle
\begin{equation}\label{Eq6C:oracle_dyson}
    \mr{Ham-}O = \sum_{l=1}^M \ket{l} \bra{l} \otimes O(t_l),
\end{equation}
where $l$ labels the discretized time $t_l$ with the partition number $M \in \order{1/\varepsilon}$ \cite{Low2018-dyson}.
The oracle $\mr{Ham-}O$ includes multiple implementation of $O(t_l)$, which is an oracle for a static instantaneous Hamiltonian $H(t_l)$ (some specific cases such as LCUs and sparse-access matrices with integrable time-dependency can be simplified \cite{Kieferova2019-dyson}).
Our algorithm for time-periodic Hamiltonians uses the oracles $\{ O_m,\ket{G_m}\}$, which gives block-encoding of each Fourier component $H_m$ as Eq. (\ref{Eq6B:block_encoding_H_m}), and the oracle for the coefficients, $G_\mr{coeff}$ given by Eq. (\ref{Eq6B:oracl_coefficient}).
The query complexity in our algorithm is roughly measured by the oracle for a static operator $H_m$ since the latter one $G_\mr{coeff}$, a quantum gate on at-most $\order{1}$ qubits, is usually negligible.
Therefore, note that our algorithm adopts essentially the same measure for the query complexity as the query complexity, while the truncated Dyson-series algorithm counts it by rather complicated oracles involving discretized time.

\subsubsection{Adiabatic-like cases}
We assess the cost for simulating time-periodic Hamiltonians in the adiabatic-like cases based on Table \ref{Table:comparison_algorithms}.
First, we compare the number of ancillary qubits with those for other algorithms. 
What should be noted in our algorithm is its scaling in the inverse error $1/\varepsilon$.
Our algorithm requires $\order{\log \log (1/\varepsilon)}$ additional qubits, whose number lies just in the middle of that for the qubitization ($\varepsilon$-independent) and that for the truncated Dyson-series algorithm [$\order{\log (1/\varepsilon)}$].
Importantly, the reduction compared to the latter one can be attributed to the faster convergence of our formalism based on the Fourier indices.
The error arising from time discretization polynomially decays in the partition number $M$; as a result, the truncated Dyson-series algorithm requires at-least $\order{\log (1/\varepsilon)}$ ancillary qubits via the oracle Eq. (\ref{Eq6C:oracle_dyson}).
In contrast, we introduce the truncation order $l_\imax$ for the Floquet-Hilbert space.
The error by this cutoff scales as $\order{(\gamma t /l_\imax)^{l_\imax}}$, whose decay is faster than the exponential function $e^{-\order{l_\imax}}$.
This leads to the reduction of ancillary qubits from $\order{\log (1/\varepsilon)}$ to $\order{\log \log (1/\varepsilon)}$.

Next, we compare the query complexity.
When we fix the inverse error $1/\varepsilon$ and increase time $t$, it scales as $\order{\alpha t}$.
The linear increase in time $t$ implies the optimality of our algorithm in time.
When we consider $1/\varepsilon \to \infty$ for given time, the query complexity scales as
\begin{equation}
    \frac{\log (1/\varepsilon)}{\log \log \log (1/\varepsilon)},
\end{equation}
where we substitute the form of the $o(\log (1/\varepsilon))$ term, Eq. (\ref{Eq3B:o_log_term_adiabatic}), into Eq. (\ref{Eq3B:query_complexity_adiabatic}).
This scaling is nearly-optimal in $1/\varepsilon$.
Importantly, their contributions affect the query complexity in an additive way as $\alpha t + o (\log (1/\varepsilon))$.
This means that our algorithm deals with time-periodic systems with the cost sufficiently close to qubitization [See Eq. (\ref{Eq2:qubitization_additive_scale})], and saves resource compared to the truncated Dyson-series algorithm requiring the multiplicative query complexity $\alpha t \times o(\log (\alpha t /\varepsilon))$.
As long as we suppose polynomial accuracy $\varepsilon \in 1/\poly{N}$, implying $\log (1/\varepsilon) \lesssim \alpha t$, the query complexity can achieve the best scaling for time-independent systems.

While we hereby provide comparison with the truncated Dyson-series algorithm, our algorithm outperforms other time-dependent Hamiltonian simulation algorithms \cite{Berry2020-time-dep,Haah2021-time-dep,Chen2021-time-dep,Watkins2022-time-dep} in the presence of time-periodicity.
For instance, Ref. \cite{Watkins2022-time-dep} can achieve the additive query complexity when the second derivative of $H(t)$ in $t$ vanishes (i.e., linear dependence), but time-periodicity prohibits vanishing derivatives.
For time-periodic Hamiltonian simulation, the additive query complexity close to that of the qubitization is unique to our algorithm.
Any of the other algorithms using the discretized time requires at-least $\order{1/\varepsilon}$ degrees of freedom, while we need $\order{\log (1/\varepsilon)}$ for the Fourier indices.
Thus, our algorithm always saves the number of ancillary qubits.

\subsubsection{Generic long-time cases}

Here, we evaluate the computational resources for generic long-time regimes.
We suppose that the frequency $\omega$ is constant and hence small compared to the whole-system energy scales $\alpha$ and $\gamma$, which usually increase polynomially in the system size $N$.
This assumption comes from the fact that $\omega$ typically represents the frequency of external drives (e.g. light), which is size-independent.
More importantly, when the frequency $\omega$ is $\poly{N}$ so that it becomes comparable to $\alpha$ and $\gamma$, the dynamics can be efficiently simulated by the methods for time-independent Hamiltonians with the help of high-frequency expansions \cite{Abanin2015-zg,Kuwahara2016-yn,Mori2016-tp,Abanin2017-li,Abanin2017-zs}, as we prove in Appendix \ref{Asec:high_frequency}.
Therefore, it is sufficient to consider the computational resources under $\omega \in \order{N^0}$ and $\alpha,\gamma \in \poly{N}$.

The number of ancillary qubits is similar to that for the adiabatic cases. 
We note that its scaling in the time [$\order{\log \log t}$] overwhelms that of the truncated Dyson-series algorithm [$\order{\log t}$] in addition to the inverse error $1/\varepsilon$.
The number of ancillary qubits for time-periodic Hamiltonians lie between those for time-independent and time-dependent Hamiltonians in terms of both $t$ and $1/\varepsilon$.

We discuss the query complexity given by Theorem \ref{Thm3:resource_long_time}.
Under the fixed allowable error $\varepsilon$, the scaling of the query complexity in sufficiently large time $t$ is given by
\begin{equation}
    \alpha t + \omega t \frac{\log (\omega t /\varepsilon)}{\log \log \log (\omega t /\varepsilon)} \leq \{\alpha + \omega \log (1/\varepsilon) \} t + \omega t \log (\omega t).
\end{equation}
Due to the second term in the right hand side, the query complexity has at least nearly-optimal dependence in $t$, which involves logarithmic correction $\log (\omega t)$.
However, reflecting the assumptions $\alpha \in \poly{N}$ and $\omega \in \order{N^0}$, the $\order{\omega t \log (\omega t)}$ term can be non-negligible compared to $\order{\alpha t}$ only when the time reaches $t \sim e^{\alpha/\omega}/\omega \in \order{e^{\poly{N}} T}$. 
Therefore, unless we consider unpractical time scales $t \in e^{\Theta (N)}$, this nearly-optimal scaling is formal, and the query complexity actually increases linearly in time $t$, showing the optimal scaling.
On the other hand, when we focus on the scaling in the inverse error $1/\varepsilon$, the query complexity scales as
\begin{equation}
    \omega t \frac{\log (1/\varepsilon)}{\log \log \log (1/\varepsilon)}.
\end{equation}
Thus, our algorithm achieves nearly-optimal scaling in $1/\varepsilon$ with slight deviation from the optimal dependence by $\log \log (1/\varepsilon) / \log \log \log (1/\varepsilon)$.

When it comes to the combined scaling of the query complexity in $t$ and $1/\varepsilon$, we note that it has an additive form as
\begin{equation}
    \alpha t + \omega t \times o(\log (1/\varepsilon)),
\end{equation}
where we neglect $\order{\omega t \log (\omega t)}$.
Although linear increase in $t$ couples with logarithmic increase in $1/\varepsilon$, the coefficient $\omega t \in \order{N^0} t$ is much smaller than the whole energy-time scale $\alpha t \in \poly{N} t$.
As a consequence, also in generic long-time cases, our algorithm achieves the query complexity close to that of qubitization [See Eq. (\ref{Eq2:qubitization_additive_scale})] and saves much cost compared to the truncated Dyson-series algorithm, where the query complexity scales as $\poly{N} t \times o(\log (\alpha t/\varepsilon))$.

\section{Illustrative examples}\label{Sec:examples}

In this section, we briefly discuss some potential applications of the algorithm.
We expect that it can be applied to nonequilibrium quantum many-body phenomena, which are often of interest in condensed matter physics and quantum chemistry.
In terms of quantum computation, it will offer an efficient protocol for adiabatic state preparation, which can be applied to quantum phase estimation for instance.
We suggest the simplest examples for both applications below.

\subsection{Nonequilibrium quantum many-body phenomena}\label{Subsec:example_Fermi_Hubbard}

The first application is to simulate nonequilibrium dynamics of periodically-driven quantum materials.
Optical responses are typical but of great interest both in condensed matter physics and quantum chemistry.
We pick up an $N$-site Fermi-Hubbard model under laser light as the simplest case;
\begin{eqnarray}
    H(t) &=& H_\text{Hub} + H_\mr{ext}(t), \\
    H_\text{Hub} &=& \sum_{k } \sum_{\sigma= \uparrow, \downarrow} \epsilon_{k} \hat{n}_{k\sigma} + U \sum_{x} \hat{n}_{x\uparrow} \hat{n}_{x\downarrow}. \label{Eq8:Fermi_Hubbard}
\end{eqnarray}
Here, $\hat{n}_{k\sigma}=\hat{c}_{k\sigma}^\dagger \hat{c}_{k\sigma}$ and $\hat{n}_{x\sigma}=\hat{c}_{x\sigma}^\dagger \hat{c}_{x\sigma}$ are number operators of electrons in the momentum and real spaces respectively, generated by fermionic annihilation operators $\hat{c}_{k\sigma}$ and $\hat{c}_{x\sigma}$.

The time-periodic term $H_\mr{ext}(t)$ represents the coupling with light.
When we shine linearly-polarized light with the frequency $\Omega$, it is given by
\begin{equation}
    H_\mr{ext}(t) = \sin (\Omega t) \sum_{x, \sigma} V_x \hat{n}_{x\sigma},
\end{equation}
which results in $T=2\pi/\Omega$ and $m_\imax=1$.
The Fourier components $H_m$ are given by $H_0=H_\mr{Hub}$ and $H_{\pm 1} = (\pm i) \sum_{x,\sigma} (V_x/2) \hat{n}_{x\sigma}$.
To evaluate the cost of simulating $\ket{\psi(t)}$, we compose the oracles for them.
We employ a unitary operation, called fermionic fast Fourier transform (FFFT) \cite{Verstraete2009FFFT,Ferris2014FFFT,Babbush2018FFFT}, which transforms the basis in the momentum space $k_i$ to that in the real space $x_i$ as $\hat{n}_{k_i \sigma} = \mr{FFFT}^\dagger \hat{n}_{x_i \sigma} \mr{FFFT}$.
We map the fermionic system to a spin system by Jordan-Wigner transformation as $\hat{n}_{x\sigma}=(1+Z_{x\sigma})/2$ (``$x\sigma$" denotes an index for qubits).
By neglecting constant terms and a conserved particle number $\sum_{x\sigma} \hat{n}_{x\sigma}$, the block-encoding for $\{H_m\}$ can be constructed by the technique for LCUs, as Eq. (\ref{Eq3B:block_encode_LCU}).
Assuming $\epsilon_{k}, U, V_x \geq 0$ without loss of generality, this leads to
\begin{eqnarray}
    O_0 &=& \ket{0}\bra{0} \otimes \sum_{x, \sigma} \ket{x,\sigma}\bra{x,\sigma} \otimes \mr{FFFT}^\dagger Z_{x\sigma} \mr{FFFT} \nonumber \\
    && \quad + \ket{1}\bra{1} \otimes \sum_{x,\sigma} \ket{x,\sigma}\bra{x,\sigma} \otimes Z_{x\uparrow} Z_{x\downarrow}, \label{Eq8:Oracle_gate_Fermi_Hubbard} \\
    \ket{G_0} &=& \sum_{x,\sigma} \left( \sqrt{\frac{\varepsilon_{x}}{\alpha_0}}\ket{0} +  \sqrt{\frac{U}{2\alpha_0}}\ket{1} \right) \ket{x,\sigma}, \label{Eq8:Oracle_state_Fermi_Hubbard} \\
    O_{\pm 1} &=& I \otimes \sum_{x,\sigma} \ket{x,\sigma}\bra{x,\sigma} \otimes (\pm i Z_{x\sigma}), \label{Eq8:Oracle_gate_external}  \\
    \ket{G_{\pm 1}} &=& \ket{0} \sum_{x,\sigma} \sqrt{\frac{V_x}{ 2 \alpha_{\pm 1}}} \ket{x,\sigma}, \label{Eq8:Oracle_state_external} 
\end{eqnarray}
with $\alpha_0 = 2 \sum_{x} \epsilon_{x} + UN$ and $\alpha_{\pm 1}= \sum_{x} V_x$.
In the above oracles, the unitary gate $O_0$ requires much cost due to $\order{1}$-times usage of FFFT, which can be implemented with at-most $\order{N \log N}$-depth quantum circuits composed of adjacent two-qubit gates \cite{Babbush2018FFFT}.
The depth needed for each oracle is at-most $\order{N \log N}$.

The number of ancillary qubits for them to express $\{ \ket{0}\ket{x,\sigma}, \ket{1}\ket{x,\sigma} \}$ amounts to $n_a \in \order{\log N}$.
The energy scales of the whole Hamiltonian, $\alpha$, determined by Eq. (\ref{Eq3B:def_alpha}) as
\begin{equation}
    \alpha =  2 \sum_{x} \epsilon_{x} + UN + 2\sum_{x} V_x.
\end{equation}
Although it is difficult to obtain $\gamma$ from its definition Eq. (\ref{Eq3A:def_gamma}), we can easily obtain its upper bound, which results in
\begin{equation}
    \gamma \leq \sum_{m = \pm 1} \norm{H_m} = \sum_x V_x.
\end{equation}
It is adequate for determining the computational resource, since $\alpha$ is rather dominant in the query complexity and $\gamma$ appears as $\order{\log \gamma}$ in the number of ancillary qubits. 
When we define the characteristic local energy scale by $\alpha_\mr{loc}=\max (\epsilon_k,U,V_x)$, they are approximately described by $\alpha, \gamma \in \order{\alpha_\mr{loc} N}$.

Finally, if we are interested in the time-evolved state $\ket{\psi(t)}$ over multiple periods, the following resources are required to achieve the precision $1-\order{\varepsilon}$;
\begin{itemize}
    \item Number of ancillary qubits; 
    \begin{equation}\label{Eq8A:ancillary_fermi_hubbard}
        \order{\log ( \alpha_\mr{loc} N/\Omega ) + \log \log (\Omega t /\varepsilon)}.
    \end{equation}
    \item Overall gate complexity;
    \begin{equation}\label{Eq8A:complexity_fermi_hubbard}
        \order{\left\{ \alpha_\mr{loc} Nt + \Omega t \log (\Omega t/\varepsilon) \right\} N \log N}.
    \end{equation}
\end{itemize}
The above results are based on Eqs. (\ref{Eq7B:cost_amp2_long_time}) and (\ref{Eq7B:ancillary_long_time}), where we neglect size-independent additional gates for each query to the oracles.
According to classical numerical calculations \cite{Silva2018-ww,Murakami2018-sp,Murakami2021-oy}, the above model is expected to host high-harmonic generation, where intense oscillation with the frequency $n\Omega$ ($n=2,3,\hdots$) arises in response to laser light with the frequency $\Omega$.
Our algorithm allows to efficiently identify such a nontrivial nonequilibrium phenomenon with guaranteed accuracy by simulating the Fourier spectrum of some observables $\braket{\psi(t)|O|\psi(t)}$ (e.g. electric current) \cite{hhg}.

Another interesting example is a discrete time crystal as a phase of matter inherent in nonequilibrium, where time-translation symmetry is spontaneously broken \cite{Khemani2016-dq,Else2016-mg,Khemani2019-pf}.
When we choose $H_\mr{ext}(t)$ by uniform circularly-polarized ac field represented by
\begin{equation}
    H_\mr{ext}(t) = V \sum_{x} \left( e^{i\Omega t} \hat{c}_{x\uparrow}^\dagger \hat{c}_{x\downarrow} + e^{-i \Omega t} \hat{c}_{x\downarrow}^\dagger \hat{c}_{x\uparrow} \right),
\end{equation}
the Fermi-Hubbard model becomes a potential platform for a time-crystalline phenomenon protected by Floquet dynamical symmetry \cite{Chinzei2020dtc}.
Its signature can be detected by subharmonic oscillations of spatio-temporal correlation functions and local observables.
These values are both efficiently computed via the time-evolution operator $U(t)$ by our algorithm, requiring the computational resources similar to Eqs. (\ref{Eq8A:ancillary_fermi_hubbard}) and (\ref{Eq8A:complexity_fermi_hubbard}). 

As nonequilibrium systems dominated by time-periodic Hamiltonians have been vigorously explored as Floquet systems, our algorithm will cover various phenomena.
We also note that our algorithm is extended to time-periodic Hamiltonians with exponentially-decaying Fourier components $\norm{H_m} \lesssim e^{-\order{|m|}}$.
Since we often face at situations where high-frequency components of $H(t)$ rapidly diminish, our result will be useful for a variety of nonequilibrium phenomena in condensed matter physics and quantum chemistry, other than the above examples (See also Appendix \ref{Asubsec3:example_expo}).

\subsection{Adiabatic state preparation}
Adiabatic state preparation is a protocol to obtain a preferable quantum state by adiabatically evolving quantum systems with time-dependent Hamiltonians $H(t)$.
While it has been originally developed in the context of adiabatic quantum computation relying on the adiabatic theorems of quantum dynamics \cite{Albash2018RevModPhys}, it can be exploited also on circuit-based quantum computers by mimicking the adiabatic dynamics under $H(t)$.
One of typical aims is to prepare initial states required for quantum simulation \cite{Aspuru-Guzik2005-nr,Du2010-hl,Rattew2022-uh}.
Here, we provide the simplest application of our algorithm in this field.

We prepare a time-periodic Hamiltonian $H(t)$, which continuously connects two different time-independent Hamiltonians $\bar{H}_0$ and $\bar{H}_1$.
A certain eigenstate of $\bar{H}_0$, denoted by $\ket{\bar{\psi}_0}$, is supposed to be easily prepared on quantum circuits, and we assume that the eigenstate of $\bar{H}_1$, which is continuously connected to $\ket{\bar{\psi}_0}$, corresponds to the target state $\ket{\bar{\psi}_1}$.
As the simplest case, we organize such a time-periodic Hamiltonian by
\begin{equation}
    H(t) = \bar{H}_0 (1-\sin \omega t) + \bar{H}_1 \sin \omega t ,
\end{equation}
with satisfies $H(0)=\bar{H}_0$ and $H(T/4)= \bar{H}_1$.
Our algorithm for the adiabatic-like cases in Section \ref{Subsec:algorithm_adiabatic} enables us to efficiently execute the adiabatic state preparation.
Since the Fourier components of $H(t)$ is simply given by $H_0 = \bar{H}_0$ and $H_{\pm 1}= \pm (\bar{H}_0 - \bar{H}_1)/2i$ with $m_\imax =1$, the following oracles are necessary;
\begin{eqnarray}
    O_0 &=& I \otimes \bar{O}_0, \quad \ket{G_0} = \ket{0} \ket{\bar{G}_0}, \\
    O_{\pm 1} &=& \ket{0}\bra{0} \otimes ( \mp i \bar{O}_0 ) + \ket{1}\bra{1} \otimes (\pm i \bar{O}_1), \\
    \ket{G_{\pm 1}} &=& \frac{\sqrt{\bar{\alpha}_0}\ket{0} \ket{\bar{G}_0}+ \sqrt{\bar{\alpha}_1}\ket{1} \ket{\bar{G}_1}}{\sqrt{\bar{\alpha}_0+\bar{\alpha}_1}}.
\end{eqnarray}
Here, $(\bar{O}_0,\ket{\bar{G}_0},\bar{\alpha}_0)$ and  $(\bar{O}_1,\ket{\bar{G}_1},\bar{\alpha}_1)$ respectively provides the block-encoding of the time-independent Hamiltonians $\bar{H}_0$ and $\bar{H}_1$ as Eq. (\ref{Eq2B:block_encode}).
The oracles $O_{\pm 1}$ and $\ket{G_{\pm 1}}$ gives the block-encoding of $H_{\pm 1}$ as $\braket{G_{\pm 1 }|O_{\pm 1}|G_{\pm 1}} = 2 H_{\pm 1}/(\bar{\alpha}_0 + \bar{\alpha}_1)$.
According to the standard adiabatic theorem \cite{Albash2018RevModPhys}, the duration required for approximating the target state $\ket{\bar{\psi}_1}$ by the adiabatically-evolved state $U(T/4) \ket{\bar{\psi}_0}$ with accuracy $1-\order{\varepsilon}$ roughly amounts to $T \gtrsim \norm{\bar{H}_0 - \bar{H}_1} / \varepsilon \Delta^2$, where $\Delta$ denotes the minimal gap upon the instantaneous eigenstate of $H(t)$ continuously connecting $\ket{\bar{\psi}_0}$ and $\ket{\bar{\psi}_1}$.
Therefore, in the adiabatic state preparation based on our algorithm, the query complexity counted by the oracles of the two static Hamiltonians $\bar{H}_0$ and $\bar{H}_1$ amounts to
\begin{equation}
    \frac{\bar{\alpha}_0 + \bar{\alpha}_1}{\varepsilon\Delta^2} \norm{\bar{H}_0 - \bar{H}_1} + o(\log (1/\varepsilon)).
\end{equation}
Although the second term is buried by the first one polynomially increasing in $1/\varepsilon$, the cost has better scaling in $1/\varepsilon$ compared to the cases where similar schedules for $H(t)$ are tackled with the Trotterization [$\sim (1/\varepsilon)^{1+2/p}$] or the truncated Dyson-series algorithm [$\sim (1/\varepsilon) \log (1/\varepsilon)$].

In adiabatic quantum computation, the required time in $1/\varepsilon$ can be improved via more sophisticated scheduling, such as local adiabatic interpolation \cite{Roland2002adiabatic}, boundary cancellation \cite{Lidar2009-bv,Wiebe2012-xx} and quasi-adiabatic processes \cite{Wan2020-at}.
In particular, the last one has achieved the query complexity poly-logarithmic in $1/\varepsilon$ with the help of the truncated Dyson-series algorithm.
If we can find time-periodic Hamiltonians which are consistent with such sophisticated schedules, it will offer a more efficient protocol for the adiabatic state preparation.

\section{Discussion and Conclusion}\label{Sec:Discussion}

We conclude our paper with summarizing the results.
In this paper, we focus on time-dependent systems with time-periodicity, and organize an efficient implementation of their time-evolution operators.
Once we prepare the oracles which embeds each Fourier component and each coefficient of the Hamiltonian, a series of unitary operations on the truncated Floquet-Hilbert space extract the time-evolved state $\ket{\psi(t)}$ with an allowable error $1/\varepsilon$, whose query complexity are both optimal or nearly-optimal in $t$ and $1/\varepsilon$.
In addition, it has an additive scaling $\alpha t + o(\log(1/\varepsilon))$ [adiabatic-like regime] or $\alpha t + \omega t \times o(\log (1/\varepsilon))$ [generic long-time regime], and hence achieves the cost sufficiently close to that of the best algorithm for time-independent systems \cite{Low2019-qubitization}, despite the existence of time-dependency.
As exemplified by nonequilibrium quantum many-body phenomena and adiabatic state preparation, our algorithm will contribute to pioneering applications of quantum computers for various aims, in condensed matter physics, quantum chemistry, and quantum computation.

We finally discuss some potential future directions of our results. 
The first one is to explore efficient implementation of meaningful functions in time-periodic systems, other than the time-evolution operator $U(t)= \mcl{T}\exp (-i \int_0^t H(t^\prime) \dd t^
\prime)$.
In our algorithm, we exploit the qubitization technique to implement the exponential function, $e^{-i \ms{H}_\eff^{l_\imax}t}$, in the effective Hamiltonian $\ms{H}_\eff^{l_\imax}$. 
As the qubitization technique allows to efficiently apply various polynomial functions, we expect that our algorithm can be extended for various aims other than the unitary time evolution discussed here.
For instance, our discussion on the Lieb-Robinson bound in Section \ref{Sec:truncated_Floquet_Hilbert} is valid also when a time-periodic Hamiltonian $H(t)$ is non-hermitian.
This suggests that our algorithm may be available also for solving time-periodic linear differential equations \cite{Fang2022-wq}, exemplified by dissipative quantum may-body systems \cite{Breuer2002-xj,Rivas2012-jb}. 
While we do not expect exponential speedup due to the non-unitarity in general, it will offer an efficient way both in time and desirable accuracy.
It would be important to clarify what kind of function is useful in time-periodic systems, how the Lieb-Robinson bound and the amplitude amplification should be modified, and then how the computational cost is affected.

Another significant direction is to seek for useful tasks that can be efficiently tackled with time-periodic Hamiltonians. 
As we have shown throughout the paper, time-periodic Hamiltonians can be simulated more efficiently generic time-dependent Hamiltonians with computational resources close to those for time-independent Hamiltonians.
This means that quantum tasks which inevitably requires time-dependent operations, such as adiabatic state preparation, can be optimized by tuning their schedules in a time-periodic way.
It will be important to clarify what kind of tasks can be addressed by time-periodic Hamiltonians and how our algorithm provides better scaling in $1/\varepsilon$ for their costs, which we leave for future work.

\section*{Acknowledgment}

We thank T. Kuwahara for fruitful discussion, especially on the Lieb-Robinson bound presented in Section \ref{Subsec:Lieb_Robinson_Floquet_Hilbert}.
This work is supported by MEXT Quantum Leap Flagship Program (MEXTQLEAP) Grant No. JPMXS0118067394, JPMXS0120319794, and JST COI-NEXT program Grant No. JPMJPF2014.

\bibliographystyle{quantum}
\bibliography{bibliography.bib}

\begin{thebibliography}{10}
\bibitem{hhg}
In order to capture nontrivial phenomena in the steady states, we should attach
  a coupling with an external bath to the time-independent term $H_0$. While it
  alters the block-encoding of $H_0$, the scaling of the computational cost
  does not severely increase.
\end{thebibliography}

\clearpage
\appendix
\begin{center}
\bf{\LARGE Appendix}
\end{center}

\section{Short explanation of Eq. (\ref{Eq2A:psi_t_infinite})}
Our algorithm is mainly based on the fact that the dynamics in the Floquet-Hilbert space,
\begin{equation}
    \ket{\psi^\infty(t)} = \lim_{l_\imax \to \infty} \sum_{l \in D^{l_\imax}} e^{-il\omega t}\bra{l} e^{-i \ms{H}_\eff^{l_\imax} t} \ket{0}\ket{\psi(0)},
\end{equation}
exactly corresponds to the solution $\ket{\psi(t)}$ \cite{Levante1995-tx}.
Here, we roughly explain this relation for this paper to be self-contained.
We differentiate the above formula in $t$;
\begin{eqnarray}
    i \dv{t} \ket{\psi^\infty(t)} &=& \sum_{l \in \bbZ} e^{-il\omega t} \bra{l} (\ms{H}_\eff+l\omega) e^{-i \ms{H}_\eff t} \ket{0}\ket{\psi(0)} \nonumber \\
    &=& \sum_{l \in \bbZ} \sum_{m \in \bbZ} e^{-il\omega t} H_{-m} \bra{l+m} e^{-i \ms{H}_\eff t} \ket{0}\ket{\psi(0)} \nonumber \\
    &=& \left( \sum_{m \in \bbZ} H_{-m} e^{im\omega t} \right)
    \left( \sum_{l \in \bbZ}  e^{-il\omega t} \bra{l} e^{-i \ms{H}_\eff t} \ket{0}\ket{\psi(0)}\right) \nonumber \\
    &=& H(t) \ket{\psi^\infty(t)}.
\end{eqnarray}
Noting that $\ket{\psi^\infty(0)}=\ket{\psi(0)}$, the uniqueness of the solution of the Schr\"{o}dinger equation suggests the relation $\ket{\psi^\infty(t)}=\ket{\psi(t)}$.

\section{Proof of the theorems for formulation}\label{Asec:Proof_theorems}

\subsection{Amplitude amplification by symmetry}\label{Asubsec2:amplification_symmetry}
In Section \ref{Subsec:amplification_symmetry}, we discuss the amplification of the time-evolved state $\ket{\psi(t)}$ with exploiting the translation symmetry of generic time-periodic Hamiltonians.
With the usage of the protocol,
\begin{equation}
     \ms{U}_\mr{amp1}^{l_\imax} (t) = (\ms{U}_\mr{ini}^{4l_\imax})^\dagger e^{-i t \sum_l l \omega \ket{l}\bra{l}} e^{-i \ms{H}_\eff^{4l_\imax} t} \ms{U}_\mr{ini}^{l_\imax},
\end{equation}
we see that it generates the time-evolved state $\ket{\psi(t)}$ with $\order{1}$ amplitude as 
\begin{equation}
    \bra{0} \ms{U}_\mr{amp1}^{l_\imax} (t) \ket{0} \ket{\psi(0)} \simeq \frac{1}{2} (\ket{\psi(t)}+\order{\varepsilon}).
\end{equation}
We rigorously prove this statement, which is summarized as Theorem \ref{Thm5:amplification_symmetry}.

In order to show Theorem \ref{Thm5:amplification_symmetry}, we begin with discussing the approximate translation symmetry. 
To be precise, we evaluate how much error appears in the approximation,
\begin{equation}
     \bra{l} e^{-i\ms{H}_\eff^{4l_\imax}t} \ket{l^\prime} \simeq e^{il^\prime \omega t}\bra{l-l^\prime} e^{-i\ms{H}_\eff^{4l_\imax}t} \ket{0},
\end{equation}
which we refer to as Eq. (\ref{Eq5A:approximate_symmetry}) in the main text.
The exact upper bound on this error is given by the following theorem.

\begin{lemma}\label{LemmaA2:translation_symmetry}
\textbf{(Approximate translation symmetry)}

We choose the truncation order $l_\imax \in \Theta (\gamma t + \log (1/\varepsilon)/\log\log (1/\varepsilon))$ by Eq. (\ref{Eq4B:l_max_choice}). 
For indices $l^\prime \in D^{l_\imax}$ and $l \in D^{4l_\imax}$, the inequality
\begin{eqnarray}
    \norm{\braket{l |e^{-i \ms{H}_\eff^{4l_\imax} t}| l^\prime} - e^{i l^\prime \omega t} \braket{l \ominus l^\prime |e^{-i \ms{H}_\eff^{4l_\imax} t}| 0}} \leq 8 \frac{(\gamma t)^{n(l^\prime,l)}}{n(l^\prime,l)!} \label{EqA2:lemma_translation_symmetry}
\end{eqnarray}
is satisfied with
\begin{equation}\label{EqA2:Dyson_order_minimum}
    n(l^\prime,l) = \left\lceil \frac{(8l_\imax - 2 m_\imax - |l| - |l^\prime|)}{m_\imax}\right\rceil.
\end{equation}
Here, $l \ominus l^\prime \in D^{4l\imax}$ denotes the difference $l-l^\prime$ defined modulo $8l_\imax$.
\end{lemma}

\textbf{Proof.---} We consider dynamics under a perturbed Hamiltonian $\ms{H}_\mr{pert}(t)$, defined by
\begin{eqnarray}
    \ms{H}_\mr{pert}(t) &=& \ms{H}_\eff^{4l_\imax} + \ms{H}_b (t), \\
    \ms{H}_b(t) &=& \sum_{(l,m) \in \partial F^{4l_\imax}} (\ket{l}\bra{l \oplus m} \otimes e^{i8l_\imax \omega t}H_{-m} + \text{h.c.}),
\end{eqnarray}
with $\partial F^{4l_\imax} = \{ (l,m) \, | \, 4l_\imax - m_\imax +1 \leq l \leq 4l_\imax, \, 4l_\imax - l  +1 \leq m \leq m_\imax\}$.
In the interaction picture based on $\ms{H}_0 = \sum_{l \in D^{4l_\imax}} \ket{l}\bra{0} \otimes (H_0 -l\omega)$, the perturbed Hamiltonian $\ms{H}_{\mr{pert},I}(t)$ is exactly translation symmetric under arbitrary shift $\ket{l} \to \ket{l \oplus m}$ as 
\begin{equation}\label{EqA3:H_eff_pert_int}
    \ms{H}_{\mr{pert},I}(t) = \sum_{m \neq 0; |m| \leq m_\imax} \mr{Add}_m^{4l_\imax} \otimes e^{- im\omega t} H_m^I (t).
\end{equation}
In the original frame, this exact symmetry implies the satisfaction of
\begin{equation}
    \braket{l|\ms{U}_\mr{pert}(t)|l^\prime} = e^{i l^\prime \omega t} \braket{l \ominus l^\prime |\ms{U}_\mr{pert}(t)|0},
\end{equation}
where $\ms{U}_\mr{pert}(t)$ indicates the time evolution operator under the perturbed Hamiltonian $\ms{H}_\mr{pert}(t)$.
Upon this relation, we evaluate the upper bound on the error by a triangle inequality,
\begin{eqnarray}
    \text{[l.h.s. of Eq. (\ref{EqA2:lemma_translation_symmetry})]} &\leq& \norm{\braket{l |e^{-i \ms{H}_\eff^{4l_\imax} t} - \ms{U}_\mr{pert}(t)|l^\prime}}+ \norm{\braket{l \ominus l^\prime |e^{-i \ms{H}_\eff^{4l_\imax} t} - \ms{U}_\mr{pert}(t)|0}}. \nonumber \\
    && \label{EqA2:Lemma_symmetry_proof_1}
\end{eqnarray}

We begin with computing the bound of the first term. 
Using the Dyson series expansion in the interaction picture, similarly to Eqs. (\ref{Eq4A:def_reference_hamiltonian})-(\ref{Eq4A:Dyson_transition_path}), it can be bounded by
\begin{eqnarray}
    && \norm{\braket{l |e^{-i \ms{H}_\eff^{4l_\imax} t} - \ms{U}_\mr{pert}(t)|l^\prime}} \nonumber \\
    && \quad \leq  \sum_{n=0}^\infty \int_0^t \dd t_n \hdots \int_0^{t_2} \dd t_1 \norm{\bra{l} \prod_{i=1}^n \ms{H}_{\mr{pert},I}(t_i) - \prod_{i=1}^n \ms{H}_I^{4l_\imax}(t_i)\ket{l^\prime}}. 
    \label{EqA2:Lemma_symmetry_proof_2}
\end{eqnarray}
As discussed in the proof of Theorem \ref{Thm4:Bound_transition_rate}, each integrand is decomposed into contributions from the transition amplitudes via paths $\ket{l^\prime} \to \ket{l_1} \to \hdots \to \ket{l}$.
The difference between the Hamiltonians $\ms{H}_\mr{pert}(t)$ and $\ms{H}_\eff^{4l_\imax}$ arises only when the path goes across $D^{4l_\imax} \backslash D^{4l_\imax - m_\imax}$, which is the support of $\ms{H}_b(t)$.
For low order terms with $n < n (l^\prime, l)$, defined by Eq. (\ref{EqA2:Dyson_order_minimum}), such a nontrivial path is absent, and we obtain 
\begin{eqnarray}
    \norm{\braket{l |e^{-i \ms{H}_\eff^{4l_\imax} t} - \ms{U}_\mr{pert}(t)|l^\prime}} \leq  \sum_{n=n(l^\prime,l)}^\infty \frac{t^n}{n!} \left\{ \left(\sup_t (\norm{\ms{H}_{\mr{pert},I}(t)}) \right)^n + \gamma^n \right\}. \label{EqA2:Lemma_symmetry_proof_3}
\end{eqnarray}
Since the Hamiltonian $\ms{H}_{\mr{pert},I}(t)$ given by Eq. (\ref{EqA3:H_eff_pert_int}) is a circulant matrix (i.e. invariant under the shift of $\ket{l}$) \cite{Gray2006-da}, its operator norm can be bounded by
\begin{eqnarray}
    \norm{\ms{H}_{\mr{pert},I}(t)} &\leq& \sup_{t^\prime} \left( \norm{\sum_{m \neq 0} e^{-i m \omega t} H_m^I(t) e^{im \omega t^\prime} }\right) \nonumber \\
    &=& \sup_{t^\prime} \left( \norm{H(t-t^\prime) -H_0} \right) = \gamma.
\end{eqnarray}
As a result, Eq. (\ref{EqA2:Lemma_symmetry_proof_3}) is further bounded by
\begin{equation}
    \norm{\braket{l |e^{-i \ms{H}_\eff^{4l_\imax} t} - \ms{U}_\mr{pert}(t)|l^\prime}} \leq 2 \sum_{n=n(l^\prime, l)}^\infty \frac{(\gamma t)^n}{n!}.
\end{equation}

In a similar way, we can obtain the bound on the second term of the right hand side of Eq. (\ref{EqA2:Lemma_symmetry_proof_1}), which results in the above formula under the replacement of $n(l^\prime,l)$ by $n(0,l \ominus l^\prime)$.
Since $n(0,l \ominus l^\prime)$ is always larger than $n(l^\prime,l)$ from their definitions, we obtain the accuracy of the approximate translation symmetry as
\begin{equation}
    [\text{l.h.s of Eq. (\ref{EqA2:lemma_translation_symmetry})}] \leq 4 \sum_{n=n(l^\prime, l)}^\infty \frac{(\gamma t)^n}{n!}
    \leq 8 \frac{(\gamma t)^{n(l,l^\prime)}}{n(l^\prime,l)!}.
\end{equation}
When we choose the truncation order $l_\imax$ by Eq. (\ref{Eq4B:l_max_choice}), the integer $n(l^\prime,l)$ is always larger than $6 \gamma t$ for indices $l \in D^{4l_\imax}$ and $l^\prime \in D^{l_\imax}$.
We use Eq. (\ref{Eq4A:sum_Taylor}) for the last inequality. $\quad \square$

This lemma ensures that the approximate translation symmetry in the truncated Floquet-Hilbert space is extremely accurate; the right hand side of Eq. (\ref{EqA2:lemma_translation_symmetry}) is approximately $\order{\varepsilon^3}$ under the choice of $l_\imax$, Eq. (\ref{Eq4B:l_max_choice}).
As we intuitively discuss in Section \ref{Subsec:amplification_symmetry}, the approximate translation symmetry ensures the amplification by symmetry as Eq. (\ref{Eq5A:approximate_amp1_result}).
We next prove the consequence of Theorem \ref{Thm5:amplification_symmetry}, which gives the exact description of Eq. (\ref{Eq5A:approximate_amp1_result});
\begin{equation}
    \norm{\bra{0}\ms{U}_\mr{amp1}^{l_\imax}(t)\ket{0}\ket{\psi(0)}-\frac12 \ket{0}\ket{\psi(t)}} \leq \frac{\varepsilon}{3}.
\end{equation}

\textbf{Proof of Theorem \ref{Thm5:amplification_symmetry}.---} We track Eq. (\ref{Eq5A:amp1_calculation}) in a rigorous way. We begin with the definition,
\begin{eqnarray}
    \bra{0}\ms{U}_\mr{amp1}^{l_\imax}(t)\ket{0}\ket{\psi(0)} =  \frac{1}{4l_\imax}\sum_{l^\prime \in D^{l_\imax}}  \sum_{l \in D^{4l_\imax}} e^{-il\omega t} \bra{l} e^{-i\ms{H}_\eff^{4l_\imax}t} \ket{l^\prime} \ket{\psi(0)}. \label{EqA2:thm_amp1_proof_1}
\end{eqnarray}
We separate the summation over $l \in D^{4l_\imax}$ in the above formula by
\begin{equation} \label{EqA2:thm_amp1_proof_2}
    \sum_{l \in D^{4l_\imax}} =  \sum_{l; l-l^\prime \in D^{3l_\imax}} + \sum_{l \in D^{4l_\imax}; l-l^\prime \notin D^{3l_\imax}}.
\end{equation}
Let us focus on the first summation. For each $l^\prime \in D^{l_\imax}$, the summation can be approximated as 
\begin{eqnarray}
    && \norm{\sum_{l; l-l^\prime \in D^{3l_\imax}} e^{-il\omega t} \bra{l} e^{-i\ms{H}_\eff^{4l_\imax}t} \ket{l^\prime} \ket{\psi(0)} - \ket{\psi(t)} } \nonumber \\
    && \quad \leq \sum_{l; l-l^\prime \in D^{3l_\imax}} 8 \frac{(\gamma t)^{n(l^\prime,l)}}{n(l^\prime,l)!} + \norm{ \sum_{l \in D^{3l_\imax}} e^{-il\omega t} \bra{l} e^{-i\ms{H}_\eff^{4l_\imax}t} \ket{0}\ket{\psi(0)} -\ket{\psi(t)}} \nonumber \\
    && \quad \leq \sum_{l \in D^{4l_\imax}} 8 \frac{(\gamma t)^{n(l_\imax,l)}}{n(l_\imax,l)!} + 10 m_\imax \left( \frac{em_\imax \gamma t }{3l_\imax} \right)^{3l_\imax / m_\imax}. \label{EqA2:thm_amp1_proof_3}
\end{eqnarray}
In the first inequality, we employ Lemma \ref{LemmaA2:translation_symmetry} with taking $l-l^\prime = l \ominus l^\prime $ for $l -l^\prime \in D^{3l_\imax}$ into consideration.
The last inequality comes from Theorem \ref{Thm4:Floquet_Hilbert_truncation}.
The first term in Eq. (\ref{EqA2:thm_amp1_proof_3}) is further bounded by
\begin{equation}
    16 \sum_{l=3l_\imax-2m_\imax}^\infty \frac{(\gamma t)^{\lceil l/m_\imax \rceil}}{(\lceil l / m_\imax \rceil)!}  \leq  16 m_\imax \left( \frac{e\gamma t}{\lceil 3l_\imax /m_\imax -2 \rceil}\right)^{\lceil 3l_\imax /m_\imax -2 \rceil}, 
\end{equation}
where we use the relation Eq. (\ref{Eq4A:sum_Taylor_modulo}) and the Stirling's formula Eq. (\ref{Eq4A:Stirling_inequality}).
This accomplishes the evaluation of Eq. (\ref{EqA2:thm_amp1_proof_3}) with the usage of Eq. (\ref{Eq4B:error_condition}) as 
\begin{eqnarray}
    [\text{Eq. (\ref{EqA2:thm_amp1_proof_3})}] &\leq& 26 m_\imax \left(\frac{  e \gamma t}{3 l_\imax /m_\imax - 3 } \right)^{3 l_\imax / m_\imax - 3} \nonumber \\
    &\leq& 26 m_\imax \left[ \left(\frac{  e \gamma t}{l_\imax /m_\imax -1} \right)^{l_\imax / m_\imax -1} \right]^3 \nonumber \\
    &\leq& \frac{13}{500 (m_\imax)^2} \varepsilon^3. \label{EqA2:thm_amp1_proof_4}
\end{eqnarray}

We next compute the second summation in Eq. (\ref{EqA2:thm_amp1_proof_2}), which is taken over $l \in D^{4l_\imax}$ satisfying $l-l^\prime \notin D^{3l_\imax}$.
The Lieb-Robinson bound, dictated by Theorem \ref{Thm4:Bound_transition_rate}, immediately concludes its upper bound by
\begin{eqnarray}
    \norm{\sum_{l \in D^{4l_\imax}; l-l^\prime \notin D^{3l_\imax}} e^{-il\omega t} \bra{l} e^{-i\ms{H}_\eff^{4l_\imax}t} \ket{l^\prime} \ket{\psi(0)}} 
    &\leq& \sum_{l; l-l^\prime \notin D^{3l_\imax}} 2 \frac{(\gamma t)^{\lceil |l-l^\prime|/m_\imax \rceil}}{(\lceil |l-l^\prime|/m_\imax \rceil)!} \nonumber \\
    &\leq& 4 \sum_{l=3l_\imax}^\infty \frac{(\gamma t)^{\lceil l /m_\imax \rceil}}{(\lceil l / m_\imax \rceil)!} \nonumber \\
    &\leq& 4 m_\imax \left( \frac{em_\imax \gamma t}{3l_\imax}\right)^{3l_\imax/m_\imax} \nonumber \\
    &\leq& \frac{\varepsilon^3}{250 (m_\imax)^2}. \label{EqA2:thm_amp1_proof_5}
\end{eqnarray}

Finally, we show that the amplification protocol relying on the symmetry reproduces the time-evolved state $\ket{\psi(t)}$. 
Combining the results of Eqs. (\ref{EqA2:thm_amp1_proof_4}) and (\ref{EqA2:thm_amp1_proof_5}), we arrive at
\begin{eqnarray}
    \norm{\bra{0}\ms{U}_\mr{amp1}^{l_\imax}(t)\ket{0}\ket{\psi(0)}-\frac12 \ket{0}\ket{\psi(t)}} 
    &\leq& \frac{1}{4l_\imax} \sum_{l^\prime \in D^{l_\imax}} \left( \frac{13 \varepsilon^3}{500 (m_\imax)^2} + \frac{\varepsilon^3}{250 (m_\imax)^2} \right) \nonumber \\
    &\leq& \frac{3}{200 (m_\imax)^2} \varepsilon^3, \label{EqA2:thm_amp1_proof_6}
\end{eqnarray}
which is smaller than $\varepsilon/3$ for $m_\imax \in \bbN$ and $\varepsilon \in [0,1]$. $\quad \square$

\subsection{Validity of refined effective Hamiltonian}\label{Asubsec2:refine_effective_Hamiltonian}
In Section \ref{Subsec:block_encode_H_eff}, we raise an efficient block-encoding protocol for implementing $e^{-i\ms{H}_\eff ^{l_\imax}t}$ by the qubitization technique. 
The strategy is to substitute the effective Hamiltonian under the periodic boundary conditions, $\ms{H}_{\eff,\mr{pbc}}^{4l_\imax}$, for the original one $\ms{H}_\eff^{4l_\imax}$; The complexity of the oracles reduces from $\order{l_\imax}$ to $\order{\log l_\imax}$.
In order to justify this replacement of $\ms{H}_{\eff}^{4l_\imax}$ with $\ms{H}_{\eff,\mr{pbc}}^{4l_\imax}$, we provide Theorem \ref{Thm6:refined_H_eff}, which says that its error is bounded from above by
\begin{eqnarray}
    \norm{\bra{0} \ms{U}_\mr{amp1,pbc}^{l_\imax} (t) \ket{0} \ket{\psi(0)} - \frac12 \ket{\psi(t)}} &\leq& \frac{\varepsilon}{3}, \label{EqA2:thm_amp1_pbc}\\
    \norm{ \ms{U}_\mr{amp2,pbc}^{l_\imax} (t) \ket{0} \ket{\psi(0)} - \ket{0}\ket{\psi(t)}} &\leq& \varepsilon. \label{EqA2:thm_amp2_pbc}
\end{eqnarray}
The amplification protocols under the refined effective Hamiltonian, $\ms{U}_\mr{amp1,pbc}^{l_\imax} (t)$ and $\ms{U}_\mr{amp2,pbc}^{l_\imax} (t)$, are designated by Eqs. (\ref{Eq6B:def_amp1_pbc}) and (\ref{Eq6B:def_amp2_pbc}).

\textbf{Proof of Theorem \ref{Thm6:refined_H_eff}.---} The validity of $\ms{U}_\mr{amp2,pbc}^{l_\imax} (t)$, i.e. Eq. (\ref{EqA2:thm_amp2_pbc}), is derived by Eq. (\ref{EqA2:thm_amp1_pbc}) following the discussion in Section \ref{Subsec:Oblivious_amplification}.
It is sufficient to prove the relation, Eq. (\ref{EqA2:thm_amp1_pbc}). 
We first evaluate the upper bound of
\begin{eqnarray}
    && \norm{\bra{0}\ms{U}_\mr{amp1,pbc}^{4l_\imax}(t)\ket{0}\ket{\psi(0)}  - \bra{0}\ms{U}_\mr{amp1}^{4l_\imax}(t)\ket{0}\ket{\psi(0)} } \nonumber \\
    && \quad \leq \frac{1}{4l_\imax} \sum_{l \in D^{4l_\imax}} \sum_{l^\prime \in D^{l_\imax}}\norm{\braket{l|e^{-i \ms{H}_{\eff,\mr{pbc}}^{4l_\imax}  t} - e^{-i \ms{H}_\eff^{4l_\imax}  t}|l^\prime}}. \label{Aeq2:thm_refined_proof_1}
\end{eqnarray}
Each transition amplitude in the summation can be computed in a similar way to the proof of Lemma \ref{LemmaA2:translation_symmetry}. 
We decompose the refined effective Hamiltonian by $\ms{H}_{\eff,\mr{pbc}}^{4l_\imax} = \ms{H}_{\eff}^{4l_\imax} + \ms{H}_b$ with the perturbation
\begin{equation}
    \ms{H}_b = \sum_{(l,m) \in \partial F^{4l_\imax}} (\ket{l}\bra{l \oplus m} \otimes H_{-m} + \text{h.c.}).
\end{equation}
Following the discussion in the proof of Lemma \ref{LemmaA2:translation_symmetry}, the perturbation $\ms{H}_b$ which locally acts on the domain $\partial F^{4l_\imax}$ hardly changes the transition amplitude from the original effective Hamiltonian $\ms{H}_\eff^{4l_\imax}$.
As a result, the deviation can be bounded by
\begin{eqnarray}
    \norm{\braket{l|e^{-i \ms{H}_{\eff,\mr{pbc}}^{4l_\imax}  t} - e^{-i \ms{H}_\eff^{4l_\imax}  t}|l^\prime}} \leq \sum_{n=n(l^\prime,l)}^\infty \frac{t^n}{n!} \left\{ \left(\sup_{t} (\norm{\ms{H}_{\mr{pbc},I}(t)}) \right)^n + \gamma^n \right\}, \label{Aeq2:thm_refined_proof_2}
\end{eqnarray}
where the Hamiltonian in the interaction picture is defined by
\begin{equation}
    \ms{H}_{\mr{pbc},I}(t) = [\ms{U}_0(t)]^\dagger \left( \sum_{m \neq 0} \mr{Add}_{m}^{4l_\imax} \otimes H_m \right) \ms{U}_0(t).
\end{equation}
We again use the property of circulant matrices, and obtain $\norm{\ms{H}_{\mr{pbc},I}(t)} \leq \gamma$, which leads to
\begin{eqnarray}
    \norm{\braket{l|e^{-i \ms{H}_{\eff,\mr{pbc}}^{4l_\imax}  t} - e^{-i \ms{H}_\eff^{4l_\imax}  t}|l^\prime}} &\leq& \sum_{n=n(l^\prime,l)}^\infty 2 \frac{(\gamma t)^n}{n!} \leq 4 \left( \frac{\gamma t}{n(l^\prime,l)} \right)^{n(l^\prime,l)}. 
\end{eqnarray}
As a result, we get the difference between the original and refined effective Hamiltonians as 
\begin{eqnarray}
    \norm{\bra{0}\ms{U}_\mr{amp1,pbc}^{4l_\imax}(t) - \ms{U}_\mr{amp1}^{4l_\imax}(t)\ket{0}\ket{\psi(0)} } &\leq& \frac12 \sum_{l \in D^{4l_\imax}} 4 \frac{(\gamma t)^{n(l_\imax,l)}}{n(l_\imax,l)!} \nonumber \\
    &\leq& 4 \sum_{l=3l_\imax-2m_\imax}^\infty \frac{(\gamma t)^{\lceil l/m_\imax \rceil}}{(\lceil l/m_\imax \rceil)!} \nonumber \\
    &\leq& 4 \left( \frac{3\gamma t }{3l_\imax/m_\imax-2} \right)^{3l_\imax/m_\imax-2} \nonumber \\
    &\leq& \frac{\varepsilon^3}{250 (m_\imax)^2}. \label{Aeq2:thm_refined_proof_3}
\end{eqnarray}
As stated by Theorem \ref{Thm5:amplification_symmetry}, the original amplification protocol $\ms{U}_\mr{amp1}^{4l_\imax}(t)$ accurately generates $\ket{\psi(t)}$ as Eq. (\ref{EqA2:thm_amp1_proof_6}).
Therefore, the one under the refined effective Hamiltonian is also justified as
\begin{eqnarray}
    &&\norm{\bra{0} \ms{U}_\mr{amp1,pbc}^{l_\imax} (t) \ket{0} \ket{\psi(0)} - \frac12 \ket{\psi(t)}} \nonumber \\
    && \quad \leq \norm{\bra{0}\ms{U}_\mr{amp1,pbc}^{4l_\imax}(t)-\ms{U}_\mr{amp1}^{4l_\imax}(t)\ket{0}\ket{\psi(0)} }  + \norm{\bra{0}\ms{U}_\mr{amp1}^{4l_\imax}(t)\ket{0}\ket{\psi(0)} - \frac12 \ket{\psi(t)}} \nonumber \\
    && \quad \leq \frac{19}{1000(m_\imax)^2} \varepsilon^3. \label{Aeq2:thm_refined_proof_4}
\end{eqnarray}
This bound is actually smaller than $\varepsilon/3$, and we complete the proof of Theorem \ref{Thm6:refined_H_eff}. $\quad \square$

\section{Extension to exponentially-decaying Fourier components}\label{Asec:Extension_expo}

In the main text, we focus on time-periodic Hamiltonians which have vanishing Fourier components $H_m=0$ for $|m| > m_\imax$.
Here, we generalize our results to time-periodic Hamiltonians $H(t)$ with exponentially-decaying Fourier components as
\begin{equation}
    \norm{H_m} \leq h e^{-|m|/\zeta}, \quad h,\zeta > 0,
\end{equation}
for $|m|>0$.
The norm of $H_0$ is arbitrary as long as it is bounded.
The effective Hamiltonian in the truncated Floquet-Hilbert space, $\ms{H}_\eff^{l_\imax}$, is the same as Eq. (\ref{Eq3A:def_effective_H_l_max}), but it is dense in the basis $\{\ket{l} \}_{l \in D^{l_\imax}}$ compared to the cases where $H_m=0$ is satisfied for $|m| > m_\imax$.  

We formulate the protocol in a similar manner to the main text, and show that the computational resources for the time-evolved state have nearly-optimal dependence both in $t$ and $1/\varepsilon$.
The difference mainly comes from the form of the Lieb-Robinson bound and the infinite series of $\{ H_m \}$ needed for designating $H(t)$.
The former one affects the truncation order $l_\imax$ for the Floquet-Hilbert space.
The latter one yields the change in block-encoding so that almost all the information about $H(t)$ can be embedded with keeping the efficiency.

\subsection{Truncation order of Floquet-Hilbert space}

We first determine the proper truncation order $l_\imax$ for the Floquet-Hilbert space, as we did in Section \ref{Sec:truncated_Floquet_Hilbert}.
To this aim, we begin with deriving the Lieb-Robinson bound on the transition rate, corresponding to Theorem \ref{Thm4:Bound_transition_rate}.

\begin{theorem}\label{Athm3:Bound_transition_rate}
\textbf{(Bound on transition rate)}

We assume $\norm{H_m} \leq h e^{-|m|/\zeta}$ with certain positive constants $h$ and $\zeta$. 
Then, for $l, l^\prime$ such that $|l|, |l^\prime| \leq l_\imax$, the transition rate is bounded from above by
\begin{equation}
    \norm{\bra{l} e^{-i \ms{H}_\eff^{l_\imax} t} \ket{l^\prime}} \leq  \exp \left( - \frac{|l-l^\prime| - 2 \beta \zeta^\prime h t}{\zeta^\prime} + 2/\beta \right).
\end{equation}
Here, $\beta$ and $\zeta^\prime$ are positive constants defined by
\begin{eqnarray}
\beta = (1-e^{-1/\zeta})^{-1}, \quad
\zeta^\prime = (1/\zeta - 1 + e^{-1/\zeta})^{-1}. \label{Aeq3:def_zeta_prime}
\end{eqnarray}
\end{theorem}

\textbf{Proof.---} We prove the theorem in a similar way to Theorem \ref{Thm4:Bound_transition_rate}. 
We start from the Dyson series expansion in the interaction picture [See Eq. (\ref{Eq4A:Dyson_transition_path})],
\begin{eqnarray}
    \norm{\bra{l} e^{-i \ms{H}_\eff^{l_\imax} t} \ket{l^\prime}} &\leq& \sum_{n=0}^\infty \int_0^t \dd t_n \hdots \int_0^{t_2} \dd t_1 \norm{\sum_{\{l_i\}}\prod_{i=1}^n \braket{l_i|\ms{H}_I(t_i)|l_{i-1}}} \nonumber \\
    &\leq& \sum_{n=0}^\infty \int_0^t \dd t_n \hdots \int_0^{t_2} \dd t_1 \sum_{\{l_i\}}\prod_{i=1}^n \norm{\braket{l_i|\ms{H}_I(t_i)|l_{i-1}}}, \label{Aeq3:thm_LRbound_proof_1}
\end{eqnarray}
where we insert the identity $\sum_{l_i \in D^{l_\imax}} \ket{l_i}\bra{l_i}=I$ for $n-1$ times.
The summation $\sum_{\{l_i\}}$ is taken over $l_i \in D^{l_\imax}$ for $i=1,2,\hdots,n-1$ under fixed $l_0=l^\prime$ and $l_n=l$.
We introduce a new variable $m_i=l_i-l_{i-1}$ instead of using $\{l_i\}$, and then the above integrand is bounded by
\begin{eqnarray}
    F_n &\equiv& \sum_{\{l_i\}}\prod_{i=1}^n \norm{\braket{l_i|\ms{H}_I(t_i)|l_{i-1}}} \nonumber \\
    &\leq& \sum_{\{m_i\}_{i=1}^n \in \bbZ^n} \delta_{l^\prime + m_1+\hdots+m_n, l}  \prod_{i=1}^n \norm{H_{-m_i}} \nonumber \\
    &\leq& \sum_{\{m_i\}_{i=1}^n \in \bbZ^n} \delta_{l^\prime + m_1+\hdots+m_n, l} h^n e^{-\sum_{i=1}^n |m_i|/\zeta}. 
\end{eqnarray}
Since $l^\prime + m_1+\hdots+m_n=l$ implies $|m_1|+\hdots+|m_n| \geq |l-l^\prime|$, it can be further bounded as follows,
\begin{eqnarray}
F_n &\leq& \sum_{\{m_i\}_{i=1}^n \in \bbZ^n} \theta \left( \sum_{i=1}^n |m_i| -|l-l^\prime| \right) h^n e^{-\sum_{i=1}^n |m_i|/\zeta} \nonumber \\
&\leq& (2h)^n \sum_{m_1,\hdots,m_n=0}^\infty  \theta \left( \sum_{i=1}^n m_i -|l-l^\prime| \right) e^{-\sum_{i=1}^n m_i/\zeta}. \nonumber \\
&\equiv& (2h)^n S_n(|l-l^\prime|).
\end{eqnarray}
Here, $\theta(x)$ is a step function, defined by $\theta (x) = 1$ for $x \geq 0$ and $\theta(x)=0$ otherwise. 
The summation $S_n (M)$ is defined by
\begin{equation}
    S_n (M) = \sum_{m_1,\hdots,m_n=0}^\infty  \theta \left( \sum_{i=1}^n m_i - M \right) e^{-\sum_i m_i/\zeta},
\end{equation}
which satisfies $S_1(M)= \sum_{m=M}^\infty e^{-m/\zeta} = \beta e^{-M/\zeta}$.

Let us evaluate the upper bound on $S_n(M)$.
We split the summation over $m_n$ into the one over $m_n \leq M-1$ and the one over $m_n \geq M$, which results in 
\begin{equation}
S_n (M) = \sum_{m_n=0}^{M-1} e^{-m_n/\zeta} S_{n-1}(M-m_n) + \beta^n e^{-M/\zeta}.
\end{equation}
We use this relation recursively until $S_1$ appears.
After the single use of this equality, we obtain
\begin{eqnarray}
    S_n(M) &\leq& \sum_{m_n=0}^{M-1} \sum_{m_{n-1}=0}^{M-m_n-1} e^{-(m_n+m_{n-1})/\zeta} S_{n-2}(M-m_n-m_{n-1}) \nonumber \\
    && \qquad + \beta^{n-1} e^{-M/\zeta} \sum_{m_n=0}^{M-1} 1 + \beta^n e^{-M/\zeta},
\end{eqnarray}
where we use $e^{-m_n/\zeta} \leq 1$ for the second term.
By repeating this calculation $n-1$ times, we arrive at
\begin{eqnarray}
    S_n(M) &\leq& \left( \sum_{m_1=0}^{M-1} \hdots \sum_{m_{n-1}=0}^{M-1-M_{n-2}} \right) e^{-M_{n-1}/\zeta} S_1 (M-M_{n-1}) \nonumber \\
    && \qquad + e^{-M/\zeta} \sum_{k=1}^{n-1} \beta^{n-k+1} \left( \sum_{m_1=0}^{M-1}  \hdots \sum_{m_{k-1}=0}^{M-1-M_{k-2}} \right) 1 \nonumber \\
    &=& e^{-M/\zeta} \sum_{k=1}^{n} \beta^{n-k+1} \left( \sum_{m_1=0}^{M-1}  \hdots \sum_{m_{k-1}=0}^{M-1-M_{k-2}} \right) 1,
\end{eqnarray}
where $\sum_{i=1}^{k} m_k$ is denoted by $M_k$.
In the above formula, the summation of $1$ over $\{ m_i \}_{i=1}^{k-1}$ represents the number of lattice points included in a $(k-1)$-dimensional pyramid $P_{k-1}^M = \{ (x_1,\hdots,x_{k-1}) \in [0,M-1]^k \, | \, 0 \leq \sum_i x_i \leq M-1 \}$, and hence it is bounded by the volume of $P_{k-1}^{M+2}$.
This leads to the relation,
\begin{eqnarray}
S_n(M) \leq \beta^n e^{-M/\zeta} \sum_{k=1}^n \frac{\{ \beta^{-1} (M+2) \}^{k-1}}{(k-1)!} \leq \beta^n e^{-(1/\zeta-1/\beta)M + 2/\beta},
\end{eqnarray}

We define a positive constant $\zeta^\prime$ by $1/\zeta^\prime = 1/\zeta - 1/\beta$ as  Eq. (\ref{Aeq3:def_zeta_prime}).
The above inequality enables to evaluate the integrand $F_n$ as
\begin{equation}
    F_n \leq (2h)^n S_n (|l-l^\prime|) \leq  (2\beta h)^n e^{-|l-l^\prime|/\zeta^\prime + 2/\beta}.
\end{equation}
Finally, going back to the inequality Eq. (\ref{Aeq3:thm_LRbound_proof_1}), we arrive at the bound on the transition rate,
\begin{eqnarray}
    \norm{\bra{l} e^{-i \ms{H}_\eff^{l_\imax} t} \ket{l^\prime}} \leq 
    \sum_{n=0}^\infty \frac{t^n}{n!}  (2\beta h)^n e^{-|l-l^\prime|/\zeta^\prime + 2/\beta} \leq e^{2\beta h t - |l-l^\prime|/\zeta^\prime + 2/\beta}.
\end{eqnarray}
This completes the proof of the theorem. $\quad \square$

Theorem \ref{Athm3:Bound_transition_rate} says that the bound on the transition amplitude from $\ket{l^\prime}$ to $\ket{l}$ exponentially decays in the distance $|l-l^\prime|$ for Hamiltonians with $\norm{H_m} \lesssim e^{-\order{|m|}}$.
The decay is relatively slow compared to the cases $H_m=0$ ($|m|>m_\imax$) showing $\order{|l-l^\prime|^{-|l-l^\prime|}}$.
In fact, this result is reminiscent of the Lieb-Robinson bound for short-ranged interacting systems \cite{Robinson1976-wd,Nachtergaele2006-ok,Nachtergaele2006-il}, which provides exponentially-decaying correlation functions under interactions $U_{ij} \sim e^{-\order{|i-j|}}$.
Based on Theorem \ref{Athm3:Bound_transition_rate}, we assess how the exact time-evolved state can be approximated by the truncated Floquet-Hilbert space, and determine the proper truncation order $l_\imax$.
We summarize the result by the following theorem, which is a counterpart of Theorem \ref{Thm4:Floquet_Hilbert_truncation}.

\begin{theorem}\label{Athm3:Floquet_Hilbert_truncation}
\textbf{(Floquet-Hilbert space truncation)}

We assume $\norm{H_m} \leq C e^{-|m|/\zeta}$.
The exact time-evolved state $\ket{\psi(t)}$ is approximated by the truncated state $\ket{\psi^{l_\imax}(t)}$ [See Eq. (\ref{Eq3A:def_psi_l_max})] as
\begin{equation}\label{Aeq3:thm_truncation_deviation}
    \norm{\ket{\psi(t)}-\ket{\psi^{l_\imax}(t)}} \leq  4 \zeta^\prime e^{2\beta h t - (l_\imax - 1 )/\zeta^\prime + 2/\beta}.
\end{equation}
\end{theorem}

\textbf{Proof.---} 
The derivation is similar to the one for Theorem \ref{Thm4:Floquet_Hilbert_truncation}.
According to Eqs. (\ref{Eq4A:thm_truncation_proof_1})-(\ref{Eq4A:thm_truncation_proof_3}), we should evaluate the following two values,
\begin{eqnarray}
    \varepsilon_1 = \sum_{l \in (D^{l_\imax^\prime} \backslash D^{l_\imax})} \norm{\bra{l} e^{i\ms{H}_\eff^{l_\imax^\prime}t} \ket{0}}, \quad
    \varepsilon_2 = \sum_{l \in D^{l_\imax}} \norm{\bra{l} e^{i\ms{H}_\eff^{l_\imax^\prime}t}-e^{i\ms{H}_\eff^{l_\imax}t} \ket{0}}, \label{Aeq3:thm_truncation_epsilon_2}
\end{eqnarray}
where the left hand side of Eq. (\ref{Aeq3:thm_truncation_deviation}) is bounded by $\varepsilon_1 + \varepsilon_2$ under $l_\imax^\prime \to \infty$.
Theorem \ref{Athm3:Bound_transition_rate} soon concludes the upper bound of $\varepsilon_1$;
\begin{eqnarray}
\varepsilon_1 &\leq& \sum_{l \in (D^{l_\imax^\prime} \backslash D^{l_\imax})} e^{2 \beta h t - |l|/\zeta^\prime + 2/\beta} \nonumber \\
&\leq& 2  \int_{l_\imax-1}^\infty \dd x e^{2\beta h t - x /\zeta^\prime + 2/\beta} \nonumber \\
&\leq& 2 \zeta^\prime e^{2\beta h t - (l_\imax-1)/\zeta^\prime + 2/\beta}. \label{Aeq3:thm_truncation_epsilon_1_result}
\end{eqnarray}

The second error $\varepsilon_2$ can be evaluated by the Dyson series expansion like Eq. (\ref{Eq4A:Dyson_transition_path}); each term of $\varepsilon_2$ in Eq. (\ref{Aeq3:thm_truncation_epsilon_2}) is bounded from above by
\begin{eqnarray}
    \sum_{n=0}^\infty \int_0^t \dd t_n \hdots \int_0^{t_2} \dd t_1 \norm{\braket{l|\prod_{i=1}^n \ms{H}_I^{l_\imax^\prime}(t_i) - \prod_{i=1}^n \ms{H}_I^{l_\imax}(t_i)|0}}. \label{Aeq3:thm_truncation_proof_1}
\end{eqnarray}
As stated in the proof of Theorem \ref{Thm4:Bound_transition_rate}, each of $\braket{l|\prod_{i=1}^n \ms{H}_I^{l_\imax^\prime}(t_i) |0}$ and $\braket{l|\prod_{i=1}^n \ms{H}_I^{l_\imax}(t_i) |0}$ is decomposed into a product of transition amplitudes via a path $\ket{0} \to \ket{l_1} \to \hdots \to \ket{l_{n-1}} \to \ket{l}$.
Their difference appears only when the path goes across $D^{l_\imax^\prime} \backslash D^{l_\imax}$.
In other words, denoting the summation over the set of $\{l_i\}$ taking such nontrivial paths by $\sum_{\{ l_i \}}^\prime$, we reach
\begin{eqnarray}
    [\text{Eq. (\ref{Aeq3:thm_truncation_proof_1})}] \leq \sum_{n=0}^\infty \int_0^t \dd t_n \hdots \int_0^{t_2} \dd t_1 \sum_{\{ l_i \}} \,^\prime \prod_{i=1}^n \norm{\braket{l_i | \ms{H}_I^{l_\imax^\prime}(t_i) | l_{i-1}}}. \label{Aeq3:thm_truncation_proof_2}
\end{eqnarray}
When we define the hopping distance $m_i$ by $m_i=l_i -l_{i-1}$, it should satisfy $\sum_{i=1}^n |m_i| \geq (l_\imax - |l|) + (l_\imax - 0)$ for $\{ l_i \}$ such that a nontrivial path is organized.
This results in the relation described by
\begin{eqnarray}
    \varepsilon_2 &\leq& \sum_{l \in D^{l_\imax}}\sum_{n=0}^\infty \frac{t^n}{n!} \sum_{\{m_i\}_{i=1}^n \in \bbZ^n} \left(\prod_{i=1}^n  \norm{H_{-m_i}} \right) \times \theta \left( \sum_{i=1}^n |m_i| - 2l_\imax + |l| \right) \nonumber \\
    &\leq&  \sum_{l \in D^{l_\imax}} \sum_{n=0}^\infty \frac{(2 h t)^n}{n!} S_n (2l_\imax - |l|) \nonumber \\
    &\leq& \sum_{l \in D^{l_\imax}} \sum_{n=0}^\infty \frac{(2 \beta h t)^n}{n!} e^{-(2l_\imax - |l|)/\zeta^\prime + 2/\beta} \nonumber \\
    &\leq& 2 \zeta^\prime e^{2\beta h t - (l_\imax - 1 )/\zeta^\prime + 2/\beta}. \label{Aeq3:thm_truncation_proof_3}
\end{eqnarray}
Combining this inequality with $\norm{\ket{\psi(t)}-\ket{\psi^{l_\imax}(t)}} \leq \varepsilon_1 + \varepsilon_2$ and Eq. (\ref{Aeq3:thm_truncation_epsilon_1_result}), we obtain Eq. (\ref{Aeq3:thm_truncation_deviation}), which completes the proof. $\quad \square$

Theorem \ref{Athm3:Floquet_Hilbert_truncation} determines the proper truncation order of the Floquet-Hilbert space for the algorithm.
When we aim at the desirable accuracy as $\norm{\ket{\psi(t)}-\ket{\psi^{l_\imax}(t)}} \leq \varepsilon$, it is sufficient to choose the truncation order $l_\imax$ by
\begin{eqnarray}
    l_\imax &=& \left\lceil 2 \beta \zeta^\prime ht + \zeta^\prime \log (1/\varepsilon) + \zeta^\prime \log (4\zeta^\prime )+ \frac{2 \zeta^\prime}{\beta}  + 1 \right\rceil \in \Theta \left( ht + \log (1/\varepsilon) \right). \label{Aeq3:l_max_choice_exp}
\end{eqnarray}
The result has similar scaling in $t$ and $1/\varepsilon$ to the one for Hamiltonians with a finite number of Fourier components, described by Eq. (\ref{Eq4B:l_max_choice}), other than the factor of $\log \log (1/\varepsilon)$.
While the actual $m_\imax$ is infinite for exponentially-decaying Fourier components as $\norm{H_m} \leq h e^{-|m|/\zeta}$, the parameters $h$ and $\zeta^\prime$ play a role of $\gamma$ and $m_\imax$ respectively.

\subsection{Amplitude amplification}

We next verify the validity of the amplitude amplification protocols, discussed in Section \ref{Sec:Amplification}, for time-periodic Hamiltonians with exponentially-decaying Fourier components $\norm{H_m} \leq h e^{-|m|/\zeta}$.

The first amplification relies on the translation symmetry of the effective Hamiltonian, as discussed in \ref{Subsec:amplification_symmetry}.
To show its extension, we begin with discussing the approximate translation symmetry in the truncated Floquet-Hilbert space, which is a counterpart of Lemma \ref{LemmaA2:translation_symmetry}.

\begin{lemma}\label{LemmaA3:translation_symmetry}
\textbf{(Approximate translation symmetry)}

We assume $\norm{H_m} \leq h e^{-|m|/\zeta}$, and consider the truncated Floquet-Hilbert space $\bbC^{8l_\imax} \otimes \mcl{H}$. 
The transition rate has an approximate translation symmetry in that it satisfies
\begin{eqnarray}
    \norm{\bra{l} e^{-i \ms{H}_\eff^{4 l_\imax}  t} \ket{l^\prime} - e^{il^\prime \omega t} \bra{l \ominus l^\prime} e^{-i \ms{H}_\eff^{4 l_\imax}  t} \ket{0}} \leq 2  e^{2\beta h t - (8l_\imax - |l|-|l^\prime|)/\zeta^\prime + 2/ \beta}. \label{Aeq3:lemma_approximate_symmetry}
\end{eqnarray}
\end{lemma}

\textbf{Proof.---} The proof is done in a similar manner to that of Lemma \ref{LemmaA3:translation_symmetry}.
We consider a perturbation $\tilde{\ms{H}}_b(t)$ designated by
\begin{eqnarray}
    \tilde{\ms{H}}_b(t) &=& \sum_{(l,m) \in \partial \tilde{F}^{4l_\imax}} \ket{l}\bra{l \oplus m} \otimes e^{8 i l_\imax \omega t} H_{-m} + \text{h.c.}, \label{Aeq3:perturbed_H_b(t)}
\end{eqnarray}
with $ \partial \tilde{F}^{4l_\imax} = \{ (l,m) \, | \, l \in D^{4_\imax}, \, 8l_\imax -l + 1 \leq m \leq 8l_\imax -1 \}$.
This Hamiltonian indicates hopping terms that go across the boundaries $\ket{4l_\imax}$ and $\ket{-4l_\imax+1}$.
Let $\tilde{\ms{U}}_\mr{pert}(t)$ denote a time evolution operator under $\ms{H}_\eff^{4l_\imax}+\tilde{\ms{H}}_b(t)$.
Then, due to the exact translation symmetry in the interaction picture, the transition amplitude $\braket{l |\tilde{\ms{U}}_\mr{pert}(t)|l^\prime}$ satisfies
\begin{equation}\label{Aeq3:lemma_exact_symmetry_int}
    \braket{l |\tilde{\ms{U}}_\mr{pert}(t)|l^\prime} = e^{i l^\prime \omega t} \braket{l \ominus l^\prime |\tilde{\ms{U}}_\mr{pert}(t)|0}.
\end{equation}

The difference of transition amplitudes between $\ms{H}_\eff^{4l_\imax}+\tilde{\ms{H}}_b(t)$ and $\ms{H}_\eff^{4l_\imax}$ is bounded in a similar way to Eqs. (\ref{Aeq3:thm_truncation_proof_1}) and (\ref{Aeq3:thm_truncation_proof_3}).
The difference survives only when the trajectory $\ket{l^\prime} \to \ket{l_1} \to \ket{l_{n-1}} \to \ket{l}$ pass through the boundaries $\ket{4 l_\imax}$ and $\ket{- 4 l_\imax +1 }$ via $\tilde{\ms{H}}_b(t)$, and then its length $\sum_{i=1}^n |m_i|$ with $m_i = l_i - l_{i-1}$ should be equal to or larger than $(4 l_\imax - |l|)+(4 l_\imax-|l^\prime|)$.
We obtain its upper bound in a similar way to Eq. (\ref{Aeq3:thm_truncation_proof_3}),
\begin{eqnarray}
    \norm{\braket{l |\tilde{\ms{U}}_\mr{pert}(t)-e^{-i \ms{H}_\eff^{4 l_\imax}t}|l^\prime}} &\leq&  \sum_{n=0}^n \frac{(2ht)^n}{n!} S_n(8 l_\imax - |l|-|l^\prime|) \nonumber \\
    &\leq& e^{2\beta h t - (8l_\imax - |l|-|l^\prime|)/\zeta^\prime + 2/ \beta}. \label{Aeq3:transition_difference_H_b(t)}
\end{eqnarray}
Using this relation twice and the symmetry Eq. (\ref{Aeq3:lemma_exact_symmetry_int}) leads to Eq. (\ref{Aeq3:lemma_approximate_symmetry}). $\quad \square$

With the usage of this approximate translation symmetry, we can organize the amplitude amplification by symmetry like Theorem \ref{Thm5:amplification_symmetry}.
We prepare the truncated Floquet-Hilbert space $\bbC^{8l_\imax} \otimes \mcl{H}$, and make the initial state uniform in $\ket{l}$ with $l_\imax \in \Theta ( h t + \log (1/\varepsilon))$.
The state resulting from the time evolution $\exp (-i \ms{H}_\eff^{4l_\imax} t)$ outputs the target time-evolved state $\ket{\psi(t)}$ with amplitude $1/2$ as follows.

\begin{theorem}\label{Athm3:amplification_symmetry}
\textbf{(Amplification by symmetry)}

We assume $\norm{H_m} \leq h e^{-|m|/\zeta}$, and choose the truncation order $l_\imax \in \order{ht+\log(1/\varepsilon)}$ by Eq. (\ref{Aeq3:l_max_choice_exp}).
Let us summarize the amplification protocol relying on the symmetry by $\ms{U}_\mr{amp1}^{l_\imax}(t)$, whose explicit formula is given by Eq. (\ref{Eq5A:U_amp1_def}).
Then, it generates the time-evolved state $\ket{\psi(t)}$ with $\order{1}$ amplitude as
\begin{equation}\label{Aeq3:amp1_deviation_result_expo}
    \norm{\braket{0|\ms{U}_\mr{amp1}^{l_\imax}(t)|0}\ket{\psi(0)} - \frac{1}{2} \ket{\psi(t)}} \leq \frac{5}{64(\zeta^\prime)^2} \varepsilon^3.
\end{equation}
It is reasonable to assume that the desirable error $\varepsilon$ is sufficiently small, not greater than the constant $8 \zeta^\prime / \sqrt{15}$. Then, the right hand side of the above formula is bounded by $\varepsilon/3$, which reproduces Theorem \ref{Thm5:amplification_symmetry}.
\end{theorem}

\textbf{Proof.---} The proof follows that of Theorem \ref{Thm5:amplification_symmetry}, described in Appendix \ref{Asubsec2:amplification_symmetry}.
The left hand side of the above inequality can be bounded by two contributions determined by Eq. (\ref{EqA2:thm_amp1_proof_2}).
The first contribution, which corresponds to Eq. (\ref{EqA2:thm_amp1_proof_3}), is given by
\begin{eqnarray}
    && \norm{\sum_{l; l-l^\prime \in D^{3l_\imax}} e^{-il\omega t} \bra{l} e^{-i\ms{H}_\eff^{4l_\imax}t} \ket{l^\prime} \ket{\psi(0)} - \ket{\psi(t)} } \nonumber \\
    && \quad \leq \sum_{l; l-l^\prime \in D^{3l_\imax}} 2 e^{2\beta h t - \frac{7 l_\imax - |l|}{\zeta^\prime} + \frac{2}{\beta}} + \norm{\ket{\psi^{3l_\imax}(t)} - \ket{\psi(t)}} \nonumber \\
    && \quad \leq 8 \zeta^\prime e^{2\beta h t - (3 l_\imax - 1 )/\zeta^\prime + 2/\beta} \leq \frac{\varepsilon^3}{8 (\zeta^\prime)^2} .
\end{eqnarray}
Here, we used Lemma \ref{LemmaA3:translation_symmetry} in the second line, Theorem \ref{Athm3:Floquet_Hilbert_truncation} in the third line, and the choice of $l_\imax$, Eq. (\ref{Aeq3:l_max_choice_exp}), in the last line.
The second contribution, which is a counterpart of Eq. (\ref{EqA2:thm_amp1_proof_5}), is bounded by
\begin{eqnarray}
     \norm{\sum_{l; l-l^\prime \notin D^{3l_\imax}} e^{-il\omega t} \bra{l} e^{-i\ms{H}_\eff^{4l_\imax}t} \ket{l^\prime} \ket{\psi(0)}} &\leq& \sum_{l; l-l^\prime \notin D^{3l_\imax}} e^{2\beta ht - |l-l^\prime|/\zeta^\prime+2/\beta} \nonumber \\
    &\leq& 2 \zeta^\prime e^{2\beta ht - (3l_\imax-1)/\zeta^\prime + 2/\beta} \leq \frac{\varepsilon^3}{32 (\zeta^\prime)^2}. \nonumber \\
    &&
\end{eqnarray}
By combining these results like Eq. (\ref{EqA2:thm_amp1_proof_6}), we complete the proof of Theorem \ref{Athm3:amplification_symmetry}. $\quad \square$

The above theorem ensures the amplification by symmetry also for time-periodic Hamiltonians with $\norm{H_m} \leq h e^{-|m|/\zeta}$; it enhances the amplitude from $\order{l_\imax^{-1}}$ to $1/2$ only with additional $\order{\log l_\imax}$ elementary gates.
In order to bring the amplitude up to $1-\order{\varepsilon}$, we need the oblivious amplitude amplification.
Its validity immediately follows from Theorem \ref{Athm3:amplification_symmetry}, as we discussed in \ref{Subsec:Oblivious_amplification}.
In other words, when we apply the protocol $\ms{U}_\mr{amp2}^{l_\imax}(t)$, defined by Eq. (\ref{Eq5B:def_amp2}), the time-evolved state $\ket{\psi(t)}$ can be obtained with accuracy $1-\order{\varepsilon}$;
\begin{equation}
    \norm{\ms{U}_\mr{amp2}^{l_\imax}(t)\ket{0}\ket{\psi(0)} - \ket{0} \ket{\psi(t)}} \leq \frac{15}{64(\zeta^\prime)^2} \varepsilon^3.
\end{equation}
The right hand side is bounded by $\varepsilon$ for sufficiently small $\varepsilon$.

\subsection{Qubitization technique for effective Hamiltonian}

Simulating $\ket{\psi(t)}$ via the amplitude amplification $\ms{U}_\mr{amp2}^{l_\imax}(t)$ requires the time evolution operators $\exp (-i \ms{H}_\mr{LP}^{4l_\imax}t)$ and $\exp (-i \ms{H}_\eff^{4l_\imax}t)$.
The former one is the same as Section \ref{Subsec:Block_encoding_linear_potential}.
We hereby present how the latter one is implemented for the cases $\norm{H_m} \leq h e^{-|m|/\zeta}$.
We take a similar strategy to Section \ref{Subsec:block_encode_H_eff}, that is, we first organize a refined effective Hamiltonian $\ms{H}_{\eff,\mr{pbc}}^{4l_\imax}$, which can accurately reproduce the dynamics under $\ms{H}_\eff^{4l_\imax}$.
After that, we compose its block-encoding which can be efficiently achievable.

\subsubsection{Refined effective Hamiltonian}

We introduce a refined effective Hamiltonian $\ms{H}_{\eff,\mr{pbc}}^{4l_\imax}$, which acts on the truncated Floquet-Hilbert space $\bbC^{8l_\imax} \otimes \mcl{H}$, by
\begin{equation}\label{Aeq3:refined_H_eff}
    \ms{H}_{\eff,\mr{pbc}}^{4l_\imax} = \ms{H}_\eff^{4l_\imax} + \tilde{\ms{H}}_b, \quad 
    \tilde{\ms{H}}_b = \sum_{(l,m) \in \partial \tilde{F}^{4l_\imax}} \ket{l}\bra{l \oplus m} \otimes H_{-m} + \text{h.c.},
\end{equation}
where $\partial \tilde{F}^{4l_\imax}$ shares the definition with the one in Eq. (\ref{Aeq3:perturbed_H_b(t)}).
Owing to the additional term $\ms{H}_b$, the hopping terms induced by $H_{-m}$ become translation symmetric in $\ket{l}$ as
\begin{equation}\label{Aeq3:refined_H_eff_result}
    \ms{H}_{\eff,\mr{pbc}}^{4l_\imax} = \sum_{m \in D^{4l_\imax}} \mr{Add}_m^{4l_\imax} \otimes H_m - \ms{H}_\mr{LP}^{4l_\imax}.
\end{equation}
Here, a full quantum adder $\mr{Add}_m^{4l_\imax}$, defined by Eq. (\ref{Eq6B:def_adder}), appears and it allows efficient implementation of block-encoding as discussed later.
Before going to its block-encoding, we prove the validity of the refined effective Hamiltonian as a counterpart of Theorem \ref{Thm6:refined_H_eff}.

\begin{theorem}\label{Athm3:refined_H_eff}
\textbf{(Refined effective Hamiltonian)}

We assume $\norm{H_m} \leq h e^{-|m|/\zeta}$, and organize the two amplification protocols $\ms{U}_\mr{amp1,pbc}^{l_\imax}(t)$ and $\ms{U}_\mr{amp2,pbc}^{l_\imax}(t)$ respectively based on Eqs. (\ref{Eq6B:def_amp1_pbc}) and (\ref{Eq6B:def_amp2_pbc}) with using the refined effective Hamiltonian $\ms{H}_{\eff,\mr{pbc}}^{4l_\imax}$ given by Eq. (\ref{Aeq3:refined_H_eff}).
When the truncation order $l_\imax \in \Theta (ht+\log(1/\varepsilon))$ is chosen by Eq. (\ref{Aeq3:l_max_choice_exp}), they also provide the exact time-evolved state $\ket{\psi(t)}$ as 
\begin{eqnarray}
    \norm{\bra{0} \ms{U}_\mr{amp1,pbc}^{l_\imax} (t) \ket{0} \ket{\psi(0)} - \frac12 \ket{\psi(t)}} &\leq& \frac{11}{128 (\zeta^\prime)^2} \varepsilon^3 , \label{EqA3:thm_amp1_pbc} \\
    \norm{ \ms{U}_\mr{amp2,pbc}^{l_\imax} (t) \ket{0} \ket{\psi(0)} - \ket{0}\ket{\psi(t)}} &\leq& \frac{33}{128 (\zeta^\prime)^2} \varepsilon^3,  \label{EqA3:thm_amp2_pbc}
\end{eqnarray}
for arbitrary initial states $\ket{\psi(0)} \in \mcl{H}$.
For a allowable error $\varepsilon$ sufficiently small compared to the constant $\zeta^\prime$, both of the left hand sides are smaller than $\varepsilon$.
\end{theorem}

\textbf{Proof.---} We prove the theorem as we do for Theorem \ref{Thm6:refined_H_eff} in Appendix \ref{Asubsec2:refine_effective_Hamiltonian}.
First, we evaluate the difference of the transition rates between $\ms{H}_{\eff,\mr{pbc}}^{4l_\imax}$ and $\ms{H}_{\eff}^{4l_\imax}$. 
We replace the perturbation $\tilde{H}_b(t)$ by the boundary term $\tilde{H}_b$ in the proof of Lemma \ref{LemmaA3:translation_symmetry} (See Appendix \ref{Asubsec2:amplification_symmetry}).
We obtain the same result as Eq. (\ref{Aeq3:transition_difference_H_b(t)}),
\begin{eqnarray}
    \norm{\bra{l} e^{-i \ms{H}_{\mr{eff,pbc}}^{4l_\imax}  t} \ket{l^\prime} - \bra{l} e^{-i \ms{H}_{\mr{eff}}^{4l_\imax}  t} \ket{l^\prime}} \leq  e^{2 \beta h t - (8l_\imax -|l|-|l^\prime|)/\zeta^\prime + 2/\beta}.
\end{eqnarray}
Once we obtain this bound, we can track the proof of Theorem \ref{Thm6:refined_H_eff}, composed of Eqs. (\ref{Aeq2:thm_refined_proof_1})-(\ref{Aeq2:thm_refined_proof_4}). 
First, we evaluate the deviation from $\ms{U}_\mr{amp1}^{l_\imax}(t)$ as
\begin{eqnarray}
     \norm{\bra{0}\ms{U}_\mr{amp1,pbc}^{4l_\imax}(t) - \ms{U}_\mr{amp1}^{4l_\imax}(t)\ket{0}\ket{\psi(0)} } &\leq&  \sum_{l \in D^{4l_\imax}, l^\prime \in D^{l_\imax}} \frac{e^{2 \beta h t - (8l_\imax -|l|-|l^\prime|)/\zeta^\prime + 2/\beta}}{4l_\imax} \nonumber \\
    &\leq& \frac{1}{2} \zeta^\prime e^{2\beta h t - (3l_\imax -1 )/\zeta^\prime + 2/\beta} \leq \frac{\varepsilon^3}{128 (\zeta^\prime)^2} . 
\end{eqnarray}
Since $\bra{0} \ms{U}_\mr{amp1}^{4l_\imax}(t)\ket{0}\ket{\psi(0)}$ accurately provides the state $\ket{\psi(t)}/2$ as Eq. (\ref{Aeq3:amp1_deviation_result_expo}) according to Theorem \ref{Athm3:amplification_symmetry}, a triangle inequality concludes Eq. (\ref{EqA3:thm_amp1_pbc}).
Since the oblivious amplitude amplification generates the error at-most three times larger than Eq. (\ref{EqA3:thm_amp1_pbc}) according to the discussion in Section \ref{Subsec:Oblivious_amplification}, Eq. (\ref{EqA3:thm_amp2_pbc}) is immediately derived. $\quad \square$

\subsubsection{Block-encoding}
We compose block-encoding of the refined effective Hamiltonian $\ms{H}_{\eff,\mr{pbc}}^{4l_\imax}$ to implement its time evolution operator by qubitization.

We begin with describing the assumption on time-periodic Hamiltonians $H(t)$.
If we adopt the one for the main text as Eq. (\ref{Eq6B:block_encoding_H_m}), we should have the knowledge about block-encoding for the infinite series $\{ H_m \}_{m \in \bbZ}$. 
To avoid this complexity, we instead consider time-periodic Hamiltonians $H(t)$ written by
\begin{equation}
    H(t) = \sum_{j=1}^{j_\imax} \alpha_j(t) M_j, \quad \alpha_j(t+T) = \alpha_j (t).
\end{equation}
with coefficients $\{ \alpha_j (t) \}_{j=1}^{j_\imax}$ and operators $\{ M_j \}_{j=1}^{j_\imax}$.
We assume the knowledge of block-encoding for each operator $M_j$ as
\begin{equation}\label{EqA3:block_encode_M_j}
    \braket{G_j|O_j|G_j} = M_j, \quad \ket{G_j}= G_j \ket{0}^{\otimes n_a}, 
\end{equation}
with an oracle state $\ket{G_j} \in \bbC^{2^{n_a}}$ and an oracle unitary gate $O_j$ on $\bbC^{2^{n_a}} \otimes \mcl{H}$.
We can always set the denominator in block-encoding by $1$ since its absolute value and its phase can be absorbed respectively into the coefficient $\alpha_j(t)$ and the oracle $O_j$.
The Fourier components are given by
\begin{equation}\label{Aeq3:block_encoding_H_m_expo}
    H_m = \sum_{j=1}^{j_\imax} \alpha_j^m M_j, \quad \alpha_j^m = \frac{1}{T}\int_0^T \dd t
\alpha_j(t) e^{im\omega t},
\end{equation}
and it is always possible to impose $\alpha_j^m \geq 0$ for every $j,m$. 
This is because, when a certain coefficient $\alpha_j^m$ is complex, we can divide it into $\alpha_j^m= \Re \alpha_j^m + i \Im \alpha_j^m$ and redefine $M_j$ with the signs of $\Re \alpha_j^m, \Im \alpha_j^m$ and the phase $i$.
In other words, adding $-M_j$, $iM_j$, and $-iM_j$ to the set $\{M_j \}_j$ reproduces the form of Eq. (\ref{Aeq3:block_encoding_H_m_expo}) with $\alpha_j^m \geq 0$, where $j_\imax$ is replaced by at-most $4l_\imax$.
In the following discussion, the oracles $O_j$ and $G_j$ are respectively supposed to be composed of at-most $C$ elementary gates.
In addition, we also assume the access to the coefficients $\alpha_j^m$ for $m \in D^{l_\imax}$ by the oracle unitary gate $G_\mr{coef}^{l_\imax}$ as
\begin{eqnarray}
    G_\mr{coef}^{l_\imax} \ket{0} \ket{0} =  \sum_{m \in D^{l_\imax}}  \sum_{j=1}^{j_\imax}\sqrt{\frac{\alpha_j^m}{\alpha^{l_\imax}}} \ket{m} \ket{j}, \quad
    \alpha^{l_\imax} = \sum_{m \in D^{l_\imax}} \sum_{j=1}^{j_\imax} \alpha_j^m. \label{Aeq3:G_coeff_expo} 
\end{eqnarray}
The oracle state $G_\mr{coef}^{l_\imax} \ket{0} \ket{0} \in \bbC^{2l_\imax+j_\imax}$ is a $\order{\log l_\imax + \log j_\imax}$-qubit ancillary state, and its preparation $G_\mr{coef}^{l_\imax}$ is supposed to require $C^{l_\imax}$ elementary gates.
The parameter $\alpha^{l_\imax}$ is bounded by the whole-system energy scale as
\begin{equation}\label{Aeq3:alpha_l_max_bound_expo}
    \alpha^{l_\imax} \leq \sum_{m \in \bbZ} \sum_{j=1}^{j_\imax} \alpha_j^m \equiv \alpha.
\end{equation}
Although the definition of $\alpha$ is different from that of the main text, it again provides the bound on the scale of time-dependent terms as
\begin{equation}
    \gamma \leq \sum_{m \in \bbZ} \sum_{j=1}^{j_\imax} |\alpha_j^m| \cdot \norm{M_j} \leq \alpha,
\end{equation}
where we use $\norm{M_j} \leq 1$ from Eq. (\ref{EqA3:block_encode_M_j}).

We move to how we organize an oracle state and oracle unitary gate for the refined $\ms{H}_{\eff,\mr{pbc}}^{4l_\imax}$.
Substituting the Fourier components $H_m$ into Eq. (\ref{Aeq3:refined_H_eff_result}) results in 
\begin{equation}
    \ms{H}_{\eff,\mr{pbc}}^{4l_\imax} = \sum_{m \in D^{4l_\imax}} \sum_{j=1}^{j_\imax} \alpha_j^m \mr{Add}_m^{4l_\imax} \otimes M_j - \ms{H}_\mr{LP}^{4l_\imax}.
\end{equation}
We introduce five kinds of ancillary systems labeled by $a,b,c,d$, and $e$.
The systems $a$, $b$, and $e$ are respectively described by $2^{n_a}$-, $8l_\imax$-, and $2$-dimensional Hilbert spaces, as we consider in Section \ref{Subsec:block_encode_H_eff}.
States of the system $c$ is spanned by $\{ \ket{m} \}_{m \in D^{4l_\imax}}$ due to the absence of $m_\imax$. 
The system $d$ is characterized by $\{\ket{j}\}_{j=1}^{j_\imax}$.
The oracle unitary gate for the refined effective Hamiltonian $\ms{H}_{\eff,\mr{pbc}}^{4l_\imax}$ is given by
\begin{eqnarray}
    \ms{O}_\eff^{4l_\imax} &=& \ket{0}\bra{0}_e \otimes O_J \otimes \mr{Add}^{4l_\imax} \otimes I_b - \ket{1}\bra{1}_e \otimes I_d \otimes I_c \otimes \ms{O}_\mr{LP}^{4l_\imax} \otimes I_a, \\
    O_J &=& \sum_{j=1}^{j_\imax} \ket{j}\bra{j}_d \otimes O_j, \quad
    \mr{Add}^{4l_\imax} = \sum_{m \in D^{4l_\imax}} \ket{m}\bra{m}_c \otimes \mr{Add}_m^{4l_\imax}.
\end{eqnarray}
Here, the unitary gates $O_J$, $\mr{Add}_m^{4l_\imax}$ and $\ms{O}_\mr{LP}^{4l_\imax}$ are respectively implemented by $\order{j_\imax C}$, $\order{\log l_\imax}$, and $\order{\log l_\imax}$ elementary gates (the second one represents addition of a variable $m \in D^{4l_\imax}$ \cite{Cuccaro2004-jp}).
Next, we provide the oracle state $\ket{G_\eff^{4l_\imax}}$ on the auxiliary systems $a,b,c,d,$ and $e$ by
\begin{eqnarray}
    \ket{G_\eff^{4l_\imax}} &=& G_\eff^{4l_\imax}\ket{w^{4l_\imax}}_e (G_\mr{coef}^{4l_\imax}\ket{0}_d \ket{0}_c) \ket{a^{4l_\imax}}_b \ket{0}_a, \\
    G_\eff^{4l_\imax}  &=& I_e \otimes \sum_{j=1}^{j_\imax} \ket{j}\bra{j}_d \otimes I_c \otimes I_b \otimes (G_j)_a, \\
    \ket{w^{4l_\imax}}_e &=& \frac{\sqrt{\alpha^{4l_\imax}}\ket{0}_e + \sqrt{4l_\imax \omega} \ket{1}_e}{\sqrt{\alpha^{4l_\imax} + 4l_\imax \omega }}. 
\end{eqnarray}
We can confirm the relation,
\begin{equation}
    \braket{G_\eff^{4l_\imax}|\ms{O}_\eff^{4l_\imax}|G_\eff^{4l_\imax}} = \frac{\ms{H}_{\eff,\mr{pbc}}^{4l_\imax}}{\alpha^{4l_\imax}+4l_\imax \omega},
\end{equation}
by directly substituting the above equations.
The cost of implementing the oracle state $\ket{G_\eff^{4l_\imax}}$ from the trivial state $\ket{0}_e \ket{0}_d \ket{0}_c \ket{0}_b \ket{0}_a$ is $\order{j_\imax C + C^{4l_\imax} + \log l_\imax}$ elementary gates.

Therefore, implementing the time evolution operator $\exp(-i \ms{H}_{\eff,\mr{pbc}} t)$ with the qubitization technique requires the following resources for time-periodic Hamiltonians satisfying $\norm{H_m} \leq h e^{-|m|/\zeta}$;
\begin{itemize}
    \item Number of ancillary qubits; $\order{\log j_\imax + \log l_\imax}$
    \item Number of overall gates;
    \begin{eqnarray}
        \order{ \left\{ (\alpha + l_\imax \omega) t +  o(\log (1/\varepsilon)) \right\} \left( j_\imax C + C^{4l_\imax} + n_a + \log j_\imax + \log l_\imax  \right) }. \nonumber
    \end{eqnarray}
\end{itemize}
We replace $\alpha^{4l_\imax}$ by $\alpha$ based on the inequality, Eq. (\ref{Aeq3:alpha_l_max_bound_expo}).

\subsection{Algorithm and comparison}\label{Asubsec3:algorithm_comparison_expo}
The algorithm for computing the time-evolved state for time-periodic Hamiltonian with $\norm{H_m} \leq h e^{-|m|/\zeta}$ is almost the same as those in Section \ref{Sec:Algorithm_cost}.
We separately consider adiabatic-like cases and generic long-time cases.
The computational resources are obtained by replacing $\gamma$ in Table \ref{Table:comparison_algorithms} by $h$.
The query complexity is determined by the coefficients of $j_\imax C$ and $C^{4l_\imax}$ in the number of overall elementary gates.
Their scaling is summarized as follows;

\begin{flushleft}
\textbf{Adiabatic-like cases.---}
\end{flushleft}
The scaling of the query complexity is optimal in the time $t$ and nearly-optimal in the inverse error $1/\varepsilon$.
\begin{itemize}
    \item Number of ancillary qubits; $n_a + \order{\log j_\imax + \log (\gamma t) + \log \log (1/\varepsilon)}$.
    \item Query complexity; $\order{\alpha t + \log (1/\varepsilon)}$.
    \item Additional gates per query; $\order{ n_a + \log j_\imax + \log (\gamma t)  + \log \log (1  /\varepsilon) }$.
\end{itemize}

\begin{flushleft}
\textbf{Generic long-time cases.---}
\end{flushleft}
The scaling of the query complexity is nearly-optimal in the time $t$ and the inverse error $1/\varepsilon$.
For practical problems up to $\poly{N}$-time, it becomes optimal in the time $t$.
\begin{itemize}
    \item Number of ancillary qubits; $n_a + \order{\log j_\imax + \log (\gamma/\omega) + \log \log (\omega t/\varepsilon)}$.
    \item Query complexity; $\order{\alpha t + \omega t \log (\omega t/\varepsilon)}$.
    \item Additional gates per query; $\order{ n_a + \log j_\imax + \log (\gamma / \omega)  + \log \log (\omega t/\varepsilon)}$.
\end{itemize}

We note that $\log \log (1/\varepsilon)$ factor in the query complexity is buried by $l_\imax$.
This difference from the cases $H_m=0$ ($|m|>m_\imax$) essentially originates from the form of the Lieb-Robinson bound dictated by \ref{Athm3:Bound_transition_rate}.
A slightly slow decay in the transition amplitude results in slightly larger cost for the cases $\norm{H_m} \leq h e^{-|m|/\zeta}$, compared to the results in the main text.
This leads to the nearly-optimal dependence of the query complexity in $1/\varepsilon$.
Anyway, since the factor $\log \log (1/\varepsilon)$ is not so large compared to others, the relation to the quantum algorithms for time-independent and generic time-dependent Hamiltonians, discussed in Section \ref{Sec:Algorithm_cost}, is maintained. 
Importantly, in the query complexity, the whole energy scales $\alpha t$ and $\gamma t$ are separated from the $\order{\log (1/\varepsilon)}$ term; the query complexity is usually much smaller than that of the truncated Dyson-series algorithm, and close to that of the qubitization technique.
The number of ancillary qubits, which scales as $\order{\log \log (1/\varepsilon)}$ for an allowable error $\varepsilon$, also lies between those for these algorithms.

We also mention about the complexity of the oracles for the cases $\norm{H_m} \leq h e^{-|m|/\zeta}$.
In contrast to the cases $H_m=0$ ($|m|>m_\imax$), we require a quantum circuit which embeds the function $\alpha_j^m$ to the amplitude of a quantum state with $\order{j_\imax (\gamma t + \log (1/\varepsilon))}$ or $\order{j_\imax (\gamma/\omega + \log (\omega t/\varepsilon))}$ degrees of freedom, by the oracle $G_\mr{coef}^{4l_\imax}$ defined by Eq. (\ref{Aeq3:G_coeff_expo}).
While integrable functions are efficiently implemented \cite{Grover2002-ry}, the number of elementary gates for the worst cases amounts to the dimension of the quantum state, i.e. $C^{4l_\imax} \in$ $\order{j_\imax (\gamma t + \log (1/\varepsilon))}$ or $\order{j_\imax (\gamma/\omega + \log (\omega t/\varepsilon))}$.
While the preparation of oracles is relatively difficult compared to the cases $H_m=0$ ($|m|>m_\imax$), we emphasize that it is much easier than that of the truncated Dyson-series algorithm.
In the truncated Dyson-series algorithm, we need more or less an oracle that embeds the information of time-dependent Hamiltonians at every discretized time into a quantum state.
Such an oracle is exemplified by Eq. (\ref{Eq6C:oracle_dyson}) for generic Hamiltonians \cite{Low2018-dyson}.
For time-dependent LCU Hamiltonians or sparse-access Hamiltonians, it requires a quantum state holding all the coefficients $\alpha_j(t)$ at every discretized time $t$ as an oracle \cite{Kieferova2019-dyson}. 
The number of elementary gates per oracle can be proportional to that of the discretized time, $\order{j_\imax \omega \gamma t/ \alpha \varepsilon}$, at the worst case.
Therefore, our algorithm also improves the complexity of the oracles as well as the query complexity, compared to the truncated Dyson-series algorithm.

\subsection{Example}\label{Asubsec3:example_expo}
For the cases $\norm{H_m} \leq h e^{-|m|/\zeta}$, we adopt the assumption of block-encoding as Eq. (\ref{Aeq3:block_encoding_H_m_expo}), in contrast to the cases $H_m=0$ ($|m|>m_\imax$).
To show that it is reasonable and versatile in condensed matter physics and quantum chemistry, we provide a simple example.

We consider an $N$-site Fermi-Hubbard model $H_\mr{Hub}$, given by Eq. (\ref{Eq8:Fermi_Hubbard}).
While we consider an idealized laser light which has a constant amplitude in Section \ref{Subsec:example_Fermi_Hubbard}, we focus on a rather realistic case where Gaussian wave packets of lights are shone to materials \cite{Murakami2021-oy}.
Then, if the time scale of the wave packets is sufficiently larger than the light with the frequency $\Omega$, the external drive $H_\mr{ext}(t)$ is well described by
\begin{equation}
    H_\mr{ext}(t) = \left( \sum_{n \in \bbZ} e^{-(t-(n+1/2)T)^2/2 \tau^2} \sin \Omega t \right) \sum_{x,\sigma} V_x \hat{n}_{x\sigma},
\end{equation}
where peaks of the wave packets are located at $t = (n+1/2)T$.
The period $T$ is designated by $T=p(2\pi/\Omega)$ with the number of the waves of the light for each wave packet $p \in \bbN$.

Let us organize the block-encoding required for simulation.
By carefully computing Fourier components of $H(t)=H_\mr{Hub}+H_\mr{ext}(t)$, we arrive at $H_0=H_\mr{Hub}$ and 
\begin{equation}
    H_m = i (-1)^{p+m} \mr{sgn}(m) A_m \sum_{x,\sigma} V_x \hat{n}_{x\sigma}, \quad 
    A_m =
    \frac{\omega \tau}{\sqrt{2\pi}} e^{-\frac12 (p^2+m^2)(\omega \tau)^2} \sinh \left\{ p |m| (\omega \tau)^2 \right\},
\end{equation}
for $m \neq 0$. We define $\mr{sgn}(m)$ by $\mr{sgn}(m)=m/|m|$ for $m \neq 0$ and $\mr{sgn}(0)=0$.
Since $A_m \geq 0$ rapidly decays as $e^{-\order{m^2}}$, this Hamiltonian is suitable for our algorithm, satisfying $\norm{H_m} \leq h e^{-|m|/\zeta}$.
This satisfies the assumption of Eq. (\ref{Aeq3:block_encoding_H_m_expo}) with non-negative coefficients $\alpha_j^m$ by choosing $j_\imax = 3$;
\begin{eqnarray}
    M_{1} &=& \frac{H_\mr{Hub}}{2\sum_x \epsilon_x +UN}, \quad \alpha_{1}^m = \left( 2\sum_x \epsilon_x +UN \right) \delta_{m0}, \\
    M_{2} &=& + i \frac{\sum_{x,\sigma} V_x \hat{n}_{x\sigma}}{2 \sum_x V_x}, \quad \alpha_{2}^m = 2 \sum_x V_x A_m \delta_{(-1)^{p+m}\mr{sgn}(m), +1}, \\
    M_{3} &=& - i \frac{\sum_{x,\sigma} V_x \hat{n}_{x\sigma}}{2 \sum_x V_x}, \quad \alpha_{3}^m = 2 \sum_x V_x A_m \delta_{(-1)^{p+m}\mr{sgn}(m), -1}. 
\end{eqnarray}
The block-encoding for $\{ M_j \}$ can be composed by Eqs. (\ref{Eq8:Oracle_gate_Fermi_Hubbard})-(\ref{Eq8:Oracle_state_external}) in Section \ref{Subsec:example_Fermi_Hubbard}.
Therefore, by resorting to our algorithm for the adiabatic-like cases, we can efficiently simulate the response to a realistic wave packet of laser light with the number of ancillary qubits and the query complexity provided in Appendix \ref{Asubsec3:algorithm_comparison_expo}.

\section{Efficient algorithm for high-frequency regimes with $\omega \simeq \alpha, \gamma$}\label{Asec:high_frequency}
In Section \ref{Subsec:algorithm_long_time}, we concentrate on the generic long-time regime with the frequency $\omega \in \order{N^0}$ for the number of sites, while the whole energy scales $\alpha,\gamma$ are given by $\poly{N}$.
Here, we show its reasonableness by showing that quantum simulation of broad time-periodic Hamiltonians under the frequency comparable to the whole energy scale can be almost attributed to that of time-independent Hamiltonians.

As a generic setup for time-periodic Hamiltonians, we assume the locality [= the system involves at-most $\order{1}$-body interactions] and the extensiveness. 
The latter one means that the energy scale per site, represented by $\lambda$, is bounded.
In our definitions, the scale $\lambda$ is approximated by $\order{\alpha/N, \gamma/N}$.
When the ratio of the frequency $\omega$ to the local energy scale $\lambda$ is sufficiently large as $\omega/\lambda \gg 1$, we can apply the perturbative expansion in $(\omega/\lambda)^{-1}$, called the high-frequency expansion \cite{Abanin2015-zg,Kuwahara2016-yn,Mori2016-tp,Abanin2017-li,Abanin2017-zs}.
Recent developments in generic local and extensive Floquet systems have revealed the accurate upper bound on the error as follows \cite{Kuwahara2016-yn};
\begin{eqnarray}
    \norm{U(nT) - e^{-i H_\mr{FM}^{l} nT}} \leq C_1 \lambda N nT e^{-C_2 l_0} + C_3 n N (\lambda T)^{l+2} (l+1)!. \label{Aeq4:high_frequency_precision}
\end{eqnarray}
In the above inequality, $C_1$, $C_2$, and $C_3$ represent some positive constants, and the integer $l_0 \in \order{\omega/\lambda}$ provides the optimal truncation order of the perturbation theory, while the actual truncation order $l$ ($\leq l_0$) is optional.
The time-independent Hamiltonian $H_\mr{FM}^l$ is the $l$-th order Floquet-Magnus expansion, given by
\begin{eqnarray}
    H_\mr{FM}^{l} = \sum_{l^\prime = 0}^l H_\mr{FM}^{(l^\prime)}, \quad H_\mr{FM}^{(0)} = H_0, \quad
    H_\mr{FM}^{(1)} = \frac{1}{2i T} \int_0^T \dd t_1 \int_0^{t_1} \dd t_2 [H(t_1),H(t_2)],
\end{eqnarray}
where each $l^\prime$-th order term in $\omega/\lambda$, represented by $H_\mr{FM}^{(l^\prime)}$, involves $l^\prime$-fold multi-commutators of $H(t)$ (See Ref. \cite{Kuwahara2016-yn} for the explicit formula).

Based on the above formalism, we see that the long-time dynamics at $t=(n+\delta)T \gg T$ with $n \in \bbN$ and $\delta \in [0,1)$ can be computed in a trivial way when $\omega$ is comparable to $\alpha,\gamma \in \poly{N}$.
In this regime, the integer $l_0$ is $\order{N}$ due to the local energy scale $\lambda \in \order{\alpha/N, \gamma/N}$.
Then, we simulate the dynamics $U(t)=U(\delta T) U(nT)$ with approximating the latter one $U(nT)$ by Eq. (\ref{Aeq4:high_frequency_precision}).
We choose the truncation order $l$ by $\order{N^0}$.
Since $H(t)$ generally has $\poly{N}$ local terms, the Floquet-Magnus expansion $H_\mr{FM}^{l}$ for such $l \in \order{N^0}$ can be identified with $\poly{N}$-time classical computation, and $H_\mr{FM}^{l}$ is also composed of $\poly{N}$ local terms.
Therefore, we can apply the qubitization technique to the implementation of $\exp (-i H_\mr{FM}^l nT)$ with the cost designated by Eq. (\ref{Eq2B:qubitization_cost}).
Since the right hand side of Eq. (\ref{Aeq4:high_frequency_precision}) is bounded by $\order{n N^{-l-1}}$, the replacement of $U(nT)$ by $\exp (-i H_\mr{FM}^l nT)$ with an allowable error $\varepsilon$ is valid as long as
\begin{equation}
    n \leq \mr{Const.} \times \varepsilon N^{l+1}.
\end{equation}
We usually employ a polynomial accuracy $\varepsilon \in \order{N^{-\nu_1}}$ in quantum computation.
The assumption for the frequency, represented by $\omega \in \order{N^{\nu_2}}$, implies the period $T \in \order{N^{-\nu_2}}$. 
Therefore, by properly choosing the $\order{N^0}$ truncation order $l \geq \nu_1 + \nu_2 -1$, we can efficiently simulate $U(nT)$ for arbitrary polynomial time $nT \in \order{N^{l+1-\nu_1-\nu_2}}$ with the qubitization technique for time-independent Hamiltonians.
The remaining micromotion $U(\delta T)$ can be implemented with our protocol for the adiabatic-like regime (See Section \ref{Subsec:algorithm_adiabatic}), but it does not affects the computational cost due to $\delta T \ll nT$.

In short, the dynamics under time-periodic Hamiltonians $H(t)$ with $\omega \sim \alpha, \gamma$ can be simulated with the time-independent Hamiltonian approaches.
As a result, it is reasonable to concentrate on the cases where $\omega$ is negligible compared to $\alpha$ and $\gamma$, as we do in Section \ref{Subsec:algorithm_long_time}.

\end{document}